\documentclass[%
 reprint,
 superscriptaddress,
 amsmath, amssymb,
 aps,
 prd,
]{revtex4-2}

\usepackage{mathtools}              
\usepackage{url}                    
\usepackage{graphicx}               
\usepackage{dcolumn}                
\usepackage{bm}                     
\usepackage{hyperref}               
\hypersetup{linkcolor=red, citecolor=blue, urlcolor=cyan}
\usepackage{xcolor}                 
\def\aap{A\&A}
\def\apj{ApJ}
\def\mnras{MNRAS}
\def\apjl{ApJL}

\def\prd{PRD}

\def\prl{PRL}

\def\ssr{SSRv}

\def\araa{ARAA}
\def\prc{PRC}
\def\jcap{JCAP}

\DeclareMathOperator{\sgn}{sgn}

\begin{document}
\preprint{APS/123-QED}

\title{Constraints on QCD-based equation of state of quark stars from neutron star maximum mass, radius, and tidal deformability observations}

\author{João V. Zastrow}
\email{joaovzastrow@gmail.com}
\affiliation{Universidade do Estado de Santa Catarina, Joinville, 89219-710, SC, Brazil}

\author{Jonas P. Pereira}
\email{jonas.pereira@unb.br}
\affiliation{Institute of Physics \& International Center of Physics, University of Brasilia, 70297-400, Brasilia, Federal District, Brazil}
\affiliation{Programa de Pós-Graduação em Astrofísica, Cosmologia e Gravitação (PPGCosmo), Federal University of Espírito Santo, Vitória-ES, 29075-910, Brazil}
\affiliation{Departamento de Astronomia, Instituto de Astronomia, Geofísica e Ciências Atmosféricas (IAG), Universidade de São Paulo, São Paulo, 05508-090, Brazil}
\affiliation{Nicolaus Copernicus Astronomical Center, Polish Academy of Sciences, Warsaw, 00-716, Poland}

\author{Rafael C. R. de Lima}
\affiliation{Universidade do Estado de Santa Catarina, Joinville, 89219-710, SC, Brazil}

\author{Jorge E. Horvath}
\affiliation{Departamento de Astronomia, Instituto de Astronomia, Geofísica e Ciências Atmosféricas (IAG), Universidade de São Paulo, São Paulo, 05508-090, Brazil}

\date{\today}

\begin{abstract}
\noindent Neutron stars (NSs), the densest known objects composed of matter, provide a unique laboratory to probe whether strange quark matter is the true ground state of matter. We investigate the range of parameters relevant for the equation of state of strange stars taking into account a quantum chromodynamics (QCD)-informed model. The sampling of the parameter space was performed using quasi-random Latin hypercube sampling, ensuring uniform coverage. The parameters are related to the energy density difference between quark matter and the QCD vacuum, the strength of strong interactions, and the gap parameter for color superconductivity. To restrict them, we incorporate observational constraints on the maximum mass of NSs (from both observations of binary systems and NS mergers), the radii of $1.4$~M$_{\odot}$ NSs (inferred from gravitational wave (GW) and electromagnetic observations), and tidal deformations (from GW170817). Our results indicate that strong quark interactions play a crucial role in the description of NSs, with a minimum deviation of at least $20\%$ from the behavior of free quarks. Additionally, we find that color superconductivity is relevant, and derive a maximum gap parameter reaching approximately $230$~MeV for a strange quark mass of $100$~MeV. Furthermore, we determine that the surface-to-vacuum energy density jump for quark stars lies within the range $(1.1-2.2)$~$\rho_{\rm{sat}}$, where $\rho_{\rm{sat}} \simeq 2.7 \times 10^{14}$~g~cm$^{-3}$ is the nuclear saturation density. As a byproduct of the observational constraints, we find that the radius of a $1.4$~M$_{\odot}$ quark star lies in the range $(10.0-12.3)$~km, while its dimensionless tidal deformability falls within the interval $(270-970)$. In addition, all the above constraints are fully consistent with the small mass and radius inferred electromagnetically for the central compact object XMMU J173203.3-344518 (NS) in the supernova remnant HESS J1731-347, with $M = 0.77^{+0.20}_{-0.17}$~M$_{\odot}$ and $R = 10.40^{+0.86}_{-0.78}$~km. The above results may offer important inputs for studies on quark and, in a complementary way, hybrid stars, serving as parameter references for more detailed investigations into their astrophysical signatures, including tidal deformations, cooling curves, quasi-normal modes, and ellipticities. We briefly discuss these implications.
\end{abstract}

\maketitle

\section{Introduction}
\label{sec:introduction}

Neutron stars (NSs), the densest extended objects in the universe, offer a unique window into regimes of matter inaccessible to terrestrial particle accelerators. With densities spanning up to fifteen orders of magnitude, they serve as natural laboratories for studying superdense matter. While nuclear physics can confidently probe densities up to nuclear saturation \cite{2013ApJ...773...11H}, our understanding beyond this limit remains much more uncertain \cite{2024ApJ...971L..19R}. This is where other NSs observables enter to guide the investigation \cite{2024ApJ...971L..19R, 2020GReGr..52..109C}.

Macroscopic NS observables such as masses, radii, and moments of inertia provide crucial constraints on the nature of ultradense matter. The interpretation of other probes (for example, the X-ray bursts related to the radii of the sources \cite{2016ARA&A..54..401O}) are still subject to uncertainties and discussion of fundamental issues. With the rise of multimessenger astronomy \cite{2017ApJ...848L..12A}, new observables have also emerged, including tidal deformations/deformabilities, ellipticities, and quasi-normal modes, among others. When combined \cite{2020PhRvD.101l3007L}, these observables may light on one of the most fundamental questions about NSs: what lies within their interiors?

Quantum Chromodynamics (QCD) suggests that at extreme densities, hadronic matter may dissolve into its fundamental constituents---quarks. It is quite possible that, if at all, ``neutron'' stars could harbor these phases. The exploration of this regime started with simple free-quark models, but today we know that various phases of quark matter could exist, such as color superconductivity in the two-flavor ($u, d$) phase (2SC) or, at higher densities, the color-flavor-locked (CFL) phase involving all three light quarks ($u, d, s$) \cite{2001ARNPS..51..131A, 2008RvMP...80.1455A}. Given that quarks are the building blocks of hadrons, it has been considered that, if deconfinement occurs, quark matter could even represent the most absolutely stable form of matter. This hypothesis has been put forth by Itoh \cite{1970PThPh..44..291I}, Bodmer \cite{1971PhRvD...4.1601B} and Witten \cite{1984PhRvD..30..272W} and is widely known as the {\it strange matter} hypothesis. In this picture, NSs would not be composed of nucleonic matter, but might actually be quark stars consisting of deconfined ($u, d, s$) quark matter because of the ground state hypothesis. The latter is motivated by presence of strange quark, which arises naturally due to weak interactions converting down quarks into strange quarks \cite{2008RvMP...80.1455A}, and after energy minimization, it is clear that the inclusion of a third quark flavor allows a more efficient redistribution of quarks across the Fermi levels, lowering the system's total energy. However, it is impossible to establish from first principles whether this energetic advantage is enough to make strange matter absolutely stable, a hypothesis in search of observational/experimental confirmation.

Strange quark stars \cite{1986A&A...160..121H, 1986ApJ...310..261A, 1989MNRAS.241...43B}, if they exist, could exhibit distinct structural properties compared to traditional NSs \cite{2021PhRvD.104d3002L, 2021IJMPD..3050016H}. Observations indicate that NSs can have masses larger than $2$~M$_{\odot}$ (PSR J0348+0432 has a mass of $2.01 \pm 0.04$~M$_{\odot}$ \cite{2013Sci...340..448A}, PSR J0740+6620 $2.08 \pm 0.07$~M$_{\odot}$ \cite{2021ApJ...915L..12F} and PSR J0952-0607 $2.35 \pm 0.17$~M$_{\odot}$ \cite{2022ApJ...934L..17R}). Their radii typically range from $10$~km to $14$~km, depending somewhat on their mass \citep{2019ApJ...887L..24M, 2019ApJ...887L..21R, 2020PhRvD.101f3007E, 2021ApJ...918L..28M, 2021ApJ...918L..27R}. When several observations are combined, for both $1.4$~M$_{\odot}$ and $2$~M$_{\odot}$ stars, the median radius is around $12$~km, with an uncertainty of approximately $\pm 5\%$ at the $95\%$ confidence level (CL) \cite{2024ApJ...971L..19R}, i.e., they are surprisingly close in spite of the large mass difference. Theoretical models suggest that quark stars with similar masses would have similar radii, but potentially smaller radii for masses $\leq 1$~M$_{\odot}$ due to the higher density of deconfined quark matter. Finally, universal relations\cite{2013Sci...341..365Y, 2014PhRvD..90f3010Y, 2017PhR...681....1Y} for quark stars differ from those of hadronic stars \cite{2025arXiv250311515L}, providing an additional way to distinguish between these two classes of compact objects, but these tests would need higher precision and firm detection of objects $\leq 1$~M$_{\odot}$.

The celebrated gravitational wave (GW) event GW170817 \cite{2018PhRvL.121p1101A} provided crucial insights into the internal composition of NSs by means of the measurement of the tidal deformability parameter \cite{2008ApJ...677.1216H, 2009PhRvD..80h4035D, 2010PhRvD..81l3016H}. The relatively small value of tidal deformability inferred from GW170817 \cite{2018PhRvL.121p1101A, 2019PhRvX...9a1001A} suggests that NSs have a more compact structure than previously expected, which is in agreement with the predictions of some quark matter EOSs because it slightly favors softer EOSs \cite{2020PhRvD.101f3007E}. The lower tidal deformability might naturally be explained if at least one of the merging stars contained a quark matter core or was entirely a quark star \cite{2021PhRvD.104d3002L}. This feature makes them slightly more compatible with the observational constraints from GW170817 than traditional NSs described solely by hadronic EOSs \cite{2018PhRvL.120q2703A}. Although these results do not conclusively prove the existence of quark stars, they provide motivation for further exploration of them. Future GW detections, particularly those involving heavier NS mergers, will be essential in testing these hypotheses and refining the constraints on the EOS of ultradense matter \cite{2020NatPh..16..907A}.

The first free-quark models of cooling suggested that strange quark stars may cool more rapidly than hadronic NSs due to the presence of enhanced neutrino emission channels \cite{2006NuPhA.777..497P}. In quark matter, the direct Urca process---where quarks directly participate in weak interactions---is highly efficient and can result in significantly faster cooling compared to standard hadronic NSs \cite{2008RvMP...80.1455A}. However, the presence of color superconducting phases in quark stars, such as the CFL phase, could suppress certain cooling mechanisms, leading to slower cooling rates in some scenarios \cite{1991PhRvD..44.3797H, 1997JPhG...23.2029S}. Theoretical models of quark star cooling predict a broader range of possible temperature evolutions compared to purely hadronic NSs \cite{2005PhRvC..71d5801G, 2006NuPhA.777..497P, 2015SSRv..191..239P}, making observational constraints particularly relevant \cite{2005PhRvC..71d5801G}. Observations of NS surface temperatures from X-ray missions such as Chandra and XMM-Newton have provided constraints on cooling models. Some sources, such as the NS in Cassiopeia A, exhibit cooling rates that are difficult to reconcile with purely hadronic models, potentially hinting at exotic components such as quark matter \cite{2013A&A...555L..10S}. For further examples, see \cite{2008RvMP...80.1455A} and references therein. Conversely, the observation of relatively warm NSs at older ages places constraints on the extent to which enhanced cooling processes, such as those expected in strange quark stars, can operate. Future observations, particularly with improved thermal modeling, more precise temperature measurements, and an extended reliable sample of extracted effective temperatures, will be important in determining whether strange quark matter plays a role in NS interiors.

As it stands, despite the abundance and relative precision of current observables, they do not yet provide definitive evidence for the existence of quark stars. What can be done, however, is to impose constraints on specific models, ensuring their compatibility with observations. In this context, it is desirable to take into account as closely as possible what QCD suggests about superdense matter. Unfortunately, first-principles calculations are not feasible in the density regime relevant to NSs. Nevertheless, effective quark models offer valuable insights into the behavior of superdense matter. One such model, developed by Alford and collaborators \cite{2005ApJ...629..969A, 2018ApJ...860...12P} (motivated by previous QCD analysis; see, e.g., \cite{2001PhRvD..64g4017A, 2001PhRvD..63l1702F, 2002PhRvD..66g4017L}), incorporates aspects of asymptotically large densities while introducing corrections to match the density range found in NSs. In this framework, the squared speed of sound deviates from the conformal $1/3$ (valid for asymptotically large energies \cite{2015PhRvL.114c1103B, 2020arXiv201110940K}), being modified by strong interaction effects, the strange quark mass, and quark superconductivity. Astrophysical and nuclear-physics constraints already exist when considering modifications to the EOS due to quark degrees of freedom within hadrons \cite{2024PhRvL.132z2701K}, but these constraints could be refined by incorporating additional observational data or reconsidering certain assumptions. The inference of an NS with a small mass and radius in the supernova remnant HESS J1731-347 \cite{2022NatAs...6.1444D, 2024ApJ...967..159D, 2023A&A...672L..11H} motivates us to explore the possibility that such objects could actually be strange stars. This is precisely one of the main goals of our work: to constrain the parameters of an effective quark EOS suitable for NS densities, while incorporating all existing astrophysical constraints.

We take into account multiple observational and theoretical constraints, including maximum masses (both observationally inferred by pulsar binaries and theoretically constrained by GW170817---the event with the highest signal-to-noise ratio of merging neutron stars), radii (determined from electromagnetic observations), and tidal deformations (also primarily from GW170817). We focused on an initial exploration of the relevant EOS parameter space using quasi-random Latin hypercube sampling \cite{McKay1979LatinHypercube}, which is simpler and less computationally expensive than a complete Bayesian analysis, although it does not provide confidence intervals for the parameters. We show that this approach enables valuable inferences about the strength of strong interactions, the gap parameter associated with quark superconductivity, and the effective bag parameter, which is closely related to the surface-to-vacuum density jump of quark stars. These results are particularly relevant, as they provide a foundation for more detailed studies incorporating these parameters, such as analyses of quasi-normal modes, GW emissions, and NS cooling. 

Our work differs from other existing ones \cite{2024JCAP...11..038W, 2018PhRvD..97h3015Z, 2021ApJ...917L..22M, 2021PhRvD.103f3018Z, 2021MNRAS.506.5916L} in some important aspects. In \cite{2018PhRvD..97h3015Z}, the authors explore a reduced set of sampling points in the parameter space. In \cite{2021PhRvD.103f3018Z}, a dimensionless form of the EOS originally proposed in \cite{2005ApJ...629..969A, 2018ApJ...860...12P} is employed. While that formulation incorporates various quark matter phases, it does not explicitly account for the 2-flavor and 3-flavor surfaces, which are essential for accurately identifying quark and hybrid stars and for properly delineating the parameter region to be sampled. Further details are provided at the end of Section~\ref{sec:quark_matter_eos}. In addition, \cite{2021PhRvD.103f3018Z} considers only a fixed sign of their $\lambda$ parameter (which is associated with our $a_2$ coefficient; see Eq.~\eqref{eq:omega_quark_matter_3f}), whereas we allow it to take either sign in light of recent multimessenger constraints \cite{2024PhRvL.132z2701K}. In practice, this modifies the space of parameters (to be sampled) considerably. Other works perform a Bayesian analysis using the same EOS model as the one used here, employing only tidal deformability measures from multiple GW events \cite{2024JCAP...11..038W}, or only the events GW170817 and GW190425 \cite{2021ApJ...917L..22M}, or even only measurements of PSR J0030+0451 from NICER \cite{2021MNRAS.506.5916L}. To the best of our knowledge, we could not find quark star studies incorporating multiple constraints from electromagnetic and GW observations in a Bayesian inference approach. Given the scarcity of data, it is reasonable to include as many sets as possible, in an analogous fashion as currently done in cosmological studies.

We structure our work as follows. Section~\ref{sec:spherical_stars} presents the relevant NS structure equations. Section~\ref{sec:coalescence} discusses tidal interactions in binary systems for both the adiabatic and perfect-fluid cases. In Section~\ref{sec:quark_matter_eos}, we examine the EOS adopted for strange quark matter. Appendix~\ref{ap:QS_parameters_constraints} provides additional details on the region of parameter space compatible with the existence of quark stars. Section~\ref{sec:methods} outlines the methods used to solve the relevant equations, incorporating observational constraints on mass, radius, and tidal deformations. The results, including the available parameter space, are presented in Sec.~\ref{sec:results} and discussed in Sec.~\ref{sec:discussion_and_conclusions}. Finally, we summarize our main results in Section~\ref{sec:summary}. Unless stated otherwise, we use geometric units and adopt a metric signature of $(-, +, +, +)$.

\section{Spherically symmetric stars}
\label{sec:spherical_stars}

The relativistic equations that describe the internal structure of spherically symmetric stars in hydrostatic equilibrium, known as Tolman-Oppenheimer-Volkoff (TOV) equations, are obtained by solving Einstein's equations \cite{1939PhRv...55..364T, 1939PhRv...55..374O, Misner2017Gravitation, Schutz2009FirstCourseGR} and are shown here for clarity. Using the metric of a static and spherically symmetric spacetime, given by

\begin{equation}
    ds^2 = - e^{\nu(r)} dt^2 + e^{\lambda(r)} dr^2 + r^2 \left( d \theta^2 + \sin^2 \theta d \phi^2 \right) ,
\label{eq:spherically_symmetric_line_element}
\end{equation}

\noindent assuming an energy-momentum tensor for a perfect fluid,

\begin{equation}
    T_{\mu \nu} = (\rho + p) U_{\mu} U_{\nu} + p g_{\mu \nu} ,
\label{eq:perfect_fluid_stress_energy_tensor}
\end{equation}

\noindent and defining the mass function

\begin{equation}
    m(r) \coloneqq \dfrac{1}{2} r \left( 1 - e^{-\lambda} \right) ,
\label{eq:mass_function_definition}
\end{equation}

\noindent it is possible to find the structure equations governing equilibrium in NSs. Applying the conservation of energy-momentum and Einstein's equations, for the case of static, local observers ($U_{\mu} \propto \delta^{0}_{\mu}$) we obtain the TOV equation and the mass conservation as:

\begin{equation}
    \dfrac{dm}{dr} = 4 \pi r^2 \rho ,
\label{eq:tov_dm_dr}
\end{equation}

\begin{equation}
    \dfrac{d \nu}{dr} = \dfrac{2 (m + 4 \pi r^3 p)}{r (r - 2m)} ,
\label{eq:tov_dnu_dr}
\end{equation}

\begin{equation}
    \dfrac{dp}{dr} = -\dfrac{(\rho + p)(m + 4 \pi r^3 p)}{r (r - 2m)} .
\label{eq:tov_dp_dr}
\end{equation}

With Eqs.~\eqref{eq:tov_dm_dr} and \eqref{eq:tov_dp_dr}, together with an EOS in the form $\rho = \rho(p)$, we have three equations involving three variables dependent on $r$: $m$, $\rho$ and $p$.

\section{NS Coalescence and tidal effects}
\label{sec:coalescence}

The coalescence of compact objects in binary systems is a clear-cut source of GWs, detected by laser interferometers such as the detectors of the LIGO-Virgo-KAGRA collaboration (LVK) \cite{LVKCollaboration}, and it carries information about the internal structure of the objects involved.

We are interested in binary NS systems, where the stars orbit the center of mass of the system, emitting GWs, and consequently losing orbital energy and angular momentum. The evolution of these binary systems can be divided into three phases: the coalescence phase, known as inspiral; the merging of the compact objects, known as merger; and the final phase, known as postmerger, where the resulting compact object, which can be a NS or a black hole, has its disturbances suppressed by emission of GWs \cite{2019arXiv190500408A, 2021GReGr..53...27D}.

This work focuses on the initial stage of the inspiral phase, where the distance between the NSs is much larger than the radius of the individual objects, and the tidal effect is non-negligible but approximately static. Thus, the system can be described using a perturbative formalism known as the post-Newtonian formalism, where a static perturbative correction is necessary and sufficient to explain the tidal deformations of NSs \cite{2019arXiv190500408A, 2020GReGr..52..109C}.

In this formalism, the dimensionless tidal deformability of a compact object is \cite{2020GReGr..52..109C}

\begin{equation}
    \Lambda = \frac{2}{3} k_2 C^{-5} ,
\label{eq:Lambda_k2_relation}
\end{equation}

\noindent where $k_2$ is the Love number, which describes the deformations of perfectly elastic bodies, and $C \coloneqq M / R$ is the compactness of the object.

The combined deformability of the binary system $\tilde{\Lambda}$ affects the phase of the GW signal and can be measured by the LVK interferometers. Since the deformability changes with the NS's EOS, this measurement can be used to probe the EOSs observationally \cite{2020GReGr..52..109C}.

\subsection{Tidal Love number}

By manipulating the equations resulting from the perturbative part of Einstein's equations up to first order, one obtains the differential equation needed to calculate the Love number, given by \cite{2008ApJ...677.1216H, 2009PhRvD..80h4035D}

\begin{equation}
    y' = - C_0 r + \left( \frac{1}{r} - C_1 \right) y - \frac{1}{r} y^2 ,
\label{eq:y_perturbation_equation}
\end{equation}

\noindent where $C_0$ and $C_1$ are given by:

\begin{equation}
    C_0 = e^{\lambda} \left[ -\frac{6}{r^2} + 4 \pi (\rho + p) \frac{d \rho}{dp} + 4 \pi (5 \rho + 9 p) \right] - \left( \nu' \right)^2 ,
\label{eq:C0_coefficient}
\end{equation}

\begin{equation}
    C_1 = \frac{2}{r} + e^{\lambda} \left[ \frac{2 m}{r^2} + 4 \pi r (p - \rho) \right] .
\label{eq:C1_coefficient}
\end{equation}

By solving Eq.~\eqref{eq:y_perturbation_equation} in the limit $r \rightarrow 0$, the following boundary condition is obtained \cite{2008ApJ...677.1216H}

\begin{equation}
    y(r \rightarrow 0) \simeq 2 .
\label{eq:y_init_value}
\end{equation}

For EOSs that have phase transitions or finite densities for zero pressures, there are discontinuities in the function $\rho(p)$. We see that the derivative $d \rho / dp$ is present in the $C_0$ expression, given by Eq.~\eqref{eq:C0_coefficient}, and therefore a discontinuity in $\rho(p)$ causes a singularity in $C_0$, and consequently in $y'$, according to Eq.~\eqref{eq:y_perturbation_equation}.

Therefore, we need to add a correction $\Delta y$, due to the discontinuity of $\rho(p)$, to the solution $y$, and separate the numerical integration of Eq.~\eqref{eq:y_perturbation_equation} into two steps, before and after the discontinuity at $\rho(p)$. When the discontinuity at $\rho$ is due to a finite density at the surface of the star, as in the case of quark stars, the correction reduces to \cite{2023PhRvD.107l4016A}:

\begin{equation}
    y_s \coloneqq y(R + \epsilon) = y(R - \epsilon) + \Delta y(R) ,
\label{eq:y_of_R_expression}
\end{equation}

\begin{equation}
    \Delta y(R) = - \frac{4 \pi R^3 \rho(R - \epsilon)}{M} ,
\label{eq:delta_y_of_R_expression}
\end{equation}

\noindent where $\epsilon \rightarrow 0^{+}$. Finally, by calculating the asymptotic solution of Eq.~\eqref{eq:y_perturbation_equation} for $r \rightarrow \infty$ and comparing it with the multipole expansion of the metric, it is possible to obtain the explicit expression for the Love number, which depends on $C$ and $y_s$, and is given by \cite{2008ApJ...677.1216H, 2009PhRvD..80h4035D}

\begin{equation}
    \begin{split}
        k_2 &= \frac{8}{5} C^5 (1 - 2 C)^2 [2 + 2C (y_s - 1) - y_s] \\
        &\quad \times \big\{ 2 C [6 - 3 y_s + 3 C (5 y_s - 8)] \\
        &\quad + 4 C^3 [13 - 11 y_s + C (3 y_s - 2) + 2 C^2 (1 + y_s)] \\
        &\quad + 3(1 - 2 C)^2 [2 - y_s + 2 C (y_s - 1)] \ln{(1 - 2 C)} \big\}^{-1} .
    \end{split}
\label{eq:k2_expression}
\end{equation}

\section{Quark matter equation of state}
\label{sec:quark_matter_eos}

We present the quark matter EOS composed of free up, down, and strange quarks. We use the natural unit system, defined by $c = \hbar = 1$. Quark matter can be parameterized to a good approximation using the Landau thermodynamic potential density $\omega$ defined by \cite{2002PhRvD..66g4017L, 2018ApJ...860...12P, 2005ApJ...629..969A, 2008RvMP...80.1455A}

\begin{equation}
    \omega = - \frac{3 a_4}{4 \pi^2} \mu^4 + \frac{3 a_2}{4 \pi^2} \mu^2 + B ,
\label{eq:omega_quark_matter_3f}
\end{equation}

\noindent where $\mu \coloneqq (\mu_u + \mu_d + \mu_s) / 3$ is the quark chemical potential and $\mu_u \simeq \mu_d \simeq \mu_s$ is the chemical potential of the up, down, and strange quarks, respectively. This approximation is valid when the electron chemical potential can be neglected, which is the case in this work \cite{2018ApJ...860...12P, 2005ApJ...629..969A}. Furthermore, the free parameters $a_2$, $a_4$, and $B$ have the following physical meanings \cite{2018ApJ...860...12P, 2005ApJ...629..969A, 2008RvMP...80.1455A}:

\begin{itemize}
    \item $a_2$: for the CFL phase, this parameter corresponds to $m_s^2 - 4 \Delta^2$, where $m_s$ is the strange quark mass and $\Delta$ is the energy gap associated with the quark pairing. To also encompass other phases, we take $a_2$ as a phenomenological parameter;

    \item $a_4$: for the CFL phase, this parameter corresponds to $1 - c$, where $c$ is believed to be greater than $0.3$ (see, e.g., \cite{2005ApJ...629..969A} and references therein) and quantifies the (strong) interactions between quarks. To encompass non-CFL phases too, we take $a_4$ as a phenomenological parameter;

    \item $B$: is the effective bag constant, related to the critical density where the quark phase appears. It bears a similar meaning to the bag constant in the original MIT-bag model, where $a_2 = m_s^2$ and $a_4 = 1$. We treat $B$ as a phenomenological parameter.
\end{itemize}

It is simple to show that Eq.~\eqref{eq:omega_quark_matter_3f} leads to the expression for the density given by \cite{2018ApJ...860...12P}

\begin{equation}
    \rho = \frac{9 a_4}{4 \pi^2} \mu^4 - \frac{3 a_2}{4 \pi^2} \mu^2 + B .
\label{eq:rho_quarks_3f}
\end{equation}

Isolating $\mu$ in this expression and substituting it into Eq.~\eqref{eq:omega_quark_matter_3f}, knowing that $p = - \omega$ by the expression of the grand thermodynamic potential, we obtain

\begin{equation}
    \begin{split}
        p = &\frac{1}{3}(\rho - 4B) \\
        &- \frac{a_2^2}{12 \pi^2 a_4} \left[ 1 + \sgn(a_2) \sqrt{1 + \frac{16 \pi^2 a_4}{a_2^2}(\rho - B)} \right] .
    \end{split}
\label{eq:quark_eos_p_of_rho}
\end{equation}

Similarly, isolating $\mu$ in Eq.~\eqref{eq:omega_quark_matter_3f}, and substituting it into Eq.~\eqref{eq:rho_quarks_3f}, we obtain

\begin{equation}
    \rho = 3p + 4B + \frac{3 a_2^2}{4 \pi^2 a_4} \left[ 1 + \sgn(a_2) \sqrt{1 + \frac{16 \pi^2 a_4}{3 a_2^2}(p + B)} \right] .
\label{eq:quark_eos_rho_of_p}
\end{equation}

Thus, we have the quark matter EOS completely defined by Eqs.~\eqref{eq:quark_eos_p_of_rho} and \eqref{eq:quark_eos_rho_of_p}. Furthermore, we need the derivatives of these expressions, given by:

\begin{equation}
    \frac{dp}{d\rho} \equiv c_s^2 = \frac{1}{3} - \frac{2}{3} \sgn(a_2) \left( 1 + \frac{16 \pi^2 a_4}{a_2^2} (\rho - B) \right)^{-1/2} ,
\label{eq:quark_eos_dp_drho}
\end{equation}

\begin{equation}
    \frac{d\rho}{dp} = 3 + 2 \sgn(a_2) \left( 1 + \frac{16 \pi^2 a_4}{3 a_2^2} (p + B) \right)^{-1/2} .
\label{eq:quark_eos_drho_dp}
\end{equation}

Note that in the EOS given by Eq.~\eqref{eq:quark_eos_p_of_rho} $a_2$ appears squared, meaning that negative values are not inherently problematic. In principle, this could occur due to its limiting expression, $m_s^2 - 4\Delta^2$, if $\Delta$ is sufficiently large. Indeed, astrophysical constraints suggest that this can be the case \cite{2024PhRvL.132z2701K}. In this work, we will assume that $a_2$ can also be negative. For details about the region of parameter space compatible with quark stars, see Appendix~\ref{ap:QS_parameters_constraints}. 

Finally, we emphasize an important aspect of the parameters to be constrained. As shown in Eq.~\eqref{eq:quark_eos_rho_of_p}, the EOS---and consequently the solutions to GR---depend only on the combination $B$ and $a_2/a_4^{1/2}$, as already noted in \cite{2021PhRvD.103f3018Z}. However, when focusing on a specific class of stars (such as strange or hybrid stars), it becomes necessary to further restrict the parameter space, as detailed in Appendix~\ref{ap:QS_parameters_constraints}. The key point is that the flavor lines/surfaces---crucial for obtaining the precise region to be sampled---cannot be fully characterized using only $B$ and $a_2/a_4^{1/2}$ in general. At first glance, this might seem to contrast with the flavor lines shown in Fig. 3 of \cite{2021PhRvD.103f3018Z}. However, their lines are valid only for a fixed value of $a_4$, as is evident from their Eq. (11). This limitation implies that a complete and accurate description requires treating $(B, a_2, a_4)$ as independent parameters. Accordingly, we adopt this approach in our analysis, which enables us to derive consistent and physically meaningful constraints for strange stars.

\section{Methods}
\label{sec:methods}

We have developed a Python program to perform the numerical analysis of NSs, available at \cite{StarStructure}. The program has the necessary algorithms to numerically solve the system composed by the TOV and perturbation equations and calculate the properties of a family of stars.

There is also a script that uses these algorithms to perform an automatic analysis of the quark matter EOS. It generates a set of points in the parameter space of the EOS using quasi-random Latin hypercube sampling \cite{McKay1979LatinHypercube}. It then selects only the points within the volume allowed by the EOS and calculates the properties of each family of stars characterized by the corresponding point in the parameter space. Finally, it uses the observational data to filter the parameterizations, setting bounds on the free parameters of the EOS and the properties of the families of stars. All the EOSs kept satisfy the causality condition, $dp/d\rho \equiv c_s^2 < 1$, and all quark star parameters lie on the stable branch of the mass-radius relation, characterized by $\partial M/\partial \rho_c \geq 0$ \cite{2018ApJ...860...12P}, where $\rho_c$ denotes the central pressure.

\subsection{Observational data}
\label{subsec:data}

Now we describe the observational data used in the analysis. The chosen observables are the maximum mass of the family of stars ($M_{\rm{max}}$), the radius of the $1.4$~M$_{\odot}$ star ($R_{1.4 M_{\odot}}$), and the deformability of the $1.4$~M$_{\odot}$ star ($\Lambda_{1.4 M_{\odot}}$), whose limits are in Tab.~\ref{tab:NSs_observables_limits}, discussed next.

\begin{table}[ht]
\caption{Limits on the observables of NSs \cite{2022ApJ...934L..17R, 2018ApJ...852L..25R, 2023NatCo..14.8352P, 2017PhRvL.119p1101A}.}
\label{tab:NSs_observables_limits}
    \begin{center}
    \begin{tabular}{c|c|c}
         \hline \hline $M_{\rm{max}}$ & $R_{1.4 M_{\odot}}$ & $\Lambda_{1.4 M_{\odot}}$ \\
         $[$M$_{\odot}]$ & $[$km$]$ & $[$dimensionless$]$ \\
         \hline $(2.13, 2.33)$ & $(10.00, 13.25)$ & $(0, 970)$ \\
         \hline
    \end{tabular}
    \end{center}
\end{table}

The lower limit for the maximum mass of NSs is $M_{\rm{max}} > 2.13$~M$_{\odot}$ at $2 \sigma$ CL, a value taken from \cite{2022ApJ...934L..17R}. This lower limit is obtained using mass estimates of NSs in pulsar-white dwarf binaries and ``spider'' binaries (pulsars that have a companion star in a close orbit). The upper limit for the maximum mass of NSs is $M_{\rm{max}} < 2.33$~M$_{\odot}$ also at $2 \sigma$ CL, taken from \cite{2018ApJ...852L..25R}. It is obtained using data from the GW event GW170817 and the gamma-ray burst event GRB 170817A.

The second observable is the radius of the most common star ($R_{1.4 M_{\odot}}$), the one with mass $1.4$~M$_{\odot}$~\footnote{Note that it is not strictly ``canonical'', as the full distribution spans a range of values \cite{2023Univ...10....3R}.}. Several works encompass different models for the EOS, data, and statistical algorithms, leading to a range of results for the limits of the radius of the $1.4$~M$_{\odot}$ star. The studies summarized in Table 1 of \cite{2023NatCo..14.8352P} were used as a basis for this work. Thus, to use more conservative constraints, we take the lowest limit for the radius, $R_{1.4 M_{\odot}} > 10.00$~km, and its maximum limit, $R_{1.4 M_{\odot}} < 13.25$~km.

Finally, we have the tidal deformability of the $1.4$~M$_{\odot}$ star ($\Lambda_{1.4 M_{\odot}}$), estimated by the LIGO-Virgo collaboration using the GW event GW170817. To be more conservative, we take its highest limit, $\Lambda_{1.4 M_{\odot}} < 970$ (at $2 \sigma$ CL). It is obtained through a deformability expansion around the mass $1.4$~M$_{\odot}$ \cite{2017PhRvL.119p1101A}.

\section{Results}
\label{sec:results}

\begin{figure}
    \centering
    \includegraphics[scale=0.57]{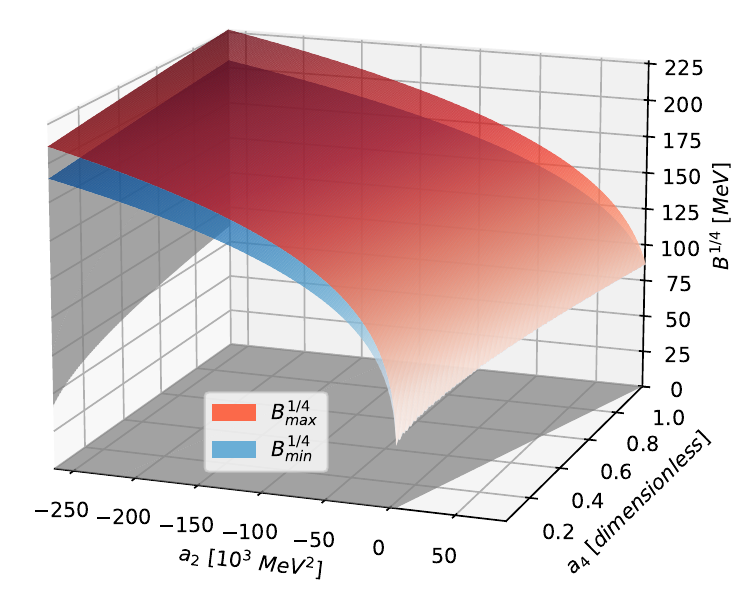}
    \caption{Parameter space of the quark EOS. Quark stars are formed in the region between the surfaces $B^{1/4}_{\rm{min}} \coloneqq \tilde{B}^{1/4}_{\rm{lim}}$ (in blue) and $B^{1/4}_{\rm{max}} \coloneqq B^{1/4}_{\rm{lim}}$ (in red) given by Eqs.~\eqref{eq:B_tilde_lim} and \eqref{eq:B_lim}. $a_4$ limits are given by Eq.~\eqref{eq:a4_limits} and $a_2$ upper limit is given by Eq.~\eqref{eq:a2_limit}, where $a_2^{\rm{max}}(a_4 = a_4^{\rm{max}}) = 88.47 \times 10^3$~MeV$^2$. The gray shades are projections of $B^{1/4}_{\rm{min}}$ and $B^{1/4}_{\rm{max}}$ surfaces at each plane.}
    \label{fig:quark_eos_parameter_space}
\end{figure}

The possible $B$s for the quark EOS [in Eq.~\eqref{eq:quark_eos_p_of_rho}] are shown in Fig.~\ref{fig:quark_eos_parameter_space} and correspond to the first condition given by Eq.~\eqref{eq:B_limits} for several $a_2$ and $a_4$. They are the ones considered in our analysis. Note here that we also allow for $a_2 < 0$. This is motivated by some limits of the effective EOS in the case the gap color-superconducting parameter $\Delta$ is large enough.

\begin{figure}
    \centering
    \includegraphics[scale=0.57]{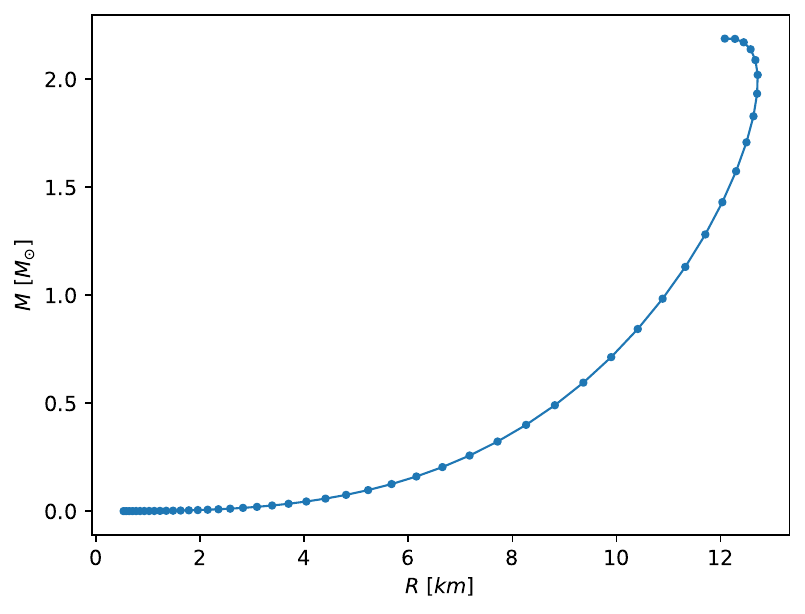}
    \caption{Quark star mass as a function of its radius for $a_2 = 14.61 \times 10^3$~MeV$^2$, $a_4 = 0.64$, and $B^{1/4} = 130.76$~MeV for different central densities. Note that these parameters fulfill all the constraints present in Tab.~\ref{tab:NSs_observables_limits}. Indeed, $M_{\rm{max}} \simeq 2.19$~M$_{\odot}$ and $R_{1.4 M_{\odot}} \simeq 11.97$~km.}
    \label{fig:quark_mass_vs_radius_curve}
\end{figure}

\begin{figure}
    \centering
    \includegraphics[scale=0.57]{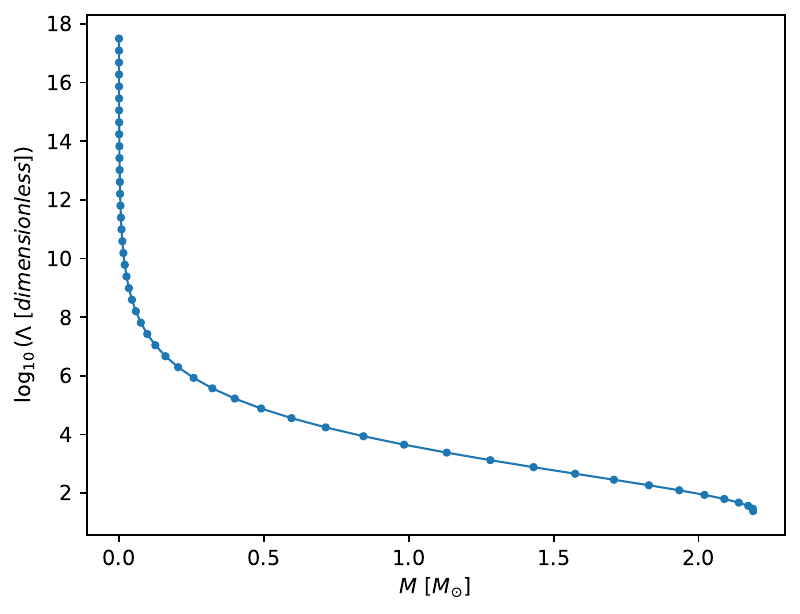}
    \caption{Logarithm of tidal deformability as a function of the mass for the same EOS of Fig.~\ref{fig:quark_mass_vs_radius_curve}. Note in this case that $\Lambda_{1.4 M_{\odot}} \simeq 852$.}
    \label{fig:quark_tidal_deformability_vs_mass_curve}
\end{figure}

Figure~\ref{fig:quark_mass_vs_radius_curve} presents the $M$-$R$ diagram for an illustrative EOS given by Eq.~\eqref{eq:quark_eos_p_of_rho} with $a_2 = 14.61 \times 10^3$~MeV$^2$, $a_4 = 0.64$, and $B^{1/4} = 130.76$~MeV, which satisfies all the criteria outlined in Sec.~\ref{subsec:data}. The corresponding tidal deformations as a function of the quark star mass are shown in Fig.~\ref{fig:quark_tidal_deformability_vs_mass_curve}.

\begin{figure}
    \centering
    \includegraphics[scale=0.57]{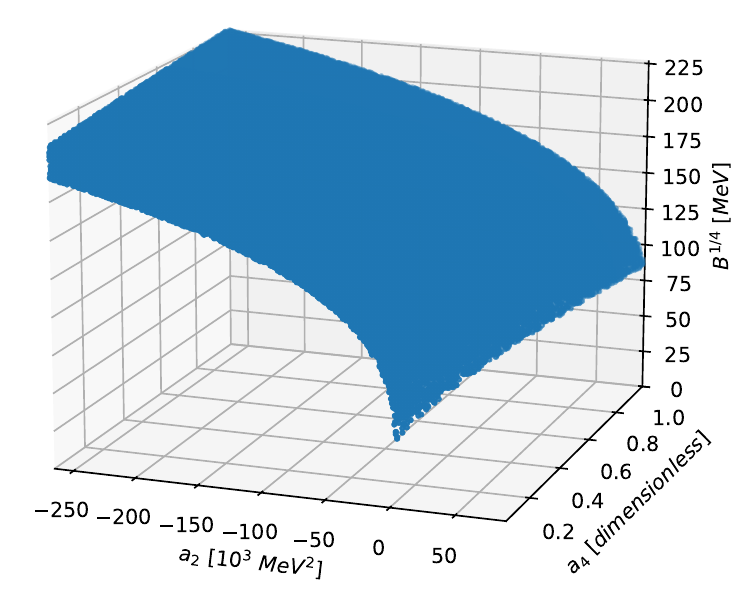}
    \caption{Sampling points in the parameter space of the quark EOS used in the analysis. Uniform coverage is guaranteed through quasi-random Latin hypercube sampling, which effectively distributes points across the parameter space. In total, approximately $260\times 10^3$ points were selected.}
    \label{fig:quark_eos_parameter_points}
\end{figure}

We now randomly sample points from the allowed $(a_2, a_4, B)$ parameter space that lead to quark star configurations. The selected points are shown in Fig.~\ref{fig:quark_eos_parameter_points}. In total, we have taken $263287$ points. For each of these, we compute the corresponding $M$-$R$ relation, determine $R_{1.4 M_{\odot}}$ and $M_{\rm{max}}$, and calculate $\Lambda_{1.4 M_{\odot}}$.

This sampling approach is conceptually similar to Markov Chain Monte Carlo (MCMC) methods used in Bayesian analysis. However, it is significantly simpler and computationally inexpensive. The main drawback is that it does not provide confidence intervals for the parameters. Nonetheless, this is not a concern at this stage, as our goal is an initial exploration of the relevant parameter space limits, paving the way for more precise future analyses.

\begin{figure}
    \centering
    \includegraphics[scale=0.57]{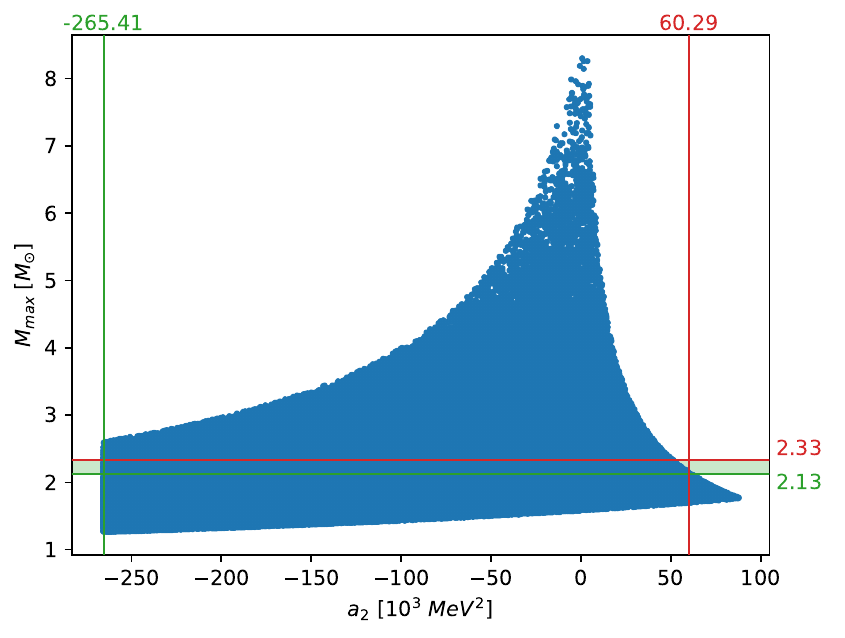}
    \caption{The maximum mass, $M_{\rm{max}}$, as a function of $a_2$ exhibits distinct behaviors depending on the sign of $a_2$. For larger negative values of $a_2$, the maximum mass decreases nonlinearly. In contrast, for positive $a_2$, the behavior is not strictly monotonic: around $a_2 \simeq 10^3$~MeV$^2$, $M_{\rm{max}}$ reaches a peak before beginning to decrease nonlinearly. Due to these decreasing trends, observational constraints on neutron star masses can be used to place limits on $a_2$, as indicated by the green and red lines.}
    \label{fig:quark_maximum_mass_vs_a2_graph}
\end{figure}

Figure~\ref{fig:quark_maximum_mass_vs_a2_graph} displays the maximum masses as a function of $a_2$ for all sampled points in Fig.~\ref{fig:quark_eos_parameter_points}. For clarity, the range of maximum masses from Tab.~\ref{tab:NSs_observables_limits} is also shown. The corresponding range of $a_2$ values that satisfy this constraint is also shown. Notably, negative values of $a_2$ can also lead to quark stars with large maximum masses. However, the dependence of $a_2$ on the mass is not straightforward. Indeed, while $a_2 \equiv -|a_2|$ results in a higher pressure compared to $a_2 = |a_2|$ for a given chemical potential [see Eq.~\eqref{eq:omega_quark_matter_3f}], the energy density also increases when $a_2 = -|a_2|$ relative to $a_2 = |a_2|$ for the same $\mu$ [see Eq.~\eqref{eq:rho_quarks_3f}]. A clear mass increase would occur if the pressure increases at a given density, which is equivalent to the EOS becoming stiffer.

\begin{figure}
    \centering
    \includegraphics[scale=0.57]{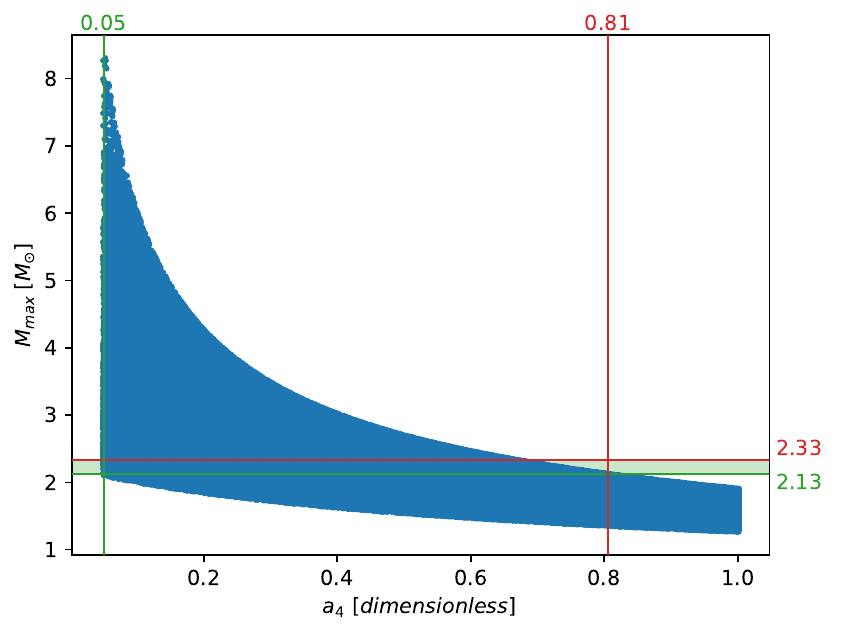}
    \caption{The maximum mass as a function of $a_4$ exhibits a nonlinear dependence. Smaller values of $a_4$, corresponding to stronger quark interactions, result in stiffer EOSs, which naturally lead to higher $M_{\rm{max}}$. However, for $a_4$ approaching $1$, the maximum mass falls below astrophysical constraints, making such cases incompatible with observations. This establishes an upper limit of $a_4 \lesssim 0.81$.}
    \label{fig:quark_maximum_mass_vs_a4_graph}
\end{figure}

Figure~\ref{fig:quark_maximum_mass_vs_a4_graph} presents the maximum mass as a function of $a_4$. The observationally constrained range of maximum masses, along with the corresponding range of $a_4$, is also shown. Notably, there exists an upper limit for $a_4$ at approximately $0.8$, providing an estimate for the minimum strength of strong interactions in highly dense stars. Additionally, the lower bound of $a_4$ can be very small, indicating that strong interactions cannot be neglected in the description of massive stars.

\begin{figure}
    \centering
    \includegraphics[scale=0.57]{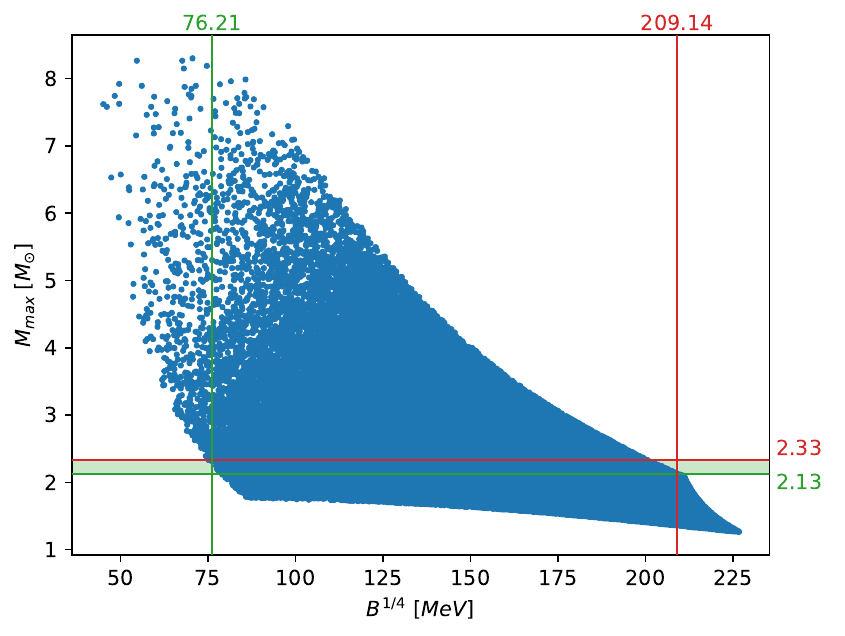}
    \caption{The maximum mass as a function of the effective gap parameter $B^{1/4}$ follows a similar trend to the simple MIT bag model, where $M_{\rm{max}} \propto B^{-1/2}=(B^{1/4})^{-2}$ \cite{1984PhRvD..30..272W}, even for the equation of state given by Eq.~\eqref{eq:quark_eos_p_of_rho}. As a result, $M_{\rm{max}}$ serves as an effective constraint on the possible values of $B$.}
    \label{fig:quark_maximum_mass_vs_b_graph}
\end{figure}

Figure~\ref{fig:quark_maximum_mass_vs_b_graph} shows the maximum mass as a function of the bag constant $B$. The observational constraints on the maximum mass impose meaningful restrictions on $B$. This is related to the well-known scaling relation $M_{\rm{max}} \propto B^{-1/2}$ in the original MIT bag model \cite{1984PhRvD..30..272W}. Indeed, $p \sim 0$ when the bag constant $B$ is proportional to the surface energy density, $B \simeq \rho_{\rm{surface}}/4$, even in the case of Eq.~\eqref{eq:quark_eos_p_of_rho}. This implies that $M_{\rm{max}}$ will still retain characteristics reminiscent of the MIT bag model. In particular, excessively large values of $B$ are disfavored, as they would correspond to high densities at which $p = 0$, leading to a decrease in the maximum mass. In addition, there exists a minimum value of $B$, corresponding to a minimum surface density that quark stars must exhibit. This has important implications for tidal deformations and stability modes, as it determines the minimum energy density discontinuity between the quark star's surface and its surrounding atmosphere.

\begin{figure}
    \centering
    \includegraphics[scale=0.57]{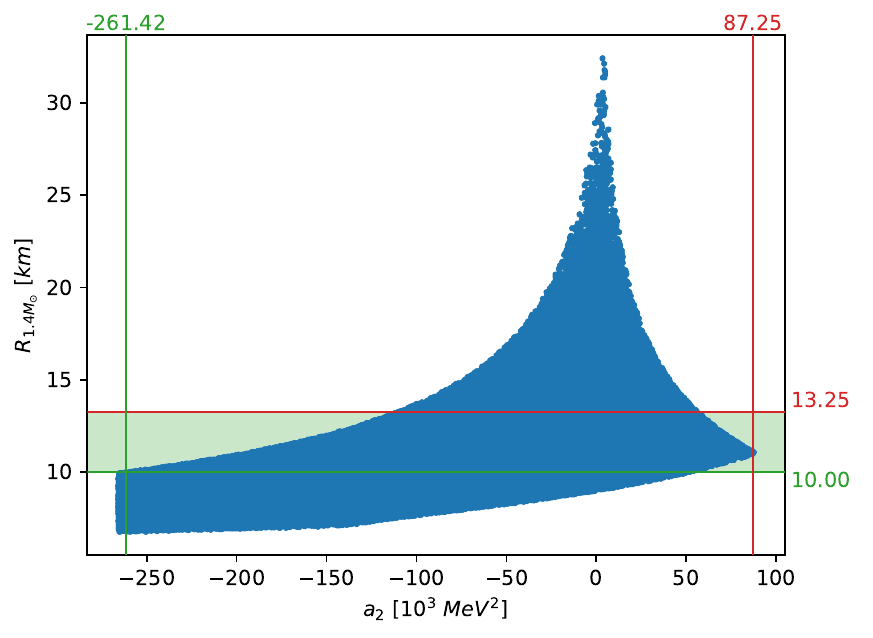}
    \caption{The radii of $1.4$~M$_{\odot}$ quark stars as a function of $a_2$. Note a similar trend to Fig.~\ref{fig:quark_maximum_mass_vs_a2_graph}.}
    \label{fig:quark_canonical_radius_vs_a2_graph}
\end{figure}

Figure~\ref{fig:quark_canonical_radius_vs_a2_graph} shows $R_{1.4 M_{\odot}}$ as a function of $a_2$. It presents a similar shape of $M_{\rm{max}}$ as a function of $a_2$, as shown in Fig.~\ref{fig:quark_maximum_mass_vs_a2_graph}. With a peak near $a_2 = 0$, $R_{1.4 M_{\odot}}$ decreases nonlinearly and asymmetrically for positive and negative values of $a_2$. This trend sets a minimum limit for $a_2$.

\begin{figure}
    \centering
    \includegraphics[scale=0.57]{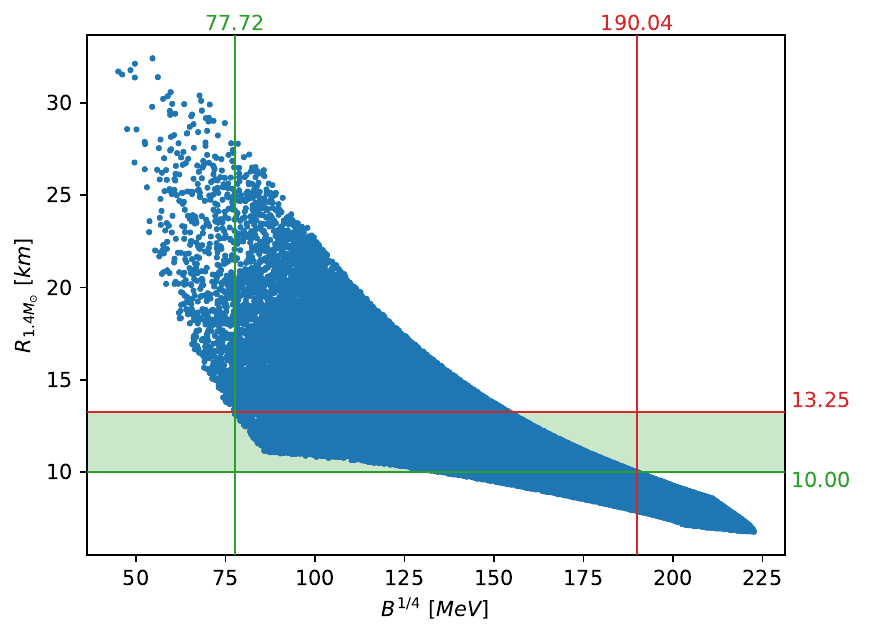}
    \caption{The radii of $1.4$~M$_{\odot}$ quark stars as a function of $B^{1/4}$ offer the most stringent constraints that radius measurements can place on the EOS. This is due to the nonlinear dependence of $R_{\rm{1.4M_\odot}}$ on $B$: larger values of $B$ lead to more compact stellar configurations.}
    \label{fig:quark_canonical_radius_vs_b_graph}
\end{figure}

The behavior of $R_{1.4 M_{\odot}}$ as a function of $B$ is shown in Fig.~\ref{fig:quark_canonical_radius_vs_b_graph}. $R_{1.4 M_{\odot}}$ decreases nonlinearly with $B$, as larger $B$ values lead to more compact stars. This decreasing trend imposes a stricter constraint on $B$ when compared to $M_{\rm{max}}$ and $\Lambda_{1.4 M_{\odot}}$.

\begin{figure}
    \centering
    \includegraphics[scale=0.57]{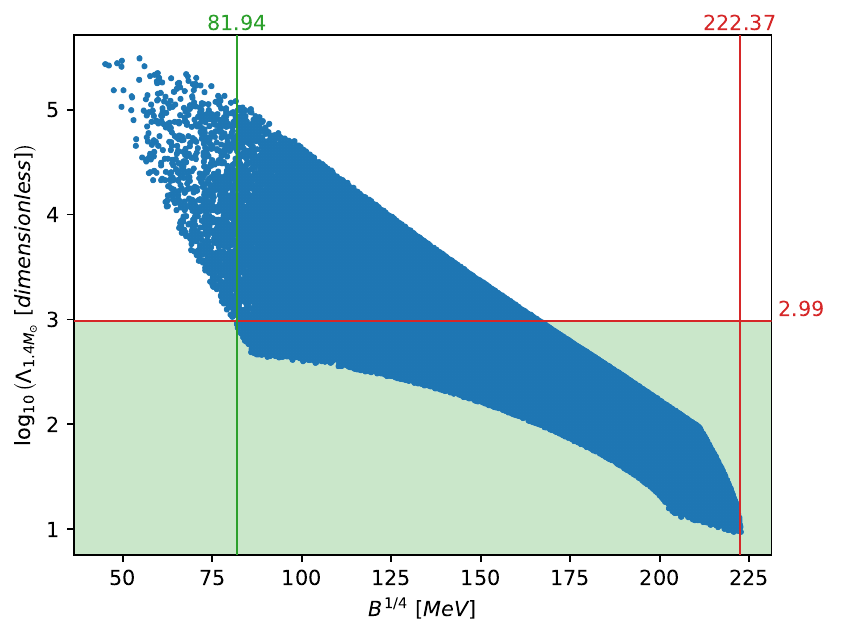}
    \caption{The tidal deformability of $1.4$~M$_{\odot}$ quark stars, $\Lambda_{1.4 M_{\odot}}$, as a function of $B^{1/4}$ decreases almost exponentially with increasing effective bag parameter. One reason for this is that larger $B$ values result in more compact quark stars, which reduce the tidal deformability. Note that the constraints on $\Lambda_{1.4 M_{\odot}}$ from GW170817 are less stringent than those imposed by the maximum mass requirement, $M_{\rm{max}}$.}
    \label{fig:quark_canonical_deformability_vs_b_graph}
\end{figure}

The effect of $B$ on the tidal deformability of $1.4$~M$_{\odot}$ stars is shown in Fig.~\ref{fig:quark_canonical_deformability_vs_b_graph}. Here, the impact of the energy density discontinuity on $\Lambda$, as described by Eq.~\eqref{eq:delta_y_of_R_expression}, has been incorporated. Notably, this allows for larger maximum values of $B$ compared to the constraints from maximum masses. This happens because larger $B$s lead to more compact quark stars, which have lower tidal deformations [see Eqs.~\eqref{eq:k2_expression} and \eqref{eq:Lambda_k2_relation}].

Table~\ref{tab:quark_eos_parameters_limits} summarizes the constraints on the parameters imposed by $M_{\rm{max}}$, $R_{1.4 M_{\odot}}$, and $\Lambda_{1.4 M_{\odot}}$, along with the combined results when all constraints are considered together.

\begin{table}[ht]
\caption{Limits on the parameters of the quark EOS.}
\label{tab:quark_eos_parameters_limits}
    \begin{center}
    \begin{tabular}{l|ccc}
        \hline \hline & $a_2$ & $a_4$ & $B^{1/4}$ \\
        & $[10^3$~MeV$^2]$ & $[$dimensionless$]$ & $[$MeV$]$ \\
        \hline $M_{\rm{max}}$ & $(-265, 60)$ & $(0.05, 0.81)$ & $(76, 209)$ \\
        $R_{1.4 M_{\odot}}$ & $(-261, 87)$ & $(0.05, 1.00)$ & $(78, 190)$ \\
        $\Lambda_{1.4 M_{\odot}}$ & $(-265, 87)$ & $(0.05, 1.00)$ & $(82, 222)$ \\
        Combined & $(-202, 42)$ & $(0.18, 0.81)$ & $(114, 185)$ \\
        \hline
    \end{tabular}
    \end{center}
\end{table}

In addition, Tab.~\ref{tab:quark_stars_properties_limits} presents the range of maximum masses, $1.4$~M$_{\odot}$ radii, and $1.4$~M$_{\odot}$ tidal deformations for quark stars, obtained by applying the combined constraints on $(a_2, a_4, B)$. Notably, $1.4$~M$_{\odot}$ quark star radii remain comparable to those of purely hadronic NSs.

\begin{table}[ht]
\caption{Limits on the properties of quark stars.}
\label{tab:quark_stars_properties_limits}
    \begin{center}
    \begin{tabular}{c|c|c}
         \hline \hline $M_{\rm{max}}$ & $R_{1.4 M_{\odot}}$ & $\Lambda_{1.4 M_{\odot}}$ \\
         $[$M$_{\odot}]$ & $[$km$]$ & $[$dimensionless$]$ \\
         \hline $(2.13, 2.33)$ & $(10.00, 12.26)$ & $(270, 970)$ \\
         \hline
    \end{tabular}
    \end{center}
\end{table}

We also applied the combined constraints on $(a_2, a_4, B)$ to determine the resulting ranges of surface densities and the corresponding ranges for the minimum and maximum values of the speed of sound, as summarized in Table~\ref{tab:quark_eos_properties_limits}. These ranges arise because $c_s^2$ depends on the density [see Eq.~\eqref{eq:quark_eos_dp_drho}], and both the density and the parameters $(a_2, a_4, B)$ are restricted to specific intervals. Note that the squared speed of sound does not remain constrained to $c_s^2 \lesssim 1/3$, as expected due to Eq.~\eqref{eq:quark_eos_dp_drho} for $a_2 < 0$. Large values of $c_s$ for the quark EOS also emerge from Bayesian inference applied to hybrid stars \citep{2018ApJ...860...12P} that are consistent with observational constraints \citep{2021PhRvC.103c5802X}. (With future observations of multiple NSs, it will also be possible to constrain the speed of sound of superdense matter using machine learning approaches, such as neural networks \citep{2020A&A...642A..78M,2025JCAP...01..073V}.)

\begin{table}[ht]
\caption{Limits on the properties of the quark EOS.}
\label{tab:quark_eos_properties_limits}
    \begin{center}
    \begin{tabular}{c|c|c}
         \hline \hline $\rho_{\rm{surface}}$ & $c_s^{\rm{min}} / c$ & $c_s^{\rm{max}} / c$ \\
         $[10^{15}$~g~cm$^{-3}]$ & $[$dimensionless$]$ & $[$dimensionless$]$ \\
         \hline $(0.29, 0.60)$ & $(0.51, 0.73)$ & $(0.55, 0.85)$ \\
         \hline
    \end{tabular}
    \end{center}
\end{table}

\begin{figure}
    \centering
    \includegraphics[scale=0.57]{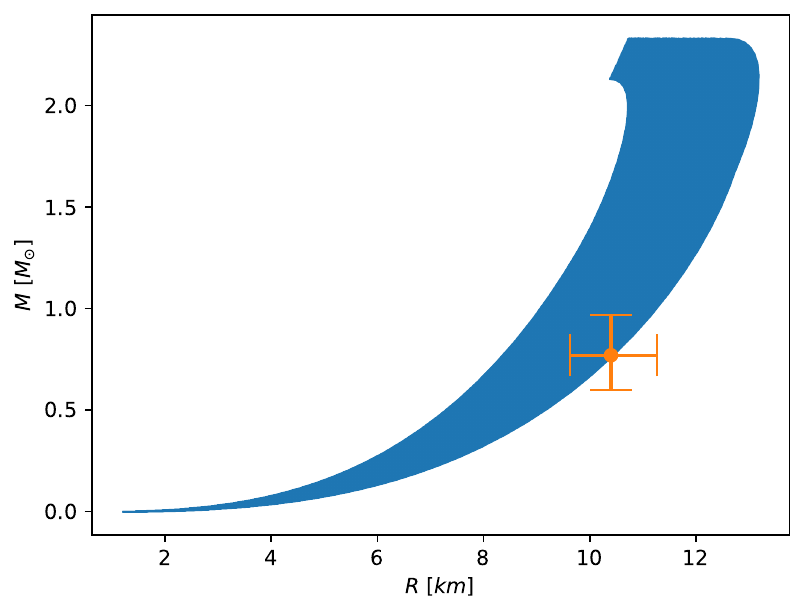}
    \caption{Mass-radius curves for the allowed quark EOS parametrizations are shown, with the quark star candidate XMMU J173203.3-344518 represented as an orange data point \cite{2023A&A...672L..11H}. Notably, this candidate remains consistent with all previously established neutron star constraints.}
    \label{fig:quark_mass_vs_radius_curves_parametrizations}
\end{figure}

Finally, in Fig.~\ref{fig:quark_mass_vs_radius_curves_parametrizations}, we present all the $M$-$R$ relations corresponding to the allowed values of $(a_2, a_4, B)$ after imposing all constraints. Additionally, we display the $2 \sigma$ CL range for the mass and radius of the central compact object XMMU J173203.3-344518. Notably, all the allowed parameter sets naturally describe this NS, reinforcing the consistency of our model with observational data.

\section{Discussion and conclusions}
\label{sec:discussion_and_conclusions}

Probing the interior of NSs is a challenging long-term task. However, the detection of GWs has helped with this issue, as their emission is directly influenced by the internal composition of the star. In the case of tidal deformations, however, a complication arises: they depend not only on the core properties but also on the entire stellar structure, and these deformations are not the dominant mechanism in the early inspiral phase of a binary merger. Regarding electromagnetic observations, the conditions at the star's surface are also crucial, as light is affected by spacetime curvature. This is precisely why NS masses and radii can be inferred using ray-tracing techniques. However, much like tidal deformations, these measurements provide only indirect insights into the star’s interior. Despite these limitations, when tidal and electromagnetic constraints are combined, they offer a more comprehensive picture of superdense matter. This guiding principle motivated our analysis.

When combined, constraints on maximum masses, tidal deformations, and radii provide significant restrictions on $(a_2, a_4, B)$ in a QCD-informed EOS for quark stars. These constraints have important implications for various observables and predictions relevant to testing the existence of such stars. We have shown that $a_4$, which serves as a proxy for the strength of strong quark interactions (in addition to that encompassed by the effective bag constant), deviates by at least $20\%$ from its asymptotic value at extremely high densities ($a_4 = 1$). This indicates that quark interactions cannot be neglected in the density range relevant to NSs of any mass. As expected, deviations from $a_4 = 1$ primarily affect the maximum mass of NSs, making it the observable most sensitive to variations in $a_4$. Additionally, we have shown that the minimum value of $a_2$ is $-202 \times 10^3$~MeV$^2$. Assuming that quark stars exist within the density range of the CFL phase, one could constrain $\Delta$ using the lower limit for $a_2$. That is the case because the relation $a_2 = m_s^2 - 4\Delta^2$ can be applied. Considering a strange quark mass of $100$~MeV, we find an upper limit for the gap parameter, $\Delta_{\rm{max}} \simeq 230$~MeV. Even if it does not provide an extremely strong upper limit, this result is fully consistent with previous astrophysical constraints ($\Delta \lesssim 216$~MeV for reasonable assumptions) \cite{2024PhRvL.132z2701K}, providing strong evidence that our simplified approach is reliable. Moreover, it agrees well with expectations from theoretical calculations \cite{1999NuPhB.537..443A}. Establishing an upper limit on the color-superconducting gap parameter is crucial for several reasons, which we discuss in the following paragraphs. Finally, we have also constrained the parameter $B$, which serves as a proxy for the energy density difference between quark matter and the QCD vacuum. Our results indicate that $B$ lies within a narrow range, significantly restricting the possible surface densities of quark stars to $(1.1-2.2)$~$\rho_{\rm{sat}}$. This constraint is particularly important as it affects perturbative quantities through nontrivial boundary conditions, among other aspects, which we explore now.

A strict range of surface densities can influence perturbative quantities, such as tidal deformations, through its effect on $y$ [see Eq.~\eqref{eq:delta_y_of_R_expression}]. For instance, a higher surface density would result in a larger jump in $y$, generally leading to a larger decrease in tidal deformability. The ellipticity $\epsilon$ of NSs could also be affected by these constraints. Indeed, in the scenario where a thin solid crust is at the top of the quark surface, the discontinuity $\Delta y$ plays a crucial role in determining $\epsilon$ \cite{2023ApJ...950..185P}. Quasi-normal modes, associated with nonradial oscillations, would also be influenced by the range of surface densities in quark stars. This is because obtaining physically consistent solutions requires properly matching internal and external spacetime perturbations. In addition, constraints on $a_4$ and $a_2$ directly affect the equation of state and, consequently, the average density of neutron stars, therefore influencing the fundamental mode ($f$-mode for $l=2$) \cite{1998MNRAS.299.1059A}. A smaller $a_4$ generally leads to a stiffer equation of state, resulting in higher fundamental mode frequencies. Once GW detectors are capable of constraining these parameters, they could also provide better insights into the properties of quark stars. In contrast, radial oscillations are generally unaffected by the surface-to-vacuum discontinuity in energy density, as the primary boundary condition at the surface is simply $p = 0$. However, radial oscillations do depend on $a_2$, meaning that a lower bound on $a_2$ would impact the eigenmodes. The same holds for the upper limit of $a_4$, as its value directly affects the stiffness of the EOS and, consequently, the oscillatory behavior of the star \cite{2019PhRvD.100k4041J, 2023EPJC...83.1065R}.

A maximum value of $\Delta$ can also have significant implications for the LOFF phase \cite{2001PhRvD..63g4016A, 2007PhRvD..76g4026M, 2008RvMP...80.1455A}. This phase is expected to exist at densities slightly lower than those required for the CFL phase and could have a non-negligible impact on various observables, including GW emission via tidal deformations \cite{2017PhRvD..95j1302L, 2019PhRvD..99b3018L}, the maximum height of mountains on NSs \cite{2007PhRvL..99w1101H}, and the overall stability of NSs \cite{2021ApJ...910..145P}. The key reason is that the shear modulus indirectly depends nonlinearly on $\Delta$, directly influencing the mechanical properties of the star’s interior. With a constraint on $\Delta$, the impact of color superconductivity on neutron star cooling curves could also be more effectively analyzed. For sufficiently large $\Delta$, neutrino emissivity is significantly suppressed, leading to slower cooling compared to purely hadronic models. Since we found that $\Delta \lesssim 230$~MeV (for $m_s=100$~MeV), which is a large upper limit, there is still plenty of room for both suppressed and unsuppressed neutrino emission to occur. Observational constraints from neutron star surface temperatures, particularly through X-ray observations, could further refine the role of color superconductivity and provide indirect constraints on $\Delta$. In fact, the cooling of the object in the supernova remnant HESS J1731-347 suggests a $\Delta$ value much smaller than the ones considered in CFL models, around $\sim 1$~MeV \cite{2023A&A...672L..11H}, illustrating the importance of measurements for the pairing problem.

For a definite constraint on $\Delta$, the quark mass must be known. We have adopted the commonly used value of $m_s = 100$~MeV, which is consistent with the $\overline{\rm{MS}}$ scheme at $\mu = 2$~GeV (see, e.g., \cite{2006EPJC...46..721C} and references therein, and Section ``Quark masses'' of \cite{2020PTEP.2020h3C01P, 2024PhRvD.110c0001N}). However, the strange quark mass depends on the renormalization scheme, the energy scale, and potentially the density \cite{2005PhRvD..71j5014F, 2007PhRvD..76g4026M, 2025arXiv250114935R}. In cases where $m_s > 100$~MeV, the upper limit on the color-superconducting gap, $\Delta_{\rm{max}}$, would also increase. For instance, for $m_s = 300$~MeV, we obtain $\Delta_{\rm{max}} \simeq 270$~MeV. As discussed above, this would alter the predictions for neutron star observables. At the same time, the upper limit $a_2 = 42 \times 10^3$~MeV$^2$ can also be used to establish a lower bound on $\Delta$ for large $m_s$. For instance, for $m_s = 300$~MeV, we obtain $\Delta_{\rm{min}} \simeq 110$~MeV. Conversely, enforcing a very small gap parameter ($\Delta \simeq 0$~MeV) constrains the strange quark mass to $m_s^{\rm{max}} = {(a_2^{\rm{max}})}^{1/2} \simeq 205$~MeV. In the absence of additional constraints on strange quark matter or the color-superconducting gap parameter, a fundamental degeneracy will persist due to the limits on $a_2$.

An important byproduct of our analysis is that, taking into account the constraints of Tab.~\ref{tab:NSs_observables_limits}, the radius of a $1.4$~M$_{\odot}$ quark star lies within the range $(10.0-12.3)$~km, while its dimensionless tidal deformability falls in the interval $(270-970)$. This suggests that radius measurements alone may not be sufficient to distinguish a quark star from an ordinary neutron star or even a hybrid star, whereas more precise constraints on the tidal deformability could provide valuable insights into the nature of quark stars. Another significant outcome of this work is that the inferred mass and radius of the central compact object XMMU J173203.3-344518 in the supernova remnant HESS J1731-347 ($M = 0.77^{+0.20}_{-0.17}$~M$_{\odot}$, $R = 10.40^{+0.86}_{-0.78}$~km) \cite{2022NatAs...6.1444D} are fully consistent with all other astrophysical constraints derived from gravitational and electromagnetic wave observations. Indeed, the constrained QCD-informed EOS for quark stars naturally reproduces the properties of this object, making the interpretation of XMMU J173203.3-344518 as a quark star particularly compelling, as suggested in several studies \cite{2022NatAs...6.1444D, 2023A&A...672L..11H, 2025EPJC...85...40J, 2024ApJ...967..159D, 2023PhRvD.108f4007C, 2023PhRvD.108f3010O, 2023PhRvD.107k4015R}. However, important caveats must be considered. First, as noted in \cite{2023A&A...672L..11H}, XMMU J173203.3-344518 could also be a standard neutron star described purely by hadronic EOSs within the framework of chiral effective field theory (see, e.g., \cite{2021ApJ...922...14P, 2024ApJ...971L..19R} and references therein). However, this scenario presents significant challenges regarding its formation via conventional core-collapse supernovae, which typically produce neutron stars with masses exceeding $1.17$~M$_{\odot}$ \cite{2025PhRvL.134g1403M}. Second, the inferred mass and radius of XMMU J173203.3-344518 are highly sensitive to assumptions about the atmosphere, temperature distribution in the emitting region, magnetic fields and other model-dependent factors (\cite{2023A&A...672L..11H, 2023ApJ...959...55V}; see also \cite{2020ApJ...889..165D, 2024JHEAp..42...52D} and references therein). Nevertheless, as emphasized in \cite{2023A&A...672L..11H}, the assumptions leading to the best-fit mass and radius estimates are robust to seriously consider the possibility that XMMU J173203.3-344518 is a quark star.

\section{Summary}
\label{sec:summary}

Our main results can be summarized as follows:

\begin{enumerate}
    \item The maximum masses, radii, and tidal deformations of neutron stars provide quite strong constraints on QCD-informed equations of state for quark stars. These astrophysical observables serve as key probes for distinguishing different EOS models.

    \item We have determined an upper limit for the color-superconducting gap parameter, $\Delta \lesssim 230$~MeV for $m_s = 100$~MeV. This follows from the inferred lower limit on $a_2$, specifically $a_2 < 0$. This result is fully consistent with previous theoretical and observational constraints and may serve as a reference value for various studies related to quasi-normal modes, ellipticities, and tidal deformations.

    \item We find that $a_4 \lesssim 0.8$, indicating that strong interaction effects cannot be neglected within the density range relevant to neutron stars. Additionally, deviations from the free-quark limit at asymptotically large densities ($a_4 = 1$) amount to at least $20\%$.

    \item The parameter $B^{1/4}$ (the effective bag constant) is in the range $(114-185)$~MeV and it constrains the surface density of neutron stars to be $(1.1-2.2)$~$\rho_{\rm{sat}}$. This sets an upper limit on the surface-to-vacuum energy density jump, a fact that could have significant implications for perturbative observables such as tidal deformations, ellipticities, and quasi-normal modes.

    \item The radius of a $1.4$~M$_{\odot}$ quark star lies within $(10.0-12.3)$~km, while its dimensionless tidal deformability is in the interval $(270-970)$.

    \item The inferred mass and radius of the central compact object XMMU J173203.3-344518 in the supernova remnant HESS J1731-347, $M = 0.77^{+0.20}_{-0.17}$~M$_{\odot}$, $R = 10.40^{+0.86}_{-0.78}$~km, are fully consistent with all other astrophysical constraints derived.
\end{enumerate}

\section{Acknowledgments}
\noindent
J.V.Z. acknowledges financial support from Udesc---Universidade do Estado de Santa Catarina and CAPES---Coordenação de Aperfeiçoamento de Pessoal de Nível Superior---under grant No. 88887.838317/2023-00.
J.P.P. acknowledges financial support from CNPq---Conselho Nacional de Desenvolvimento Cient\'ifico e Tecnol\'ogico---under grant No. 174612/2023-0.
R.C.R.L. thanks the FAPESC---Fundação de Amparo à Pesquisa e Inovação do Estado de Santa Catarina for the support under grant number FAPESC 2023TR000338.
J.E.H. thanks the CNPq---Conselho Nacional de Desenvolvimento Cient\'ifico e Tecnol\'ogico for a Research Scholarship and the FAPESP Agency supporting his participation through the Grant 2024/16892-2.

\appendix

\section{Constraints on quark star parameters}
\label{ap:QS_parameters_constraints}

According to the strange quark matter hypothesis, quark matter is the stable ground state of matter, while hadronic matter is a metastable, albeit long-lived, state \cite{1971PhRvD...4.1601B, 1984PhRvD..30..272W}. For this hypothesis to be true the phase transition from quark matter to hadronic matter must not occur inside the NS, i.e., there must not be hybrid stars, but rather quark stars. Numerically, this means that the quark-hadron interface must be at $p < 0$. Defining $g_i \coloneqq (\rho + p) / n_i$, it is possible to show that $g_i = 3 \mu$ \cite{2018ApJ...860...12P}. We can now evaluate the boundary condition at which the quark-hadron phase transition occurs at $p = 0$. By Maxwell's construction, we have that $\mu^{(q)}(0) = \mu^{(h)}(0)$, or equivalently, $g(0) \coloneqq g_i^{(q)}(0) = g_i^{(h)}(0)$, where $g(0)$ is the Gibbs free energy per baryon of the hadronic matter at zero pressure, here assumed to be that of iron, of approximately $930$~MeV.

Substituting $p = 0$ and $\mu = g(0) / 3$ into Eq.~\eqref{eq:omega_quark_matter_3f}, recalling that $p = - \omega$, we obtain

\begin{equation}
    B_{\rm{lim}} = \frac{g^2(0)}{108 \pi^2} \left[ g^2(0) a_4 - 9 a_2 \right] .
\label{eq:B_lim}
\end{equation}

If $g_i^{(q)}(0) < g(0)$, $B < B_{\rm{lim}}$, we have that quark matter is the ground state of matter, and the strange quark matter hypothesis is true. Otherwise, the quark-hadron phase transition occurs at a non-zero positive pressure inside the NS, resulting in a hybrid star.

We also need to ensure that the hadronic matter is not in a metastable equilibrium state. For this, we have $(g(0) = g_i^{(h)}(0)) < g_i^{(ud)}(0)$, where $g_i^{(ud)}(0)$ is the Gibbs free energy per baryon of the quark matter composed of only up and down quarks. The grand thermodynamic potential density for the up-down quark matter is given by \cite{2018ApJ...860...12P}

\begin{equation}
    \tilde{\omega} = - \frac{6 a_4}{\pi^2 (1 + 2^{1 / 3})^3} \tilde{\mu}^4 + \frac{a_2}{2 \pi^2} \tilde{\mu}^2 + \tilde{B} ,
\label{eq:omega_quark_matter_2f}
\end{equation}

\noindent where $\tilde{\mu} \coloneqq (\mu_u + \mu_d) / 2 = (1 + 2^{1/3}) \mu_u$ due to charge neutrality. Substituting $\tilde{p} = 0$ and $\tilde{\mu} = g(0) / 3$, recalling that $\tilde{p} = - \tilde{\omega}$, we obtain

\begin{equation}
    \tilde{B}_{\rm{lim}} = \frac{g^2(0)}{54 \pi^2} \left[ \frac{4 g^2(0) a_4}{(1 + 2^{1 / 3})^3} - 3 a_2 \right] .
\label{eq:B_tilde_lim}
\end{equation}

Thus, if $g(0) < g_i^{(ud)}(0)$, $\tilde{B} > \tilde{B}_{\rm{lim}}$, we have that hadronic matter is stable, and not up-down quark matter. We consider this condition, although recent studies suggest the possibility of up-down quark matter may be energetically more favorable under certain conditions \cite{2018PhRvL.120v2001H}.

We now have all the conditions to set the bounds on the free parameters of the quark matter EOS for the case of quark stars and hybrid stars. Starting with the parameter $B$, we can summarize the above discussion as follows:

\begin{equation}
    \begin{cases}
        (\tilde{B}_{\rm{lim}} < B < B_{\rm{lim}}) ~ \rightarrow ~ \text{Quark star} , \\
        (B > \tilde{B}_{\rm{lim}}) ~ \text{and} ~ (B > B_{\rm{lim}}) ~ \rightarrow ~ \text{Hybrid star} .
    \end{cases}
\label{eq:B_limits}
\end{equation}

For the case of quark stars, we must require that $B_{\rm{lim}} > \tilde{B}_{\rm{lim}}$. Substituting Eqs.~\eqref{eq:B_lim} and \eqref{eq:B_tilde_lim} into this condition, and isolating $a_2$, we obtain the upper bound of the parameter $a_2$ in terms of the parameter $a_4$, so that

\begin{equation}
    a_2 < \left( \frac{1}{3} - \frac{8}{3 (1 + 2^{1/3})^3} \right) g^2(0) a_4 .
\label{eq:a2_limit}
\end{equation}

Assuming that $a_4$ is positive and considering that in the CFL phase $a_4 = 1 - c$, we conclude that $a_4^{\rm{max}} = 1$, which corresponds to the case of non-interacting quarks, so that the limits of $a_4$ boil down to

\begin{equation}
    0 < a_4 < 1 .
\label{eq:a4_limits}
\end{equation}

\bibliography{bibliography}

\begin{thebibliography}{101}%
\makeatletter
\providecommand \@ifxundefined [1]{%
 \@ifx{#1\undefined}
}%
\providecommand \@ifnum [1]{%
 \ifnum #1\expandafter \@firstoftwo
 \else \expandafter \@secondoftwo
 \fi
}%
\providecommand \@ifx [1]{%
 \ifx #1\expandafter \@firstoftwo
 \else \expandafter \@secondoftwo
 \fi
}%
\providecommand \natexlab [1]{#1}%
\providecommand \enquote  [1]{``#1''}%
\providecommand \bibnamefont  [1]{#1}%
\providecommand \bibfnamefont [1]{#1}%
\providecommand \citenamefont [1]{#1}%
\providecommand \href@noop [0]{\@secondoftwo}%
\providecommand \href [0]{\begingroup \@sanitize@url \@href}%
\providecommand \@href[1]{\@@startlink{#1}\@@href}%
\providecommand \@@href[1]{\endgroup#1\@@endlink}%
\providecommand \@sanitize@url [0]{\catcode `\\12\catcode `\$12\catcode `\&12\catcode `\#12\catcode `\^12\catcode `\_12\catcode `\%12\relax}%
\providecommand \@@startlink[1]{}%
\providecommand \@@endlink[0]{}%
\providecommand \url  [0]{\begingroup\@sanitize@url \@url }%
\providecommand \@url [1]{\endgroup\@href {#1}{\urlprefix }}%
\providecommand \urlprefix  [0]{URL }%
\providecommand \Eprint [0]{\href }%
\providecommand \doibase [0]{https://doi.org/}%
\providecommand \selectlanguage [0]{\@gobble}%
\providecommand \bibinfo  [0]{\@secondoftwo}%
\providecommand \bibfield  [0]{\@secondoftwo}%
\providecommand \translation [1]{[#1]}%
\providecommand \BibitemOpen [0]{}%
\providecommand \bibitemStop [0]{}%
\providecommand \bibitemNoStop [0]{.\EOS\space}%
\providecommand \EOS [0]{\spacefactor3000\relax}%
\providecommand \BibitemShut  [1]{\csname bibitem#1\endcsname}%
\let\auto@bib@innerbib\@empty
\bibitem [{\citenamefont {{Hebeler}}\ \emph {et~al.}(2013)\citenamefont {{Hebeler}}, \citenamefont {{Lattimer}}, \citenamefont {{Pethick}},\ and\ \citenamefont {{Schwenk}}}]{2013ApJ...773...11H}%
  \BibitemOpen
  \bibfield  {author} {\bibinfo {author} {\bibfnamefont {K.}~\bibnamefont {{Hebeler}}}, \bibinfo {author} {\bibfnamefont {J.~M.}\ \bibnamefont {{Lattimer}}}, \bibinfo {author} {\bibfnamefont {C.~J.}\ \bibnamefont {{Pethick}}},\ and\ \bibinfo {author} {\bibfnamefont {A.}~\bibnamefont {{Schwenk}}},\ }\href {https://doi.org/10.1088/0004-637X/773/1/11} {\bibfield  {journal} {\bibinfo  {journal} {\apj}\ }\textbf {\bibinfo {volume} {773}},\ \bibinfo {eid} {11} (\bibinfo {year} {2013})},\ \Eprint {https://arxiv.org/abs/1303.4662} {arXiv:1303.4662 [astro-ph.SR]} \BibitemShut {NoStop}%
\bibitem [{\citenamefont {{Rutherford}}\ \emph {et~al.}(2024)\citenamefont {{Rutherford}}, \citenamefont {{Mendes}}, \citenamefont {{Svensson}}, \citenamefont {{Schwenk}}, \citenamefont {{Watts}}, \citenamefont {{Hebeler}}, \citenamefont {{Keller}}, \citenamefont {{Prescod-Weinstein}}, \citenamefont {{Choudhury}}, \citenamefont {{Raaijmakers}}, \citenamefont {{Salmi}}, \citenamefont {{Timmerman}}, \citenamefont {{Vinciguerra}}, \citenamefont {{Guillot}},\ and\ \citenamefont {{Lattimer}}}]{2024ApJ...971L..19R}%
  \BibitemOpen
  \bibfield  {author} {\bibinfo {author} {\bibfnamefont {N.}~\bibnamefont {{Rutherford}}}, \bibinfo {author} {\bibfnamefont {M.}~\bibnamefont {{Mendes}}}, \bibinfo {author} {\bibfnamefont {I.}~\bibnamefont {{Svensson}}}, \bibinfo {author} {\bibfnamefont {A.}~\bibnamefont {{Schwenk}}}, \bibinfo {author} {\bibfnamefont {A.~L.}\ \bibnamefont {{Watts}}}, \bibinfo {author} {\bibfnamefont {K.}~\bibnamefont {{Hebeler}}}, \bibinfo {author} {\bibfnamefont {J.}~\bibnamefont {{Keller}}}, \bibinfo {author} {\bibfnamefont {C.}~\bibnamefont {{Prescod-Weinstein}}}, \bibinfo {author} {\bibfnamefont {D.}~\bibnamefont {{Choudhury}}}, \bibinfo {author} {\bibfnamefont {G.}~\bibnamefont {{Raaijmakers}}}, \bibinfo {author} {\bibfnamefont {T.}~\bibnamefont {{Salmi}}}, \bibinfo {author} {\bibfnamefont {P.}~\bibnamefont {{Timmerman}}}, \bibinfo {author} {\bibfnamefont {S.}~\bibnamefont {{Vinciguerra}}}, \bibinfo {author} {\bibfnamefont {S.}~\bibnamefont {{Guillot}}},\ and\ \bibinfo {author} {\bibfnamefont {J.~M.}\ \bibnamefont
  {{Lattimer}}},\ }\href {https://doi.org/10.3847/2041-8213/ad5f02} {\bibfield  {journal} {\bibinfo  {journal} {\apjl}\ }\textbf {\bibinfo {volume} {971}},\ \bibinfo {eid} {L19} (\bibinfo {year} {2024})},\ \Eprint {https://arxiv.org/abs/2407.06790} {arXiv:2407.06790 [astro-ph.HE]} \BibitemShut {NoStop}%
\bibitem [{\citenamefont {{Chatziioannou}}(2020)}]{2020GReGr..52..109C}%
  \BibitemOpen
  \bibfield  {author} {\bibinfo {author} {\bibfnamefont {K.}~\bibnamefont {{Chatziioannou}}},\ }\href {https://doi.org/10.1007/s10714-020-02754-3} {\bibfield  {journal} {\bibinfo  {journal} {General Relativity and Gravitation}\ }\textbf {\bibinfo {volume} {52}},\ \bibinfo {eid} {109} (\bibinfo {year} {2020})},\ \Eprint {https://arxiv.org/abs/2006.03168} {arXiv:2006.03168 [gr-qc]} \BibitemShut {NoStop}%
\bibitem [{\citenamefont {{{\"O}zel}}\ and\ \citenamefont {{Freire}}(2016)}]{2016ARA&A..54..401O}%
  \BibitemOpen
  \bibfield  {author} {\bibinfo {author} {\bibfnamefont {F.}~\bibnamefont {{{\"O}zel}}}\ and\ \bibinfo {author} {\bibfnamefont {P.}~\bibnamefont {{Freire}}},\ }\href {https://doi.org/10.1146/annurev-astro-081915-023322} {\bibfield  {journal} {\bibinfo  {journal} {\araa}\ }\textbf {\bibinfo {volume} {54}},\ \bibinfo {pages} {401} (\bibinfo {year} {2016})},\ \Eprint {https://arxiv.org/abs/1603.02698} {arXiv:1603.02698 [astro-ph.HE]} \BibitemShut {NoStop}%
\bibitem [{\citenamefont {{The LIGO and VIRGO Scientific Collaboration}}\ \emph {et~al.}(2017)\citenamefont {{The LIGO and VIRGO Scientific Collaboration}}, \citenamefont {{Abbott}}, \citenamefont {{Abbott}}, \citenamefont {{Abbott}}, \citenamefont {{Acernese}}, \citenamefont {{Ackley}}, \citenamefont {{Adams}}, \citenamefont {{Adams}}, \citenamefont {{Addesso}} \emph {et~al.}}]{2017ApJ...848L..12A}%
  \BibitemOpen
  \bibfield  {author} {\bibinfo {author} {\bibnamefont {{The LIGO and VIRGO Scientific Collaboration}}}, \bibinfo {author} {\bibfnamefont {B.~P.}\ \bibnamefont {{Abbott}}}, \bibinfo {author} {\bibfnamefont {R.}~\bibnamefont {{Abbott}}}, \bibinfo {author} {\bibfnamefont {T.~D.}\ \bibnamefont {{Abbott}}}, \bibinfo {author} {\bibfnamefont {F.}~\bibnamefont {{Acernese}}}, \bibinfo {author} {\bibfnamefont {K.}~\bibnamefont {{Ackley}}}, \bibinfo {author} {\bibfnamefont {C.}~\bibnamefont {{Adams}}}, \bibinfo {author} {\bibfnamefont {T.}~\bibnamefont {{Adams}}}, \bibinfo {author} {\bibfnamefont {P.}~\bibnamefont {{Addesso}}}, \emph {et~al.},\ }\href {https://doi.org/10.3847/2041-8213/aa91c9} {\bibfield  {journal} {\bibinfo  {journal} {\apjl}\ }\textbf {\bibinfo {volume} {848}},\ \bibinfo {eid} {L12} (\bibinfo {year} {2017})},\ \Eprint {https://arxiv.org/abs/1710.05833} {arXiv:1710.05833 [astro-ph.HE]} \BibitemShut {NoStop}%
\bibitem [{\citenamefont {{Landry}}\ \emph {et~al.}(2020)\citenamefont {{Landry}}, \citenamefont {{Essick}},\ and\ \citenamefont {{Chatziioannou}}}]{2020PhRvD.101l3007L}%
  \BibitemOpen
  \bibfield  {author} {\bibinfo {author} {\bibfnamefont {P.}~\bibnamefont {{Landry}}}, \bibinfo {author} {\bibfnamefont {R.}~\bibnamefont {{Essick}}},\ and\ \bibinfo {author} {\bibfnamefont {K.}~\bibnamefont {{Chatziioannou}}},\ }\href {https://doi.org/10.1103/PhysRevD.101.123007} {\bibfield  {journal} {\bibinfo  {journal} {\prd}\ }\textbf {\bibinfo {volume} {101}},\ \bibinfo {eid} {123007} (\bibinfo {year} {2020})},\ \Eprint {https://arxiv.org/abs/2003.04880} {arXiv:2003.04880 [astro-ph.HE]} \BibitemShut {NoStop}%
\bibitem [{\citenamefont {{Alford}}(2001)}]{2001ARNPS..51..131A}%
  \BibitemOpen
  \bibfield  {author} {\bibinfo {author} {\bibfnamefont {M.}~\bibnamefont {{Alford}}},\ }\href {https://doi.org/10.1146/annurev.nucl.51.101701.132449} {\bibfield  {journal} {\bibinfo  {journal} {Annual Review of Nuclear and Particle Science}\ }\textbf {\bibinfo {volume} {51}},\ \bibinfo {pages} {131} (\bibinfo {year} {2001})},\ \Eprint {https://arxiv.org/abs/hep-ph/0102047} {arXiv:hep-ph/0102047 [hep-ph]} \BibitemShut {NoStop}%
\bibitem [{\citenamefont {{Alford}}\ \emph {et~al.}(2008)\citenamefont {{Alford}}, \citenamefont {{Schmitt}}, \citenamefont {{Rajagopal}},\ and\ \citenamefont {{Sch{\"a}fer}}}]{2008RvMP...80.1455A}%
  \BibitemOpen
  \bibfield  {author} {\bibinfo {author} {\bibfnamefont {M.~G.}\ \bibnamefont {{Alford}}}, \bibinfo {author} {\bibfnamefont {A.}~\bibnamefont {{Schmitt}}}, \bibinfo {author} {\bibfnamefont {K.}~\bibnamefont {{Rajagopal}}},\ and\ \bibinfo {author} {\bibfnamefont {T.}~\bibnamefont {{Sch{\"a}fer}}},\ }\href {https://doi.org/10.1103/RevModPhys.80.1455} {\bibfield  {journal} {\bibinfo  {journal} {Reviews of Modern Physics}\ }\textbf {\bibinfo {volume} {80}},\ \bibinfo {pages} {1455} (\bibinfo {year} {2008})},\ \Eprint {https://arxiv.org/abs/0709.4635} {arXiv:0709.4635 [hep-ph]} \BibitemShut {NoStop}%
\bibitem [{\citenamefont {{Itoh}}(1970)}]{1970PThPh..44..291I}%
  \BibitemOpen
  \bibfield  {author} {\bibinfo {author} {\bibfnamefont {N.}~\bibnamefont {{Itoh}}},\ }\href {https://doi.org/10.1143/PTP.44.291} {\bibfield  {journal} {\bibinfo  {journal} {Progress of Theoretical Physics}\ }\textbf {\bibinfo {volume} {44}},\ \bibinfo {pages} {291} (\bibinfo {year} {1970})}\BibitemShut {NoStop}%
\bibitem [{\citenamefont {{Bodmer}}(1971)}]{1971PhRvD...4.1601B}%
  \BibitemOpen
  \bibfield  {author} {\bibinfo {author} {\bibfnamefont {A.~R.}\ \bibnamefont {{Bodmer}}},\ }\href {https://doi.org/10.1103/PhysRevD.4.1601} {\bibfield  {journal} {\bibinfo  {journal} {\prd}\ }\textbf {\bibinfo {volume} {4}},\ \bibinfo {pages} {1601} (\bibinfo {year} {1971})}\BibitemShut {NoStop}%
\bibitem [{\citenamefont {{Witten}}(1984)}]{1984PhRvD..30..272W}%
  \BibitemOpen
  \bibfield  {author} {\bibinfo {author} {\bibfnamefont {E.}~\bibnamefont {{Witten}}},\ }\href {https://doi.org/10.1103/PhysRevD.30.272} {\bibfield  {journal} {\bibinfo  {journal} {\prd}\ }\textbf {\bibinfo {volume} {30}},\ \bibinfo {pages} {272} (\bibinfo {year} {1984})}\BibitemShut {NoStop}%
\bibitem [{\citenamefont {{Haensel}}\ \emph {et~al.}(1986)\citenamefont {{Haensel}}, \citenamefont {{Zdunik}},\ and\ \citenamefont {{Schaefer}}}]{1986A&A...160..121H}%
  \BibitemOpen
  \bibfield  {author} {\bibinfo {author} {\bibfnamefont {P.}~\bibnamefont {{Haensel}}}, \bibinfo {author} {\bibfnamefont {J.~L.}\ \bibnamefont {{Zdunik}}},\ and\ \bibinfo {author} {\bibfnamefont {R.}~\bibnamefont {{Schaefer}}},\ }\href@noop {} {\bibfield  {journal} {\bibinfo  {journal} {\aap}\ }\textbf {\bibinfo {volume} {160}},\ \bibinfo {pages} {121} (\bibinfo {year} {1986})}\BibitemShut {NoStop}%
\bibitem [{\citenamefont {{Alcock}}\ \emph {et~al.}(1986)\citenamefont {{Alcock}}, \citenamefont {{Farhi}},\ and\ \citenamefont {{Olinto}}}]{1986ApJ...310..261A}%
  \BibitemOpen
  \bibfield  {author} {\bibinfo {author} {\bibfnamefont {C.}~\bibnamefont {{Alcock}}}, \bibinfo {author} {\bibfnamefont {E.}~\bibnamefont {{Farhi}}},\ and\ \bibinfo {author} {\bibfnamefont {A.}~\bibnamefont {{Olinto}}},\ }\href {https://doi.org/10.1086/164679} {\bibfield  {journal} {\bibinfo  {journal} {\apj}\ }\textbf {\bibinfo {volume} {310}},\ \bibinfo {pages} {261} (\bibinfo {year} {1986})}\BibitemShut {NoStop}%
\bibitem [{\citenamefont {{Benvenuto}}\ and\ \citenamefont {{Horvath}}(1989)}]{1989MNRAS.241...43B}%
  \BibitemOpen
  \bibfield  {author} {\bibinfo {author} {\bibfnamefont {O.~G.}\ \bibnamefont {{Benvenuto}}}\ and\ \bibinfo {author} {\bibfnamefont {J.~E.}\ \bibnamefont {{Horvath}}},\ }\href {https://doi.org/10.1093/mnras/241.1.43} {\bibfield  {journal} {\bibinfo  {journal} {\mnras}\ }\textbf {\bibinfo {volume} {241}},\ \bibinfo {pages} {43} (\bibinfo {year} {1989})}\BibitemShut {NoStop}%
\bibitem [{\citenamefont {{Li}}\ \emph {et~al.}(2021{\natexlab{a}})\citenamefont {{Li}}, \citenamefont {{Yan}},\ and\ \citenamefont {{Ping}}}]{2021PhRvD.104d3002L}%
  \BibitemOpen
  \bibfield  {author} {\bibinfo {author} {\bibfnamefont {B.-L.}\ \bibnamefont {{Li}}}, \bibinfo {author} {\bibfnamefont {Y.}~\bibnamefont {{Yan}}},\ and\ \bibinfo {author} {\bibfnamefont {J.-L.}\ \bibnamefont {{Ping}}},\ }\href {https://doi.org/10.1103/PhysRevD.104.043002} {\bibfield  {journal} {\bibinfo  {journal} {\prd}\ }\textbf {\bibinfo {volume} {104}},\ \bibinfo {eid} {043002} (\bibinfo {year} {2021}{\natexlab{a}})}\BibitemShut {NoStop}%
\bibitem [{\citenamefont {{Horvath}}\ and\ \citenamefont {{Moraes}}(2021)}]{2021IJMPD..3050016H}%
  \BibitemOpen
  \bibfield  {author} {\bibinfo {author} {\bibfnamefont {J.~E.}\ \bibnamefont {{Horvath}}}\ and\ \bibinfo {author} {\bibfnamefont {P.~H.~R.~S.}\ \bibnamefont {{Moraes}}},\ }\href {https://doi.org/10.1142/S0218271821500164} {\bibfield  {journal} {\bibinfo  {journal} {International Journal of Modern Physics D}\ }\textbf {\bibinfo {volume} {30}},\ \bibinfo {eid} {2150016} (\bibinfo {year} {2021})},\ \Eprint {https://arxiv.org/abs/2012.00917} {arXiv:2012.00917 [astro-ph.HE]} \BibitemShut {NoStop}%
\bibitem [{\citenamefont {{Antoniadis}}\ \emph {et~al.}(2013)\citenamefont {{Antoniadis}}, \citenamefont {{Freire}}, \citenamefont {{Wex}}, \citenamefont {{Tauris}}, \citenamefont {{Lynch}}, \citenamefont {{van Kerkwijk}}, \citenamefont {{Kramer}}, \citenamefont {{Bassa}}, \citenamefont {{Dhillon}}, \citenamefont {{Driebe}}, \citenamefont {{Hessels}}, \citenamefont {{Kaspi}}, \citenamefont {{Kondratiev}}, \citenamefont {{Langer}}, \citenamefont {{Marsh}}, \citenamefont {{McLaughlin}}, \citenamefont {{Pennucci}}, \citenamefont {{Ransom}}, \citenamefont {{Stairs}}, \citenamefont {{van Leeuwen}}, \citenamefont {{Verbiest}},\ and\ \citenamefont {{Whelan}}}]{2013Sci...340..448A}%
  \BibitemOpen
  \bibfield  {author} {\bibinfo {author} {\bibfnamefont {J.}~\bibnamefont {{Antoniadis}}}, \bibinfo {author} {\bibfnamefont {P.~C.~C.}\ \bibnamefont {{Freire}}}, \bibinfo {author} {\bibfnamefont {N.}~\bibnamefont {{Wex}}}, \bibinfo {author} {\bibfnamefont {T.~M.}\ \bibnamefont {{Tauris}}}, \bibinfo {author} {\bibfnamefont {R.~S.}\ \bibnamefont {{Lynch}}}, \bibinfo {author} {\bibfnamefont {M.~H.}\ \bibnamefont {{van Kerkwijk}}}, \bibinfo {author} {\bibfnamefont {M.}~\bibnamefont {{Kramer}}}, \bibinfo {author} {\bibfnamefont {C.}~\bibnamefont {{Bassa}}}, \bibinfo {author} {\bibfnamefont {V.~S.}\ \bibnamefont {{Dhillon}}}, \bibinfo {author} {\bibfnamefont {T.}~\bibnamefont {{Driebe}}}, \bibinfo {author} {\bibfnamefont {J.~W.~T.}\ \bibnamefont {{Hessels}}}, \bibinfo {author} {\bibfnamefont {V.~M.}\ \bibnamefont {{Kaspi}}}, \bibinfo {author} {\bibfnamefont {V.~I.}\ \bibnamefont {{Kondratiev}}}, \bibinfo {author} {\bibfnamefont {N.}~\bibnamefont {{Langer}}}, \bibinfo {author} {\bibfnamefont {T.~R.}\ \bibnamefont
  {{Marsh}}}, \bibinfo {author} {\bibfnamefont {M.~A.}\ \bibnamefont {{McLaughlin}}}, \bibinfo {author} {\bibfnamefont {T.~T.}\ \bibnamefont {{Pennucci}}}, \bibinfo {author} {\bibfnamefont {S.~M.}\ \bibnamefont {{Ransom}}}, \bibinfo {author} {\bibfnamefont {I.~H.}\ \bibnamefont {{Stairs}}}, \bibinfo {author} {\bibfnamefont {J.}~\bibnamefont {{van Leeuwen}}}, \bibinfo {author} {\bibfnamefont {J.~P.~W.}\ \bibnamefont {{Verbiest}}},\ and\ \bibinfo {author} {\bibfnamefont {D.~G.}\ \bibnamefont {{Whelan}}},\ }\href {https://doi.org/10.1126/science.1233232} {\bibfield  {journal} {\bibinfo  {journal} {Science}\ }\textbf {\bibinfo {volume} {340}},\ \bibinfo {pages} {448} (\bibinfo {year} {2013})},\ \Eprint {https://arxiv.org/abs/1304.6875} {arXiv:1304.6875 [astro-ph.HE]} \BibitemShut {NoStop}%
\bibitem [{\citenamefont {{Fonseca}}\ \emph {et~al.}(2021)\citenamefont {{Fonseca}}, \citenamefont {{Cromartie}}, \citenamefont {{Pennucci}}, \citenamefont {{Ray}}, \citenamefont {{Kirichenko}}, \citenamefont {{Ransom}}, \citenamefont {{Demorest}}, \citenamefont {{Stairs}}, \citenamefont {{Arzoumanian}}, \citenamefont {{Guillemot}}, \citenamefont {{Parthasarathy}}, \citenamefont {{Kerr}}, \citenamefont {{Cognard}}, \citenamefont {{Baker}}, \citenamefont {{Blumer}}, \citenamefont {{Brook}}, \citenamefont {{DeCesar}}, \citenamefont {{Dolch}}, \citenamefont {{Dong}}, \citenamefont {{Ferrara}}, \citenamefont {{Fiore}}, \citenamefont {{Garver-Daniels}}, \citenamefont {{Good}}, \citenamefont {{Jennings}}, \citenamefont {{Jones}}, \citenamefont {{Kaspi}}, \citenamefont {{Lam}}, \citenamefont {{Lorimer}}, \citenamefont {{Luo}}, \citenamefont {{McEwen}}, \citenamefont {{McKee}}, \citenamefont {{McLaughlin}}, \citenamefont {{McMann}}, \citenamefont {{Meyers}}, \citenamefont {{Naidu}}, \citenamefont {{Ng}},
  \citenamefont {{Nice}}, \citenamefont {{Pol}}, \citenamefont {{Radovan}}, \citenamefont {{Shapiro-Albert}}, \citenamefont {{Tan}}, \citenamefont {{Tendulkar}}, \citenamefont {{Swiggum}}, \citenamefont {{Wahl}},\ and\ \citenamefont {{Zhu}}}]{2021ApJ...915L..12F}%
  \BibitemOpen
  \bibfield  {author} {\bibinfo {author} {\bibfnamefont {E.}~\bibnamefont {{Fonseca}}}, \bibinfo {author} {\bibfnamefont {H.~T.}\ \bibnamefont {{Cromartie}}}, \bibinfo {author} {\bibfnamefont {T.~T.}\ \bibnamefont {{Pennucci}}}, \bibinfo {author} {\bibfnamefont {P.~S.}\ \bibnamefont {{Ray}}}, \bibinfo {author} {\bibfnamefont {A.~Y.}\ \bibnamefont {{Kirichenko}}}, \bibinfo {author} {\bibfnamefont {S.~M.}\ \bibnamefont {{Ransom}}}, \bibinfo {author} {\bibfnamefont {P.~B.}\ \bibnamefont {{Demorest}}}, \bibinfo {author} {\bibfnamefont {I.~H.}\ \bibnamefont {{Stairs}}}, \bibinfo {author} {\bibfnamefont {Z.}~\bibnamefont {{Arzoumanian}}}, \bibinfo {author} {\bibfnamefont {L.}~\bibnamefont {{Guillemot}}}, \bibinfo {author} {\bibfnamefont {A.}~\bibnamefont {{Parthasarathy}}}, \bibinfo {author} {\bibfnamefont {M.}~\bibnamefont {{Kerr}}}, \bibinfo {author} {\bibfnamefont {I.}~\bibnamefont {{Cognard}}}, \bibinfo {author} {\bibfnamefont {P.~T.}\ \bibnamefont {{Baker}}}, \bibinfo {author} {\bibfnamefont {H.}~\bibnamefont
  {{Blumer}}}, \bibinfo {author} {\bibfnamefont {P.~R.}\ \bibnamefont {{Brook}}}, \bibinfo {author} {\bibfnamefont {M.}~\bibnamefont {{DeCesar}}}, \bibinfo {author} {\bibfnamefont {T.}~\bibnamefont {{Dolch}}}, \bibinfo {author} {\bibfnamefont {F.~A.}\ \bibnamefont {{Dong}}}, \bibinfo {author} {\bibfnamefont {E.~C.}\ \bibnamefont {{Ferrara}}}, \bibinfo {author} {\bibfnamefont {W.}~\bibnamefont {{Fiore}}}, \bibinfo {author} {\bibfnamefont {N.}~\bibnamefont {{Garver-Daniels}}}, \bibinfo {author} {\bibfnamefont {D.~C.}\ \bibnamefont {{Good}}}, \bibinfo {author} {\bibfnamefont {R.}~\bibnamefont {{Jennings}}}, \bibinfo {author} {\bibfnamefont {M.~L.}\ \bibnamefont {{Jones}}}, \bibinfo {author} {\bibfnamefont {V.~M.}\ \bibnamefont {{Kaspi}}}, \bibinfo {author} {\bibfnamefont {M.~T.}\ \bibnamefont {{Lam}}}, \bibinfo {author} {\bibfnamefont {D.~R.}\ \bibnamefont {{Lorimer}}}, \bibinfo {author} {\bibfnamefont {J.}~\bibnamefont {{Luo}}}, \bibinfo {author} {\bibfnamefont {A.}~\bibnamefont {{McEwen}}}, \bibinfo {author}
  {\bibfnamefont {J.~W.}\ \bibnamefont {{McKee}}}, \bibinfo {author} {\bibfnamefont {M.~A.}\ \bibnamefont {{McLaughlin}}}, \bibinfo {author} {\bibfnamefont {N.}~\bibnamefont {{McMann}}}, \bibinfo {author} {\bibfnamefont {B.~W.}\ \bibnamefont {{Meyers}}}, \bibinfo {author} {\bibfnamefont {A.}~\bibnamefont {{Naidu}}}, \bibinfo {author} {\bibfnamefont {C.}~\bibnamefont {{Ng}}}, \bibinfo {author} {\bibfnamefont {D.~J.}\ \bibnamefont {{Nice}}}, \bibinfo {author} {\bibfnamefont {N.}~\bibnamefont {{Pol}}}, \bibinfo {author} {\bibfnamefont {H.~A.}\ \bibnamefont {{Radovan}}}, \bibinfo {author} {\bibfnamefont {B.}~\bibnamefont {{Shapiro-Albert}}}, \bibinfo {author} {\bibfnamefont {C.~M.}\ \bibnamefont {{Tan}}}, \bibinfo {author} {\bibfnamefont {S.~P.}\ \bibnamefont {{Tendulkar}}}, \bibinfo {author} {\bibfnamefont {J.~K.}\ \bibnamefont {{Swiggum}}}, \bibinfo {author} {\bibfnamefont {H.~M.}\ \bibnamefont {{Wahl}}},\ and\ \bibinfo {author} {\bibfnamefont {W.~W.}\ \bibnamefont {{Zhu}}},\ }\href
  {https://doi.org/10.3847/2041-8213/ac03b8} {\bibfield  {journal} {\bibinfo  {journal} {\apjl}\ }\textbf {\bibinfo {volume} {915}},\ \bibinfo {eid} {L12} (\bibinfo {year} {2021})},\ \Eprint {https://arxiv.org/abs/2104.00880} {arXiv:2104.00880 [astro-ph.HE]} \BibitemShut {NoStop}%
\bibitem [{\citenamefont {{Romani}}\ \emph {et~al.}(2022)\citenamefont {{Romani}}, \citenamefont {{Kandel}}, \citenamefont {{Filippenko}}, \citenamefont {{Brink}},\ and\ \citenamefont {{Zheng}}}]{2022ApJ...934L..17R}%
  \BibitemOpen
  \bibfield  {author} {\bibinfo {author} {\bibfnamefont {R.~W.}\ \bibnamefont {{Romani}}}, \bibinfo {author} {\bibfnamefont {D.}~\bibnamefont {{Kandel}}}, \bibinfo {author} {\bibfnamefont {A.~V.}\ \bibnamefont {{Filippenko}}}, \bibinfo {author} {\bibfnamefont {T.~G.}\ \bibnamefont {{Brink}}},\ and\ \bibinfo {author} {\bibfnamefont {W.}~\bibnamefont {{Zheng}}},\ }\href {https://doi.org/10.3847/2041-8213/ac8007} {\bibfield  {journal} {\bibinfo  {journal} {\apjl}\ }\textbf {\bibinfo {volume} {934}},\ \bibinfo {eid} {L17} (\bibinfo {year} {2022})},\ \Eprint {https://arxiv.org/abs/2207.05124} {arXiv:2207.05124 [astro-ph.HE]} \BibitemShut {NoStop}%
\bibitem [{\citenamefont {{Miller}}\ \emph {et~al.}(2019)\citenamefont {{Miller}}, \citenamefont {{Lamb}}, \citenamefont {{Dittmann}}, \citenamefont {{Bogdanov}}, \citenamefont {{Arzoumanian}}, \citenamefont {{Gendreau}}, \citenamefont {{Guillot}}, \citenamefont {{Harding}}, \citenamefont {{Ho}}, \citenamefont {{Lattimer}}, \citenamefont {{Ludlam}}, \citenamefont {{Mahmoodifar}}, \citenamefont {{Morsink}}, \citenamefont {{Ray}}, \citenamefont {{Strohmayer}}, \citenamefont {{Wood}}, \citenamefont {{Enoto}}, \citenamefont {{Foster}}, \citenamefont {{Okajima}}, \citenamefont {{Prigozhin}},\ and\ \citenamefont {{Soong}}}]{2019ApJ...887L..24M}%
  \BibitemOpen
  \bibfield  {author} {\bibinfo {author} {\bibfnamefont {M.~C.}\ \bibnamefont {{Miller}}}, \bibinfo {author} {\bibfnamefont {F.~K.}\ \bibnamefont {{Lamb}}}, \bibinfo {author} {\bibfnamefont {A.~J.}\ \bibnamefont {{Dittmann}}}, \bibinfo {author} {\bibfnamefont {S.}~\bibnamefont {{Bogdanov}}}, \bibinfo {author} {\bibfnamefont {Z.}~\bibnamefont {{Arzoumanian}}}, \bibinfo {author} {\bibfnamefont {K.~C.}\ \bibnamefont {{Gendreau}}}, \bibinfo {author} {\bibfnamefont {S.}~\bibnamefont {{Guillot}}}, \bibinfo {author} {\bibfnamefont {A.~K.}\ \bibnamefont {{Harding}}}, \bibinfo {author} {\bibfnamefont {W.~C.~G.}\ \bibnamefont {{Ho}}}, \bibinfo {author} {\bibfnamefont {J.~M.}\ \bibnamefont {{Lattimer}}}, \bibinfo {author} {\bibfnamefont {R.~M.}\ \bibnamefont {{Ludlam}}}, \bibinfo {author} {\bibfnamefont {S.}~\bibnamefont {{Mahmoodifar}}}, \bibinfo {author} {\bibfnamefont {S.~M.}\ \bibnamefont {{Morsink}}}, \bibinfo {author} {\bibfnamefont {P.~S.}\ \bibnamefont {{Ray}}}, \bibinfo {author} {\bibfnamefont {T.~E.}\
  \bibnamefont {{Strohmayer}}}, \bibinfo {author} {\bibfnamefont {K.~S.}\ \bibnamefont {{Wood}}}, \bibinfo {author} {\bibfnamefont {T.}~\bibnamefont {{Enoto}}}, \bibinfo {author} {\bibfnamefont {R.}~\bibnamefont {{Foster}}}, \bibinfo {author} {\bibfnamefont {T.}~\bibnamefont {{Okajima}}}, \bibinfo {author} {\bibfnamefont {G.}~\bibnamefont {{Prigozhin}}},\ and\ \bibinfo {author} {\bibfnamefont {Y.}~\bibnamefont {{Soong}}},\ }\href {https://doi.org/10.3847/2041-8213/ab50c5} {\bibfield  {journal} {\bibinfo  {journal} {\apjl}\ }\textbf {\bibinfo {volume} {887}},\ \bibinfo {eid} {L24} (\bibinfo {year} {2019})},\ \Eprint {https://arxiv.org/abs/1912.05705} {arXiv:1912.05705 [astro-ph.HE]} \BibitemShut {NoStop}%
\bibitem [{\citenamefont {{Riley}}\ \emph {et~al.}(2019)\citenamefont {{Riley}}, \citenamefont {{Watts}}, \citenamefont {{Bogdanov}}, \citenamefont {{Ray}}, \citenamefont {{Ludlam}}, \citenamefont {{Guillot}}, \citenamefont {{Arzoumanian}}, \citenamefont {{Baker}}, \citenamefont {{Bilous}}, \citenamefont {{Chakrabarty}}, \citenamefont {{Gendreau}}, \citenamefont {{Harding}}, \citenamefont {{Ho}}, \citenamefont {{Lattimer}}, \citenamefont {{Morsink}},\ and\ \citenamefont {{Strohmayer}}}]{2019ApJ...887L..21R}%
  \BibitemOpen
  \bibfield  {author} {\bibinfo {author} {\bibfnamefont {T.~E.}\ \bibnamefont {{Riley}}}, \bibinfo {author} {\bibfnamefont {A.~L.}\ \bibnamefont {{Watts}}}, \bibinfo {author} {\bibfnamefont {S.}~\bibnamefont {{Bogdanov}}}, \bibinfo {author} {\bibfnamefont {P.~S.}\ \bibnamefont {{Ray}}}, \bibinfo {author} {\bibfnamefont {R.~M.}\ \bibnamefont {{Ludlam}}}, \bibinfo {author} {\bibfnamefont {S.}~\bibnamefont {{Guillot}}}, \bibinfo {author} {\bibfnamefont {Z.}~\bibnamefont {{Arzoumanian}}}, \bibinfo {author} {\bibfnamefont {C.~L.}\ \bibnamefont {{Baker}}}, \bibinfo {author} {\bibfnamefont {A.~V.}\ \bibnamefont {{Bilous}}}, \bibinfo {author} {\bibfnamefont {D.}~\bibnamefont {{Chakrabarty}}}, \bibinfo {author} {\bibfnamefont {K.~C.}\ \bibnamefont {{Gendreau}}}, \bibinfo {author} {\bibfnamefont {A.~K.}\ \bibnamefont {{Harding}}}, \bibinfo {author} {\bibfnamefont {W.~C.~G.}\ \bibnamefont {{Ho}}}, \bibinfo {author} {\bibfnamefont {J.~M.}\ \bibnamefont {{Lattimer}}}, \bibinfo {author} {\bibfnamefont {S.~M.}\ \bibnamefont
  {{Morsink}}},\ and\ \bibinfo {author} {\bibfnamefont {T.~E.}\ \bibnamefont {{Strohmayer}}},\ }\href {https://doi.org/10.3847/2041-8213/ab481c} {\bibfield  {journal} {\bibinfo  {journal} {\apjl}\ }\textbf {\bibinfo {volume} {887}},\ \bibinfo {eid} {L21} (\bibinfo {year} {2019})},\ \Eprint {https://arxiv.org/abs/1912.05702} {arXiv:1912.05702 [astro-ph.HE]} \BibitemShut {NoStop}%
\bibitem [{\citenamefont {{Essick}}\ \emph {et~al.}(2020)\citenamefont {{Essick}}, \citenamefont {{Landry}},\ and\ \citenamefont {{Holz}}}]{2020PhRvD.101f3007E}%
  \BibitemOpen
  \bibfield  {author} {\bibinfo {author} {\bibfnamefont {R.}~\bibnamefont {{Essick}}}, \bibinfo {author} {\bibfnamefont {P.}~\bibnamefont {{Landry}}},\ and\ \bibinfo {author} {\bibfnamefont {D.~E.}\ \bibnamefont {{Holz}}},\ }\href {https://doi.org/10.1103/PhysRevD.101.063007} {\bibfield  {journal} {\bibinfo  {journal} {\prd}\ }\textbf {\bibinfo {volume} {101}},\ \bibinfo {eid} {063007} (\bibinfo {year} {2020})},\ \Eprint {https://arxiv.org/abs/1910.09740} {arXiv:1910.09740 [astro-ph.HE]} \BibitemShut {NoStop}%
\bibitem [{\citenamefont {{Miller}}\ \emph {et~al.}(2021)\citenamefont {{Miller}}, \citenamefont {{Lamb}}, \citenamefont {{Dittmann}}, \citenamefont {{Bogdanov}}, \citenamefont {{Arzoumanian}}, \citenamefont {{Gendreau}}, \citenamefont {{Guillot}}, \citenamefont {{Ho}}, \citenamefont {{Lattimer}}, \citenamefont {{Loewenstein}}, \citenamefont {{Morsink}}, \citenamefont {{Ray}}, \citenamefont {{Wolff}}, \citenamefont {{Baker}}, \citenamefont {{Cazeau}}, \citenamefont {{Manthripragada}}, \citenamefont {{Markwardt}}, \citenamefont {{Okajima}}, \citenamefont {{Pollard}}, \citenamefont {{Cognard}}, \citenamefont {{Cromartie}}, \citenamefont {{Fonseca}}, \citenamefont {{Guillemot}}, \citenamefont {{Kerr}}, \citenamefont {{Parthasarathy}}, \citenamefont {{Pennucci}}, \citenamefont {{Ransom}},\ and\ \citenamefont {{Stairs}}}]{2021ApJ...918L..28M}%
  \BibitemOpen
  \bibfield  {author} {\bibinfo {author} {\bibfnamefont {M.~C.}\ \bibnamefont {{Miller}}}, \bibinfo {author} {\bibfnamefont {F.~K.}\ \bibnamefont {{Lamb}}}, \bibinfo {author} {\bibfnamefont {A.~J.}\ \bibnamefont {{Dittmann}}}, \bibinfo {author} {\bibfnamefont {S.}~\bibnamefont {{Bogdanov}}}, \bibinfo {author} {\bibfnamefont {Z.}~\bibnamefont {{Arzoumanian}}}, \bibinfo {author} {\bibfnamefont {K.~C.}\ \bibnamefont {{Gendreau}}}, \bibinfo {author} {\bibfnamefont {S.}~\bibnamefont {{Guillot}}}, \bibinfo {author} {\bibfnamefont {W.~C.~G.}\ \bibnamefont {{Ho}}}, \bibinfo {author} {\bibfnamefont {J.~M.}\ \bibnamefont {{Lattimer}}}, \bibinfo {author} {\bibfnamefont {M.}~\bibnamefont {{Loewenstein}}}, \bibinfo {author} {\bibfnamefont {S.~M.}\ \bibnamefont {{Morsink}}}, \bibinfo {author} {\bibfnamefont {P.~S.}\ \bibnamefont {{Ray}}}, \bibinfo {author} {\bibfnamefont {M.~T.}\ \bibnamefont {{Wolff}}}, \bibinfo {author} {\bibfnamefont {C.~L.}\ \bibnamefont {{Baker}}}, \bibinfo {author} {\bibfnamefont {T.}~\bibnamefont
  {{Cazeau}}}, \bibinfo {author} {\bibfnamefont {S.}~\bibnamefont {{Manthripragada}}}, \bibinfo {author} {\bibfnamefont {C.~B.}\ \bibnamefont {{Markwardt}}}, \bibinfo {author} {\bibfnamefont {T.}~\bibnamefont {{Okajima}}}, \bibinfo {author} {\bibfnamefont {S.}~\bibnamefont {{Pollard}}}, \bibinfo {author} {\bibfnamefont {I.}~\bibnamefont {{Cognard}}}, \bibinfo {author} {\bibfnamefont {H.~T.}\ \bibnamefont {{Cromartie}}}, \bibinfo {author} {\bibfnamefont {E.}~\bibnamefont {{Fonseca}}}, \bibinfo {author} {\bibfnamefont {L.}~\bibnamefont {{Guillemot}}}, \bibinfo {author} {\bibfnamefont {M.}~\bibnamefont {{Kerr}}}, \bibinfo {author} {\bibfnamefont {A.}~\bibnamefont {{Parthasarathy}}}, \bibinfo {author} {\bibfnamefont {T.~T.}\ \bibnamefont {{Pennucci}}}, \bibinfo {author} {\bibfnamefont {S.}~\bibnamefont {{Ransom}}},\ and\ \bibinfo {author} {\bibfnamefont {I.}~\bibnamefont {{Stairs}}},\ }\href {https://doi.org/10.3847/2041-8213/ac089b} {\bibfield  {journal} {\bibinfo  {journal} {\apjl}\ }\textbf {\bibinfo {volume}
  {918}},\ \bibinfo {eid} {L28} (\bibinfo {year} {2021})},\ \Eprint {https://arxiv.org/abs/2105.06979} {arXiv:2105.06979 [astro-ph.HE]} \BibitemShut {NoStop}%
\bibitem [{\citenamefont {{Riley}}\ \emph {et~al.}(2021)\citenamefont {{Riley}}, \citenamefont {{Watts}}, \citenamefont {{Ray}}, \citenamefont {{Bogdanov}}, \citenamefont {{Guillot}}, \citenamefont {{Morsink}}, \citenamefont {{Bilous}}, \citenamefont {{Arzoumanian}}, \citenamefont {{Choudhury}}, \citenamefont {{Deneva}}, \citenamefont {{Gendreau}}, \citenamefont {{Harding}}, \citenamefont {{Ho}}, \citenamefont {{Lattimer}}, \citenamefont {{Loewenstein}}, \citenamefont {{Ludlam}}, \citenamefont {{Markwardt}}, \citenamefont {{Okajima}}, \citenamefont {{Prescod-Weinstein}}, \citenamefont {{Remillard}}, \citenamefont {{Wolff}}, \citenamefont {{Fonseca}}, \citenamefont {{Cromartie}}, \citenamefont {{Kerr}}, \citenamefont {{Pennucci}}, \citenamefont {{Parthasarathy}}, \citenamefont {{Ransom}}, \citenamefont {{Stairs}}, \citenamefont {{Guillemot}},\ and\ \citenamefont {{Cognard}}}]{2021ApJ...918L..27R}%
  \BibitemOpen
  \bibfield  {author} {\bibinfo {author} {\bibfnamefont {T.~E.}\ \bibnamefont {{Riley}}}, \bibinfo {author} {\bibfnamefont {A.~L.}\ \bibnamefont {{Watts}}}, \bibinfo {author} {\bibfnamefont {P.~S.}\ \bibnamefont {{Ray}}}, \bibinfo {author} {\bibfnamefont {S.}~\bibnamefont {{Bogdanov}}}, \bibinfo {author} {\bibfnamefont {S.}~\bibnamefont {{Guillot}}}, \bibinfo {author} {\bibfnamefont {S.~M.}\ \bibnamefont {{Morsink}}}, \bibinfo {author} {\bibfnamefont {A.~V.}\ \bibnamefont {{Bilous}}}, \bibinfo {author} {\bibfnamefont {Z.}~\bibnamefont {{Arzoumanian}}}, \bibinfo {author} {\bibfnamefont {D.}~\bibnamefont {{Choudhury}}}, \bibinfo {author} {\bibfnamefont {J.~S.}\ \bibnamefont {{Deneva}}}, \bibinfo {author} {\bibfnamefont {K.~C.}\ \bibnamefont {{Gendreau}}}, \bibinfo {author} {\bibfnamefont {A.~K.}\ \bibnamefont {{Harding}}}, \bibinfo {author} {\bibfnamefont {W.~C.~G.}\ \bibnamefont {{Ho}}}, \bibinfo {author} {\bibfnamefont {J.~M.}\ \bibnamefont {{Lattimer}}}, \bibinfo {author} {\bibfnamefont {M.}~\bibnamefont
  {{Loewenstein}}}, \bibinfo {author} {\bibfnamefont {R.~M.}\ \bibnamefont {{Ludlam}}}, \bibinfo {author} {\bibfnamefont {C.~B.}\ \bibnamefont {{Markwardt}}}, \bibinfo {author} {\bibfnamefont {T.}~\bibnamefont {{Okajima}}}, \bibinfo {author} {\bibfnamefont {C.}~\bibnamefont {{Prescod-Weinstein}}}, \bibinfo {author} {\bibfnamefont {R.~A.}\ \bibnamefont {{Remillard}}}, \bibinfo {author} {\bibfnamefont {M.~T.}\ \bibnamefont {{Wolff}}}, \bibinfo {author} {\bibfnamefont {E.}~\bibnamefont {{Fonseca}}}, \bibinfo {author} {\bibfnamefont {H.~T.}\ \bibnamefont {{Cromartie}}}, \bibinfo {author} {\bibfnamefont {M.}~\bibnamefont {{Kerr}}}, \bibinfo {author} {\bibfnamefont {T.~T.}\ \bibnamefont {{Pennucci}}}, \bibinfo {author} {\bibfnamefont {A.}~\bibnamefont {{Parthasarathy}}}, \bibinfo {author} {\bibfnamefont {S.}~\bibnamefont {{Ransom}}}, \bibinfo {author} {\bibfnamefont {I.}~\bibnamefont {{Stairs}}}, \bibinfo {author} {\bibfnamefont {L.}~\bibnamefont {{Guillemot}}},\ and\ \bibinfo {author} {\bibfnamefont
  {I.}~\bibnamefont {{Cognard}}},\ }\href {https://doi.org/10.3847/2041-8213/ac0a81} {\bibfield  {journal} {\bibinfo  {journal} {\apjl}\ }\textbf {\bibinfo {volume} {918}},\ \bibinfo {eid} {L27} (\bibinfo {year} {2021})},\ \Eprint {https://arxiv.org/abs/2105.06980} {arXiv:2105.06980 [astro-ph.HE]} \BibitemShut {NoStop}%
\bibitem [{\citenamefont {{Yagi}}\ and\ \citenamefont {{Yunes}}(2013)}]{2013Sci...341..365Y}%
  \BibitemOpen
  \bibfield  {author} {\bibinfo {author} {\bibfnamefont {K.}~\bibnamefont {{Yagi}}}\ and\ \bibinfo {author} {\bibfnamefont {N.}~\bibnamefont {{Yunes}}},\ }\href {https://doi.org/10.1126/science.1236462} {\bibfield  {journal} {\bibinfo  {journal} {Science}\ }\textbf {\bibinfo {volume} {341}},\ \bibinfo {pages} {365} (\bibinfo {year} {2013})},\ \Eprint {https://arxiv.org/abs/1302.4499} {arXiv:1302.4499 [gr-qc]} \BibitemShut {NoStop}%
\bibitem [{\citenamefont {{Yagi}}\ \emph {et~al.}(2014)\citenamefont {{Yagi}}, \citenamefont {{Stein}}, \citenamefont {{Pappas}}, \citenamefont {{Yunes}},\ and\ \citenamefont {{Apostolatos}}}]{2014PhRvD..90f3010Y}%
  \BibitemOpen
  \bibfield  {author} {\bibinfo {author} {\bibfnamefont {K.}~\bibnamefont {{Yagi}}}, \bibinfo {author} {\bibfnamefont {L.~C.}\ \bibnamefont {{Stein}}}, \bibinfo {author} {\bibfnamefont {G.}~\bibnamefont {{Pappas}}}, \bibinfo {author} {\bibfnamefont {N.}~\bibnamefont {{Yunes}}},\ and\ \bibinfo {author} {\bibfnamefont {T.~A.}\ \bibnamefont {{Apostolatos}}},\ }\href {https://doi.org/10.1103/PhysRevD.90.063010} {\bibfield  {journal} {\bibinfo  {journal} {\prd}\ }\textbf {\bibinfo {volume} {90}},\ \bibinfo {eid} {063010} (\bibinfo {year} {2014})},\ \Eprint {https://arxiv.org/abs/1406.7587} {arXiv:1406.7587 [gr-qc]} \BibitemShut {NoStop}%
\bibitem [{\citenamefont {{Yagi}}\ and\ \citenamefont {{Yunes}}(2017)}]{2017PhR...681....1Y}%
  \BibitemOpen
  \bibfield  {author} {\bibinfo {author} {\bibfnamefont {K.}~\bibnamefont {{Yagi}}}\ and\ \bibinfo {author} {\bibfnamefont {N.}~\bibnamefont {{Yunes}}},\ }\href {https://doi.org/10.1016/j.physrep.2017.03.002} {\bibfield  {journal} {\bibinfo  {journal} {Phys. Reports}\ }\textbf {\bibinfo {volume} {681}},\ \bibinfo {pages} {1} (\bibinfo {year} {2017})}\BibitemShut {NoStop}%
\bibitem [{\citenamefont {{Lugones}}\ and\ \citenamefont {{Grunfeld}}(2025)}]{2025arXiv250311515L}%
  \BibitemOpen
  \bibfield  {author} {\bibinfo {author} {\bibfnamefont {G.}~\bibnamefont {{Lugones}}}\ and\ \bibinfo {author} {\bibfnamefont {A.~G.}\ \bibnamefont {{Grunfeld}}},\ }\href@noop {} {\bibfield  {journal} {\bibinfo  {journal} {arXiv e-prints}\ ,\ \bibinfo {eid} {arXiv:2503.11515}} (\bibinfo {year} {2025})},\ \Eprint {https://arxiv.org/abs/2503.11515} {arXiv:2503.11515 [nucl-th]} \BibitemShut {NoStop}%
\bibitem [{\citenamefont {{The LIGO and Virgo Scientific Collaboration}}\ \emph {et~al.}(2018)\citenamefont {{The LIGO and Virgo Scientific Collaboration}}, \citenamefont {{Abbott}}, \citenamefont {{Abbott}}, \citenamefont {{Abbott}}, \citenamefont {{Acernese}}, \citenamefont {{Ackley}}, \citenamefont {{Adams}}, \citenamefont {{Adams}}, \citenamefont {{Addesso}}, \citenamefont {{Adhikari}}, \citenamefont {{Adya}} \emph {et~al.}}]{2018PhRvL.121p1101A}%
  \BibitemOpen
  \bibfield  {author} {\bibinfo {author} {\bibnamefont {{The LIGO and Virgo Scientific Collaboration}}}, \bibinfo {author} {\bibfnamefont {B.~P.}\ \bibnamefont {{Abbott}}}, \bibinfo {author} {\bibfnamefont {R.}~\bibnamefont {{Abbott}}}, \bibinfo {author} {\bibfnamefont {T.~D.}\ \bibnamefont {{Abbott}}}, \bibinfo {author} {\bibfnamefont {F.}~\bibnamefont {{Acernese}}}, \bibinfo {author} {\bibfnamefont {K.}~\bibnamefont {{Ackley}}}, \bibinfo {author} {\bibfnamefont {C.}~\bibnamefont {{Adams}}}, \bibinfo {author} {\bibfnamefont {T.}~\bibnamefont {{Adams}}}, \bibinfo {author} {\bibfnamefont {P.}~\bibnamefont {{Addesso}}}, \bibinfo {author} {\bibfnamefont {R.~X.}\ \bibnamefont {{Adhikari}}}, \bibinfo {author} {\bibfnamefont {V.~B.}\ \bibnamefont {{Adya}}}, \emph {et~al.},\ }\href {https://doi.org/10.1103/PhysRevLett.121.161101} {\bibfield  {journal} {\bibinfo  {journal} {\prl}\ }\textbf {\bibinfo {volume} {121}},\ \bibinfo {eid} {161101} (\bibinfo {year} {2018})},\ \Eprint {https://arxiv.org/abs/1805.11581}
  {arXiv:1805.11581 [gr-qc]} \BibitemShut {NoStop}%
\bibitem [{\citenamefont {{Hinderer}}(2008)}]{2008ApJ...677.1216H}%
  \BibitemOpen
  \bibfield  {author} {\bibinfo {author} {\bibfnamefont {T.}~\bibnamefont {{Hinderer}}},\ }\href {https://doi.org/10.1086/533487} {\bibfield  {journal} {\bibinfo  {journal} {\apj}\ }\textbf {\bibinfo {volume} {677}},\ \bibinfo {pages} {1216} (\bibinfo {year} {2008})},\ \Eprint {https://arxiv.org/abs/0711.2420} {arXiv:0711.2420 [astro-ph]} \BibitemShut {NoStop}%
\bibitem [{\citenamefont {{Damour}}\ and\ \citenamefont {{Nagar}}(2009)}]{2009PhRvD..80h4035D}%
  \BibitemOpen
  \bibfield  {author} {\bibinfo {author} {\bibfnamefont {T.}~\bibnamefont {{Damour}}}\ and\ \bibinfo {author} {\bibfnamefont {A.}~\bibnamefont {{Nagar}}},\ }\href {https://doi.org/10.1103/PhysRevD.80.084035} {\bibfield  {journal} {\bibinfo  {journal} {\prd}\ }\textbf {\bibinfo {volume} {80}},\ \bibinfo {eid} {084035} (\bibinfo {year} {2009})},\ \Eprint {https://arxiv.org/abs/0906.0096} {arXiv:0906.0096 [gr-qc]} \BibitemShut {NoStop}%
\bibitem [{\citenamefont {{Hinderer}}\ \emph {et~al.}(2010)\citenamefont {{Hinderer}}, \citenamefont {{Lackey}}, \citenamefont {{Lang}},\ and\ \citenamefont {{Read}}}]{2010PhRvD..81l3016H}%
  \BibitemOpen
  \bibfield  {author} {\bibinfo {author} {\bibfnamefont {T.}~\bibnamefont {{Hinderer}}}, \bibinfo {author} {\bibfnamefont {B.~D.}\ \bibnamefont {{Lackey}}}, \bibinfo {author} {\bibfnamefont {R.~N.}\ \bibnamefont {{Lang}}},\ and\ \bibinfo {author} {\bibfnamefont {J.~S.}\ \bibnamefont {{Read}}},\ }\href {https://doi.org/10.1103/PhysRevD.81.123016} {\bibfield  {journal} {\bibinfo  {journal} {\prd}\ }\textbf {\bibinfo {volume} {81}},\ \bibinfo {eid} {123016} (\bibinfo {year} {2010})},\ \Eprint {https://arxiv.org/abs/0911.3535} {arXiv:0911.3535 [astro-ph.HE]} \BibitemShut {NoStop}%
\bibitem [{\citenamefont {{The LIGO and Virgo Scientific Collaboration}}\ \emph {et~al.}(2019)\citenamefont {{The LIGO and Virgo Scientific Collaboration}}, \citenamefont {{Abbott}}, \citenamefont {{Abbott}}, \citenamefont {{Abbott}}, \citenamefont {{Acernese}}, \citenamefont {{Ackley}}, \citenamefont {{Adams}}, \citenamefont {{Adams}} \emph {et~al.}}]{2019PhRvX...9a1001A}%
  \BibitemOpen
  \bibfield  {author} {\bibinfo {author} {\bibnamefont {{The LIGO and Virgo Scientific Collaboration}}}, \bibinfo {author} {\bibfnamefont {B.~P.}\ \bibnamefont {{Abbott}}}, \bibinfo {author} {\bibfnamefont {R.}~\bibnamefont {{Abbott}}}, \bibinfo {author} {\bibfnamefont {T.~D.}\ \bibnamefont {{Abbott}}}, \bibinfo {author} {\bibfnamefont {F.}~\bibnamefont {{Acernese}}}, \bibinfo {author} {\bibfnamefont {K.}~\bibnamefont {{Ackley}}}, \bibinfo {author} {\bibfnamefont {C.}~\bibnamefont {{Adams}}}, \bibinfo {author} {\bibfnamefont {T.}~\bibnamefont {{Adams}}}, \emph {et~al.},\ }\href {https://doi.org/10.1103/PhysRevX.9.011001} {\bibfield  {journal} {\bibinfo  {journal} {Physical Review X}\ }\textbf {\bibinfo {volume} {9}},\ \bibinfo {eid} {011001} (\bibinfo {year} {2019})},\ \Eprint {https://arxiv.org/abs/1805.11579} {arXiv:1805.11579 [gr-qc]} \BibitemShut {NoStop}%
\bibitem [{\citenamefont {{Annala}}\ \emph {et~al.}(2018)\citenamefont {{Annala}}, \citenamefont {{Gorda}}, \citenamefont {{Kurkela}},\ and\ \citenamefont {{Vuorinen}}}]{2018PhRvL.120q2703A}%
  \BibitemOpen
  \bibfield  {author} {\bibinfo {author} {\bibfnamefont {E.}~\bibnamefont {{Annala}}}, \bibinfo {author} {\bibfnamefont {T.}~\bibnamefont {{Gorda}}}, \bibinfo {author} {\bibfnamefont {A.}~\bibnamefont {{Kurkela}}},\ and\ \bibinfo {author} {\bibfnamefont {A.}~\bibnamefont {{Vuorinen}}},\ }\href {https://doi.org/10.1103/PhysRevLett.120.172703} {\bibfield  {journal} {\bibinfo  {journal} {\prl}\ }\textbf {\bibinfo {volume} {120}},\ \bibinfo {eid} {172703} (\bibinfo {year} {2018})},\ \Eprint {https://arxiv.org/abs/1711.02644} {arXiv:1711.02644 [astro-ph.HE]} \BibitemShut {NoStop}%
\bibitem [{\citenamefont {{Annala}}\ \emph {et~al.}(2020)\citenamefont {{Annala}}, \citenamefont {{Gorda}}, \citenamefont {{Kurkela}}, \citenamefont {{N{\"a}ttil{\"a}}},\ and\ \citenamefont {{Vuorinen}}}]{2020NatPh..16..907A}%
  \BibitemOpen
  \bibfield  {author} {\bibinfo {author} {\bibfnamefont {E.}~\bibnamefont {{Annala}}}, \bibinfo {author} {\bibfnamefont {T.}~\bibnamefont {{Gorda}}}, \bibinfo {author} {\bibfnamefont {A.}~\bibnamefont {{Kurkela}}}, \bibinfo {author} {\bibfnamefont {J.}~\bibnamefont {{N{\"a}ttil{\"a}}}},\ and\ \bibinfo {author} {\bibfnamefont {A.}~\bibnamefont {{Vuorinen}}},\ }\href {https://doi.org/10.1038/s41567-020-0914-9} {\bibfield  {journal} {\bibinfo  {journal} {Nature Physics}\ }\textbf {\bibinfo {volume} {16}},\ \bibinfo {pages} {907} (\bibinfo {year} {2020})},\ \Eprint {https://arxiv.org/abs/1903.09121} {arXiv:1903.09121 [astro-ph.HE]} \BibitemShut {NoStop}%
\bibitem [{\citenamefont {{Page}}\ \emph {et~al.}(2006)\citenamefont {{Page}}, \citenamefont {{Geppert}},\ and\ \citenamefont {{Weber}}}]{2006NuPhA.777..497P}%
  \BibitemOpen
  \bibfield  {author} {\bibinfo {author} {\bibfnamefont {D.}~\bibnamefont {{Page}}}, \bibinfo {author} {\bibfnamefont {U.}~\bibnamefont {{Geppert}}},\ and\ \bibinfo {author} {\bibfnamefont {F.}~\bibnamefont {{Weber}}},\ }\href {https://doi.org/10.1016/j.nuclphysa.2005.09.019} {\bibfield  {journal} {\bibinfo  {journal} {Nucl. Phys. A.}\ }\textbf {\bibinfo {volume} {777}},\ \bibinfo {pages} {497} (\bibinfo {year} {2006})},\ \Eprint {https://arxiv.org/abs/astro-ph/0508056} {arXiv:astro-ph/0508056 [astro-ph]} \BibitemShut {NoStop}%
\bibitem [{\citenamefont {{Horvath}}\ \emph {et~al.}(1991)\citenamefont {{Horvath}}, \citenamefont {{Benvenuto}},\ and\ \citenamefont {{Vucetich}}}]{1991PhRvD..44.3797H}%
  \BibitemOpen
  \bibfield  {author} {\bibinfo {author} {\bibfnamefont {J.~E.}\ \bibnamefont {{Horvath}}}, \bibinfo {author} {\bibfnamefont {O.~G.}\ \bibnamefont {{Benvenuto}}},\ and\ \bibinfo {author} {\bibfnamefont {H.}~\bibnamefont {{Vucetich}}},\ }\href {https://doi.org/10.1103/PhysRevD.44.3797} {\bibfield  {journal} {\bibinfo  {journal} {\prd}\ }\textbf {\bibinfo {volume} {44}},\ \bibinfo {pages} {3797} (\bibinfo {year} {1991})}\BibitemShut {NoStop}%
\bibitem [{\citenamefont {{Schaab}}\ \emph {et~al.}(1997)\citenamefont {{Schaab}}, \citenamefont {{Hermann}}, \citenamefont {{Weber}},\ and\ \citenamefont {{Weigel}}}]{1997JPhG...23.2029S}%
  \BibitemOpen
  \bibfield  {author} {\bibinfo {author} {\bibfnamefont {C.}~\bibnamefont {{Schaab}}}, \bibinfo {author} {\bibfnamefont {B.}~\bibnamefont {{Hermann}}}, \bibinfo {author} {\bibfnamefont {F.}~\bibnamefont {{Weber}}},\ and\ \bibinfo {author} {\bibfnamefont {M.~K.}\ \bibnamefont {{Weigel}}},\ }\href {https://doi.org/10.1088/0954-3899/23/12/027} {\bibfield  {journal} {\bibinfo  {journal} {Journal of Physics G Nuclear Physics}\ }\textbf {\bibinfo {volume} {23}},\ \bibinfo {pages} {2029} (\bibinfo {year} {1997})},\ \Eprint {https://arxiv.org/abs/astro-ph/9708092} {arXiv:astro-ph/9708092 [astro-ph]} \BibitemShut {NoStop}%
\bibitem [{\citenamefont {{Grigorian}}\ \emph {et~al.}(2005)\citenamefont {{Grigorian}}, \citenamefont {{Blaschke}},\ and\ \citenamefont {{Voskresensky}}}]{2005PhRvC..71d5801G}%
  \BibitemOpen
  \bibfield  {author} {\bibinfo {author} {\bibfnamefont {H.}~\bibnamefont {{Grigorian}}}, \bibinfo {author} {\bibfnamefont {D.}~\bibnamefont {{Blaschke}}},\ and\ \bibinfo {author} {\bibfnamefont {D.}~\bibnamefont {{Voskresensky}}},\ }\href {https://doi.org/10.1103/PhysRevC.71.045801} {\bibfield  {journal} {\bibinfo  {journal} {\prc}\ }\textbf {\bibinfo {volume} {71}},\ \bibinfo {eid} {045801} (\bibinfo {year} {2005})},\ \Eprint {https://arxiv.org/abs/astro-ph/0411619} {arXiv:astro-ph/0411619 [astro-ph]} \BibitemShut {NoStop}%
\bibitem [{\citenamefont {{Potekhin}}\ \emph {et~al.}(2015)\citenamefont {{Potekhin}}, \citenamefont {{Pons}},\ and\ \citenamefont {{Page}}}]{2015SSRv..191..239P}%
  \BibitemOpen
  \bibfield  {author} {\bibinfo {author} {\bibfnamefont {A.~Y.}\ \bibnamefont {{Potekhin}}}, \bibinfo {author} {\bibfnamefont {J.~A.}\ \bibnamefont {{Pons}}},\ and\ \bibinfo {author} {\bibfnamefont {D.}~\bibnamefont {{Page}}},\ }\href {https://doi.org/10.1007/s11214-015-0180-9} {\bibfield  {journal} {\bibinfo  {journal} {\ssr}\ }\textbf {\bibinfo {volume} {191}},\ \bibinfo {pages} {239} (\bibinfo {year} {2015})},\ \Eprint {https://arxiv.org/abs/1507.06186} {arXiv:1507.06186 [astro-ph.HE]} \BibitemShut {NoStop}%
\bibitem [{\citenamefont {{Sedrakian}}(2013)}]{2013A&A...555L..10S}%
  \BibitemOpen
  \bibfield  {author} {\bibinfo {author} {\bibfnamefont {A.}~\bibnamefont {{Sedrakian}}},\ }\href {https://doi.org/10.1051/0004-6361/201321541} {\bibfield  {journal} {\bibinfo  {journal} {Astron. Astrophys. Lett.}\ }\textbf {\bibinfo {volume} {555}},\ \bibinfo {eid} {L10} (\bibinfo {year} {2013})},\ \Eprint {https://arxiv.org/abs/1303.5380} {arXiv:1303.5380 [astro-ph.HE]} \BibitemShut {NoStop}%
\bibitem [{\citenamefont {{Alford}}\ \emph {et~al.}(2005)\citenamefont {{Alford}}, \citenamefont {{Braby}}, \citenamefont {{Paris}},\ and\ \citenamefont {{Reddy}}}]{2005ApJ...629..969A}%
  \BibitemOpen
  \bibfield  {author} {\bibinfo {author} {\bibfnamefont {M.}~\bibnamefont {{Alford}}}, \bibinfo {author} {\bibfnamefont {M.}~\bibnamefont {{Braby}}}, \bibinfo {author} {\bibfnamefont {M.}~\bibnamefont {{Paris}}},\ and\ \bibinfo {author} {\bibfnamefont {S.}~\bibnamefont {{Reddy}}},\ }\href {https://doi.org/10.1086/430902} {\bibfield  {journal} {\bibinfo  {journal} {\apj}\ }\textbf {\bibinfo {volume} {629}},\ \bibinfo {pages} {969} (\bibinfo {year} {2005})},\ \Eprint {https://arxiv.org/abs/nucl-th/0411016} {arXiv:nucl-th/0411016 [nucl-th]} \BibitemShut {NoStop}%
\bibitem [{\citenamefont {{Pereira}}\ \emph {et~al.}(2018)\citenamefont {{Pereira}}, \citenamefont {{Flores}},\ and\ \citenamefont {{Lugones}}}]{2018ApJ...860...12P}%
  \BibitemOpen
  \bibfield  {author} {\bibinfo {author} {\bibfnamefont {J.~P.}\ \bibnamefont {{Pereira}}}, \bibinfo {author} {\bibfnamefont {C.~V.}\ \bibnamefont {{Flores}}},\ and\ \bibinfo {author} {\bibfnamefont {G.}~\bibnamefont {{Lugones}}},\ }\href {https://doi.org/10.3847/1538-4357/aabfbf} {\bibfield  {journal} {\bibinfo  {journal} {\apj}\ }\textbf {\bibinfo {volume} {860}},\ \bibinfo {eid} {12} (\bibinfo {year} {2018})},\ \Eprint {https://arxiv.org/abs/1706.09371} {arXiv:1706.09371 [gr-qc]} \BibitemShut {NoStop}%
\bibitem [{\citenamefont {{Alford}}\ \emph {et~al.}(2001{\natexlab{a}})\citenamefont {{Alford}}, \citenamefont {{Rajagopal}}, \citenamefont {{Reddy}},\ and\ \citenamefont {{Wilczek}}}]{2001PhRvD..64g4017A}%
  \BibitemOpen
  \bibfield  {author} {\bibinfo {author} {\bibfnamefont {M.}~\bibnamefont {{Alford}}}, \bibinfo {author} {\bibfnamefont {K.}~\bibnamefont {{Rajagopal}}}, \bibinfo {author} {\bibfnamefont {S.}~\bibnamefont {{Reddy}}},\ and\ \bibinfo {author} {\bibfnamefont {F.}~\bibnamefont {{Wilczek}}},\ }\href {https://doi.org/10.1103/PhysRevD.64.074017} {\bibfield  {journal} {\bibinfo  {journal} {\prd}\ }\textbf {\bibinfo {volume} {64}},\ \bibinfo {eid} {074017} (\bibinfo {year} {2001}{\natexlab{a}})},\ \Eprint {https://arxiv.org/abs/hep-ph/0105009} {arXiv:hep-ph/0105009 [hep-ph]} \BibitemShut {NoStop}%
\bibitem [{\citenamefont {{Fraga}}\ \emph {et~al.}(2001)\citenamefont {{Fraga}}, \citenamefont {{Pisarski}},\ and\ \citenamefont {{Schaffner-Bielich}}}]{2001PhRvD..63l1702F}%
  \BibitemOpen
  \bibfield  {author} {\bibinfo {author} {\bibfnamefont {E.~S.}\ \bibnamefont {{Fraga}}}, \bibinfo {author} {\bibfnamefont {R.~D.}\ \bibnamefont {{Pisarski}}},\ and\ \bibinfo {author} {\bibfnamefont {J.}~\bibnamefont {{Schaffner-Bielich}}},\ }\href {https://doi.org/10.1103/PhysRevD.63.121702} {\bibfield  {journal} {\bibinfo  {journal} {\prd}\ }\textbf {\bibinfo {volume} {63}},\ \bibinfo {eid} {121702} (\bibinfo {year} {2001})},\ \Eprint {https://arxiv.org/abs/hep-ph/0101143} {arXiv:hep-ph/0101143 [hep-ph]} \BibitemShut {NoStop}%
\bibitem [{\citenamefont {{Lugones}}\ and\ \citenamefont {{Horvath}}(2002)}]{2002PhRvD..66g4017L}%
  \BibitemOpen
  \bibfield  {author} {\bibinfo {author} {\bibfnamefont {G.}~\bibnamefont {{Lugones}}}\ and\ \bibinfo {author} {\bibfnamefont {J.~E.}\ \bibnamefont {{Horvath}}},\ }\href {https://doi.org/10.1103/PhysRevD.66.074017} {\bibfield  {journal} {\bibinfo  {journal} {\prd}\ }\textbf {\bibinfo {volume} {66}},\ \bibinfo {eid} {074017} (\bibinfo {year} {2002})},\ \Eprint {https://arxiv.org/abs/hep-ph/0211070} {arXiv:hep-ph/0211070 [hep-ph]} \BibitemShut {NoStop}%
\bibitem [{\citenamefont {{Bedaque}}\ and\ \citenamefont {{Steiner}}(2015)}]{2015PhRvL.114c1103B}%
  \BibitemOpen
  \bibfield  {author} {\bibinfo {author} {\bibfnamefont {P.}~\bibnamefont {{Bedaque}}}\ and\ \bibinfo {author} {\bibfnamefont {A.~W.}\ \bibnamefont {{Steiner}}},\ }\href {https://doi.org/10.1103/PhysRevLett.114.031103} {\bibfield  {journal} {\bibinfo  {journal} {\prl}\ }\textbf {\bibinfo {volume} {114}},\ \bibinfo {eid} {031103} (\bibinfo {year} {2015})},\ \Eprint {https://arxiv.org/abs/1408.5116} {arXiv:1408.5116 [nucl-th]} \BibitemShut {NoStop}%
\bibitem [{\citenamefont {{Kojo}}(2020)}]{2020arXiv201110940K}%
  \BibitemOpen
  \bibfield  {author} {\bibinfo {author} {\bibfnamefont {T.}~\bibnamefont {{Kojo}}},\ }\href {https://doi.org/10.48550/arXiv.2011.10940} {\bibfield  {journal} {\bibinfo  {journal} {arXiv e-prints}\ ,\ \bibinfo {eid} {arXiv:2011.10940}} (\bibinfo {year} {2020})},\ \Eprint {https://arxiv.org/abs/2011.10940} {arXiv:2011.10940 [nucl-th]} \BibitemShut {NoStop}%
\bibitem [{\citenamefont {{Kurkela}}\ \emph {et~al.}(2024)\citenamefont {{Kurkela}}, \citenamefont {{Rajagopal}},\ and\ \citenamefont {{Steinhorst}}}]{2024PhRvL.132z2701K}%
  \BibitemOpen
  \bibfield  {author} {\bibinfo {author} {\bibfnamefont {A.}~\bibnamefont {{Kurkela}}}, \bibinfo {author} {\bibfnamefont {K.}~\bibnamefont {{Rajagopal}}},\ and\ \bibinfo {author} {\bibfnamefont {R.}~\bibnamefont {{Steinhorst}}},\ }\href {https://doi.org/10.1103/PhysRevLett.132.262701} {\bibfield  {journal} {\bibinfo  {journal} {\prl}\ }\textbf {\bibinfo {volume} {132}},\ \bibinfo {eid} {262701} (\bibinfo {year} {2024})},\ \Eprint {https://arxiv.org/abs/2401.16253} {arXiv:2401.16253 [astro-ph.HE]} \BibitemShut {NoStop}%
\bibitem [{\citenamefont {{Doroshenko}}\ \emph {et~al.}(2022)\citenamefont {{Doroshenko}}, \citenamefont {{Suleimanov}}, \citenamefont {{P{\"u}hlhofer}},\ and\ \citenamefont {{Santangelo}}}]{2022NatAs...6.1444D}%
  \BibitemOpen
  \bibfield  {author} {\bibinfo {author} {\bibfnamefont {V.}~\bibnamefont {{Doroshenko}}}, \bibinfo {author} {\bibfnamefont {V.}~\bibnamefont {{Suleimanov}}}, \bibinfo {author} {\bibfnamefont {G.}~\bibnamefont {{P{\"u}hlhofer}}},\ and\ \bibinfo {author} {\bibfnamefont {A.}~\bibnamefont {{Santangelo}}},\ }\href {https://doi.org/10.1038/s41550-022-01800-1} {\bibfield  {journal} {\bibinfo  {journal} {Nature Astronomy}\ }\textbf {\bibinfo {volume} {6}},\ \bibinfo {pages} {1444} (\bibinfo {year} {2022})}\BibitemShut {NoStop}%
\bibitem [{\citenamefont {{Di Clemente}}\ \emph {et~al.}(2024)\citenamefont {{Di Clemente}}, \citenamefont {{Drago}},\ and\ \citenamefont {{Pagliara}}}]{2024ApJ...967..159D}%
  \BibitemOpen
  \bibfield  {author} {\bibinfo {author} {\bibfnamefont {F.}~\bibnamefont {{Di Clemente}}}, \bibinfo {author} {\bibfnamefont {A.}~\bibnamefont {{Drago}}},\ and\ \bibinfo {author} {\bibfnamefont {G.}~\bibnamefont {{Pagliara}}},\ }\href {https://doi.org/10.3847/1538-4357/ad445b} {\bibfield  {journal} {\bibinfo  {journal} {\apj}\ }\textbf {\bibinfo {volume} {967}},\ \bibinfo {eid} {159} (\bibinfo {year} {2024})},\ \Eprint {https://arxiv.org/abs/2211.07485} {arXiv:2211.07485 [astro-ph.HE]} \BibitemShut {NoStop}%
\bibitem [{\citenamefont {{Horvath}}\ \emph {et~al.}(2023)\citenamefont {{Horvath}}, \citenamefont {{Rocha}}, \citenamefont {{de S{\'a}}}, \citenamefont {{Moraes}}, \citenamefont {{Bar{\~a}o}}, \citenamefont {{de Avellar}}, \citenamefont {{Bernardo}},\ and\ \citenamefont {{Bachega}}}]{2023A&A...672L..11H}%
  \BibitemOpen
  \bibfield  {author} {\bibinfo {author} {\bibfnamefont {J.~E.}\ \bibnamefont {{Horvath}}}, \bibinfo {author} {\bibfnamefont {L.~S.}\ \bibnamefont {{Rocha}}}, \bibinfo {author} {\bibfnamefont {L.~M.}\ \bibnamefont {{de S{\'a}}}}, \bibinfo {author} {\bibfnamefont {P.~H.~R.~S.}\ \bibnamefont {{Moraes}}}, \bibinfo {author} {\bibfnamefont {L.~G.}\ \bibnamefont {{Bar{\~a}o}}}, \bibinfo {author} {\bibfnamefont {M.~G.~B.}\ \bibnamefont {{de Avellar}}}, \bibinfo {author} {\bibfnamefont {A.}~\bibnamefont {{Bernardo}}},\ and\ \bibinfo {author} {\bibfnamefont {R.~R.~A.}\ \bibnamefont {{Bachega}}},\ }\href {https://doi.org/10.1051/0004-6361/202345885} {\bibfield  {journal} {\bibinfo  {journal} {\aap}\ }\textbf {\bibinfo {volume} {672}},\ \bibinfo {eid} {L11} (\bibinfo {year} {2023})},\ \Eprint {https://arxiv.org/abs/2303.10264} {arXiv:2303.10264 [astro-ph.HE]} \BibitemShut {NoStop}%
\bibitem [{\citenamefont {{McKay}}\ \emph {et~al.}(1979)\citenamefont {{McKay}}, \citenamefont {{Beckman}},\ and\ \citenamefont {{Conover}}}]{McKay1979LatinHypercube}%
  \BibitemOpen
  \bibfield  {author} {\bibinfo {author} {\bibfnamefont {M.~D.}\ \bibnamefont {{McKay}}}, \bibinfo {author} {\bibfnamefont {R.~J.}\ \bibnamefont {{Beckman}}},\ and\ \bibinfo {author} {\bibfnamefont {W.~J.}\ \bibnamefont {{Conover}}},\ }\href {http://www.jstor.org/stable/1268522} {\bibfield  {journal} {\bibinfo  {journal} {Technometrics}\ }\textbf {\bibinfo {volume} {21}},\ \bibinfo {pages} {239} (\bibinfo {year} {1979})}\BibitemShut {NoStop}%
\bibitem [{\citenamefont {{Wang}}\ \emph {et~al.}(2024)\citenamefont {{Wang}}, \citenamefont {{Gao}}, \citenamefont {{Liang}}, \citenamefont {{Zhao}},\ and\ \citenamefont {{Shao}}}]{2024JCAP...11..038W}%
  \BibitemOpen
  \bibfield  {author} {\bibinfo {author} {\bibfnamefont {Z.}~\bibnamefont {{Wang}}}, \bibinfo {author} {\bibfnamefont {Y.}~\bibnamefont {{Gao}}}, \bibinfo {author} {\bibfnamefont {D.}~\bibnamefont {{Liang}}}, \bibinfo {author} {\bibfnamefont {J.}~\bibnamefont {{Zhao}}},\ and\ \bibinfo {author} {\bibfnamefont {L.}~\bibnamefont {{Shao}}},\ }\href {https://doi.org/10.1088/1475-7516/2024/11/038} {\bibfield  {journal} {\bibinfo  {journal} {\jcap}\ }\textbf {\bibinfo {volume} {2024}},\ \bibinfo {eid} {038} (\bibinfo {year} {2024})},\ \Eprint {https://arxiv.org/abs/2409.11103} {arXiv:2409.11103 [astro-ph.HE]} \BibitemShut {NoStop}%
\bibitem [{\citenamefont {{Zhou}}\ \emph {et~al.}(2018)\citenamefont {{Zhou}}, \citenamefont {{Zhou}},\ and\ \citenamefont {{Li}}}]{2018PhRvD..97h3015Z}%
  \BibitemOpen
  \bibfield  {author} {\bibinfo {author} {\bibfnamefont {E.-P.}\ \bibnamefont {{Zhou}}}, \bibinfo {author} {\bibfnamefont {X.}~\bibnamefont {{Zhou}}},\ and\ \bibinfo {author} {\bibfnamefont {A.}~\bibnamefont {{Li}}},\ }\href {https://doi.org/10.1103/PhysRevD.97.083015} {\bibfield  {journal} {\bibinfo  {journal} {\prd}\ }\textbf {\bibinfo {volume} {97}},\ \bibinfo {eid} {083015} (\bibinfo {year} {2018})},\ \Eprint {https://arxiv.org/abs/1711.04312} {arXiv:1711.04312 [astro-ph.HE]} \BibitemShut {NoStop}%
\bibitem [{\citenamefont {{Miao}}\ \emph {et~al.}(2021)\citenamefont {{Miao}}, \citenamefont {{Jiang}}, \citenamefont {{Li}},\ and\ \citenamefont {{Chen}}}]{2021ApJ...917L..22M}%
  \BibitemOpen
  \bibfield  {author} {\bibinfo {author} {\bibfnamefont {Z.}~\bibnamefont {{Miao}}}, \bibinfo {author} {\bibfnamefont {J.-L.}\ \bibnamefont {{Jiang}}}, \bibinfo {author} {\bibfnamefont {A.}~\bibnamefont {{Li}}},\ and\ \bibinfo {author} {\bibfnamefont {L.-W.}\ \bibnamefont {{Chen}}},\ }\href {https://doi.org/10.3847/2041-8213/ac194d} {\bibfield  {journal} {\bibinfo  {journal} {\apjl}\ }\textbf {\bibinfo {volume} {917}},\ \bibinfo {eid} {L22} (\bibinfo {year} {2021})},\ \Eprint {https://arxiv.org/abs/2107.13997} {arXiv:2107.13997 [astro-ph.HE]} \BibitemShut {NoStop}%
\bibitem [{\citenamefont {{Zhang}}\ and\ \citenamefont {{Mann}}(2021)}]{2021PhRvD.103f3018Z}%
  \BibitemOpen
  \bibfield  {author} {\bibinfo {author} {\bibfnamefont {C.}~\bibnamefont {{Zhang}}}\ and\ \bibinfo {author} {\bibfnamefont {R.~B.}\ \bibnamefont {{Mann}}},\ }\href {https://doi.org/10.1103/PhysRevD.103.063018} {\bibfield  {journal} {\bibinfo  {journal} {\prd}\ }\textbf {\bibinfo {volume} {103}},\ \bibinfo {eid} {063018} (\bibinfo {year} {2021})},\ \Eprint {https://arxiv.org/abs/2009.07182} {arXiv:2009.07182 [astro-ph.HE]} \BibitemShut {NoStop}%
\bibitem [{\citenamefont {{Li}}\ \emph {et~al.}(2021{\natexlab{b}})\citenamefont {{Li}}, \citenamefont {{Miao}}, \citenamefont {{Jiang}}, \citenamefont {{Tang}},\ and\ \citenamefont {{Xu}}}]{2021MNRAS.506.5916L}%
  \BibitemOpen
  \bibfield  {author} {\bibinfo {author} {\bibfnamefont {A.}~\bibnamefont {{Li}}}, \bibinfo {author} {\bibfnamefont {Z.~Q.}\ \bibnamefont {{Miao}}}, \bibinfo {author} {\bibfnamefont {J.~L.}\ \bibnamefont {{Jiang}}}, \bibinfo {author} {\bibfnamefont {S.~P.}\ \bibnamefont {{Tang}}},\ and\ \bibinfo {author} {\bibfnamefont {R.~X.}\ \bibnamefont {{Xu}}},\ }\href {https://doi.org/10.1093/mnras/stab2029} {\bibfield  {journal} {\bibinfo  {journal} {\mnras}\ }\textbf {\bibinfo {volume} {506}},\ \bibinfo {pages} {5916} (\bibinfo {year} {2021}{\natexlab{b}})},\ \Eprint {https://arxiv.org/abs/2009.12571} {arXiv:2009.12571 [astro-ph.HE]} \BibitemShut {NoStop}%
\bibitem [{\citenamefont {{Tolman}}(1939)}]{1939PhRv...55..364T}%
  \BibitemOpen
  \bibfield  {author} {\bibinfo {author} {\bibfnamefont {R.~C.}\ \bibnamefont {{Tolman}}},\ }\href {https://doi.org/10.1103/PhysRev.55.364} {\bibfield  {journal} {\bibinfo  {journal} {Physical Review}\ }\textbf {\bibinfo {volume} {55}},\ \bibinfo {pages} {364} (\bibinfo {year} {1939})}\BibitemShut {NoStop}%
\bibitem [{\citenamefont {{Oppenheimer}}\ and\ \citenamefont {{Volkoff}}(1939)}]{1939PhRv...55..374O}%
  \BibitemOpen
  \bibfield  {author} {\bibinfo {author} {\bibfnamefont {J.~R.}\ \bibnamefont {{Oppenheimer}}}\ and\ \bibinfo {author} {\bibfnamefont {G.~M.}\ \bibnamefont {{Volkoff}}},\ }\href {https://doi.org/10.1103/PhysRev.55.374} {\bibfield  {journal} {\bibinfo  {journal} {Physical Review}\ }\textbf {\bibinfo {volume} {55}},\ \bibinfo {pages} {374} (\bibinfo {year} {1939})}\BibitemShut {NoStop}%
\bibitem [{\citenamefont {{Misner}}\ \emph {et~al.}(2017)\citenamefont {{Misner}}, \citenamefont {{Thorne}}, \citenamefont {{Wheeler}},\ and\ \citenamefont {{Kaiser}}}]{Misner2017Gravitation}%
  \BibitemOpen
  \bibfield  {author} {\bibinfo {author} {\bibfnamefont {C.~W.}\ \bibnamefont {{Misner}}}, \bibinfo {author} {\bibfnamefont {K.~S.}\ \bibnamefont {{Thorne}}}, \bibinfo {author} {\bibfnamefont {J.~A.}\ \bibnamefont {{Wheeler}}},\ and\ \bibinfo {author} {\bibfnamefont {D.~I.}\ \bibnamefont {{Kaiser}}},\ }\href@noop {} {\emph {\bibinfo {title} {{Gravitation}}}}\ (\bibinfo  {publisher} {Princeton University Press},\ \bibinfo {year} {2017})\BibitemShut {NoStop}%
\bibitem [{\citenamefont {{Schutz}}(2009)}]{Schutz2009FirstCourseGR}%
  \BibitemOpen
  \bibfield  {author} {\bibinfo {author} {\bibfnamefont {B.~F.}\ \bibnamefont {{Schutz}}},\ }\href@noop {} {\emph {\bibinfo {title} {{A first course in general relativity}}}}\ (\bibinfo  {publisher} {Cambridge University Press},\ \bibinfo {year} {2009})\BibitemShut {NoStop}%
\bibitem [{\citenamefont {LVK}(2025)}]{LVKCollaboration}%
  \BibitemOpen
  \bibfield  {author} {\bibinfo {author} {\bibnamefont {LVK}},\ }\href@noop {} {\bibinfo {title} {{The LIGO-Virgo-KAGRA (LVK) Collaboration}}},\ \bibinfo {howpublished} {\url{https://ligo.org/who-we-are/}} (\bibinfo {year} {2025})\BibitemShut {NoStop}%
\bibitem [{\citenamefont {{Abdelsalhin}}(2019)}]{2019arXiv190500408A}%
  \BibitemOpen
  \bibfield  {author} {\bibinfo {author} {\bibfnamefont {T.}~\bibnamefont {{Abdelsalhin}}},\ }\href {https://doi.org/10.48550/arXiv.1905.00408} {\bibfield  {journal} {\bibinfo  {journal} {arXiv e-prints}\ ,\ \bibinfo {eid} {arXiv:1905.00408}} (\bibinfo {year} {2019})},\ \Eprint {https://arxiv.org/abs/1905.00408} {arXiv:1905.00408 [gr-qc]} \BibitemShut {NoStop}%
\bibitem [{\citenamefont {{Dietrich}}\ \emph {et~al.}(2021)\citenamefont {{Dietrich}}, \citenamefont {{Hinderer}},\ and\ \citenamefont {{Samajdar}}}]{2021GReGr..53...27D}%
  \BibitemOpen
  \bibfield  {author} {\bibinfo {author} {\bibfnamefont {T.}~\bibnamefont {{Dietrich}}}, \bibinfo {author} {\bibfnamefont {T.}~\bibnamefont {{Hinderer}}},\ and\ \bibinfo {author} {\bibfnamefont {A.}~\bibnamefont {{Samajdar}}},\ }\href {https://doi.org/10.1007/s10714-020-02751-6} {\bibfield  {journal} {\bibinfo  {journal} {General Relativity and Gravitation}\ }\textbf {\bibinfo {volume} {53}},\ \bibinfo {eid} {27} (\bibinfo {year} {2021})},\ \Eprint {https://arxiv.org/abs/2004.02527} {arXiv:2004.02527 [gr-qc]} \BibitemShut {NoStop}%
\bibitem [{\citenamefont {{Arba{\~n}il}}\ \emph {et~al.}(2023)\citenamefont {{Arba{\~n}il}}, \citenamefont {{Flores}}, \citenamefont {{Lenzi}},\ and\ \citenamefont {{Pretel}}}]{2023PhRvD.107l4016A}%
  \BibitemOpen
  \bibfield  {author} {\bibinfo {author} {\bibfnamefont {J.~D.~V.}\ \bibnamefont {{Arba{\~n}il}}}, \bibinfo {author} {\bibfnamefont {C.~V.}\ \bibnamefont {{Flores}}}, \bibinfo {author} {\bibfnamefont {C.~H.}\ \bibnamefont {{Lenzi}}},\ and\ \bibinfo {author} {\bibfnamefont {J.~M.~Z.}\ \bibnamefont {{Pretel}}},\ }\href {https://doi.org/10.1103/PhysRevD.107.124016} {\bibfield  {journal} {\bibinfo  {journal} {\prd}\ }\textbf {\bibinfo {volume} {107}},\ \bibinfo {eid} {124016} (\bibinfo {year} {2023})},\ \Eprint {https://arxiv.org/abs/2305.13468} {arXiv:2305.13468 [astro-ph.HE]} \BibitemShut {NoStop}%
\bibitem [{\citenamefont {{Zastrow}}(2025)}]{StarStructure}%
  \BibitemOpen
  \bibfield  {author} {\bibinfo {author} {\bibfnamefont {J.~V.}\ \bibnamefont {{Zastrow}}},\ }\href@noop {} {\bibinfo {title} {{Star Structure}}},\ \bibinfo {howpublished} {\url{https://github.com/joaovz/Star_structure}} (\bibinfo {year} {2025})\BibitemShut {NoStop}%
\bibitem [{\citenamefont {{Rezzolla}}\ \emph {et~al.}(2018)\citenamefont {{Rezzolla}}, \citenamefont {{Most}},\ and\ \citenamefont {{Weih}}}]{2018ApJ...852L..25R}%
  \BibitemOpen
  \bibfield  {author} {\bibinfo {author} {\bibfnamefont {L.}~\bibnamefont {{Rezzolla}}}, \bibinfo {author} {\bibfnamefont {E.~R.}\ \bibnamefont {{Most}}},\ and\ \bibinfo {author} {\bibfnamefont {L.~R.}\ \bibnamefont {{Weih}}},\ }\href {https://doi.org/10.3847/2041-8213/aaa401} {\bibfield  {journal} {\bibinfo  {journal} {\apjl}\ }\textbf {\bibinfo {volume} {852}},\ \bibinfo {eid} {L25} (\bibinfo {year} {2018})},\ \Eprint {https://arxiv.org/abs/1711.00314} {arXiv:1711.00314 [astro-ph.HE]} \BibitemShut {NoStop}%
\bibitem [{\citenamefont {{Pang}}\ \emph {et~al.}(2023)\citenamefont {{Pang}}, \citenamefont {{Dietrich}}, \citenamefont {{Coughlin}}, \citenamefont {{Bulla}}, \citenamefont {{Tews}}, \citenamefont {{Almualla}}, \citenamefont {{Barna}}, \citenamefont {{Kiendrebeogo}}, \citenamefont {{Kunert}}, \citenamefont {{Mansingh}}, \citenamefont {{Reed}}, \citenamefont {{Sravan}}, \citenamefont {{Toivonen}}, \citenamefont {{Antier}}, \citenamefont {{VandenBerg}}, \citenamefont {{Heinzel}}, \citenamefont {{Nedora}}, \citenamefont {{Salehi}}, \citenamefont {{Sharma}}, \citenamefont {{Somasundaram}},\ and\ \citenamefont {{Van Den Broeck}}}]{2023NatCo..14.8352P}%
  \BibitemOpen
  \bibfield  {author} {\bibinfo {author} {\bibfnamefont {P.~T.~H.}\ \bibnamefont {{Pang}}}, \bibinfo {author} {\bibfnamefont {T.}~\bibnamefont {{Dietrich}}}, \bibinfo {author} {\bibfnamefont {M.~W.}\ \bibnamefont {{Coughlin}}}, \bibinfo {author} {\bibfnamefont {M.}~\bibnamefont {{Bulla}}}, \bibinfo {author} {\bibfnamefont {I.}~\bibnamefont {{Tews}}}, \bibinfo {author} {\bibfnamefont {M.}~\bibnamefont {{Almualla}}}, \bibinfo {author} {\bibfnamefont {T.}~\bibnamefont {{Barna}}}, \bibinfo {author} {\bibfnamefont {R.~W.}\ \bibnamefont {{Kiendrebeogo}}}, \bibinfo {author} {\bibfnamefont {N.}~\bibnamefont {{Kunert}}}, \bibinfo {author} {\bibfnamefont {G.}~\bibnamefont {{Mansingh}}}, \bibinfo {author} {\bibfnamefont {B.}~\bibnamefont {{Reed}}}, \bibinfo {author} {\bibfnamefont {N.}~\bibnamefont {{Sravan}}}, \bibinfo {author} {\bibfnamefont {A.}~\bibnamefont {{Toivonen}}}, \bibinfo {author} {\bibfnamefont {S.}~\bibnamefont {{Antier}}}, \bibinfo {author} {\bibfnamefont {R.~O.}\ \bibnamefont {{VandenBerg}}}, \bibinfo
  {author} {\bibfnamefont {J.}~\bibnamefont {{Heinzel}}}, \bibinfo {author} {\bibfnamefont {V.}~\bibnamefont {{Nedora}}}, \bibinfo {author} {\bibfnamefont {P.}~\bibnamefont {{Salehi}}}, \bibinfo {author} {\bibfnamefont {R.}~\bibnamefont {{Sharma}}}, \bibinfo {author} {\bibfnamefont {R.}~\bibnamefont {{Somasundaram}}},\ and\ \bibinfo {author} {\bibfnamefont {C.}~\bibnamefont {{Van Den Broeck}}},\ }\href {https://doi.org/10.1038/s41467-023-43932-6} {\bibfield  {journal} {\bibinfo  {journal} {Nature Communications}\ }\textbf {\bibinfo {volume} {14}},\ \bibinfo {eid} {8352} (\bibinfo {year} {2023})},\ \Eprint {https://arxiv.org/abs/2205.08513} {arXiv:2205.08513 [astro-ph.HE]} \BibitemShut {NoStop}%
\bibitem [{\citenamefont {{The LIGO and Virgo Scientific Collaboration}}\ \emph {et~al.}(2017)\citenamefont {{The LIGO and Virgo Scientific Collaboration}}, \citenamefont {{Abbott}}, \citenamefont {{Abbott}}, \citenamefont {{Abbott}}, \citenamefont {{Acernese}}, \citenamefont {{Ackley}}, \citenamefont {{Adams}}, \citenamefont {{Adams}} \emph {et~al.}}]{2017PhRvL.119p1101A}%
  \BibitemOpen
  \bibfield  {author} {\bibinfo {author} {\bibnamefont {{The LIGO and Virgo Scientific Collaboration}}}, \bibinfo {author} {\bibfnamefont {B.~P.}\ \bibnamefont {{Abbott}}}, \bibinfo {author} {\bibfnamefont {R.}~\bibnamefont {{Abbott}}}, \bibinfo {author} {\bibfnamefont {T.~D.}\ \bibnamefont {{Abbott}}}, \bibinfo {author} {\bibfnamefont {F.}~\bibnamefont {{Acernese}}}, \bibinfo {author} {\bibfnamefont {K.}~\bibnamefont {{Ackley}}}, \bibinfo {author} {\bibfnamefont {C.}~\bibnamefont {{Adams}}}, \bibinfo {author} {\bibfnamefont {T.}~\bibnamefont {{Adams}}}, \emph {et~al.},\ }\href {https://doi.org/10.1103/PhysRevLett.119.161101} {\bibfield  {journal} {\bibinfo  {journal} {\prl}\ }\textbf {\bibinfo {volume} {119}},\ \bibinfo {eid} {161101} (\bibinfo {year} {2017})},\ \Eprint {https://arxiv.org/abs/1710.05832} {arXiv:1710.05832 [gr-qc]} \BibitemShut {NoStop}%
\bibitem [{Note1()}]{Note1}%
  \BibitemOpen
  \bibinfo {note} {Note that it is not strictly ``canonical'', as the full distribution spans a range of values \cite {2023Univ...10....3R}.}\BibitemShut {Stop}%
\bibitem [{\citenamefont {{Xie}}\ and\ \citenamefont {{Li}}(2021)}]{2021PhRvC.103c5802X}%
  \BibitemOpen
  \bibfield  {author} {\bibinfo {author} {\bibfnamefont {W.-J.}\ \bibnamefont {{Xie}}}\ and\ \bibinfo {author} {\bibfnamefont {B.-A.}\ \bibnamefont {{Li}}},\ }\href {https://doi.org/10.1103/PhysRevC.103.035802} {\bibfield  {journal} {\bibinfo  {journal} {\prc}\ }\textbf {\bibinfo {volume} {103}},\ \bibinfo {eid} {035802} (\bibinfo {year} {2021})},\ \Eprint {https://arxiv.org/abs/2009.13653} {arXiv:2009.13653 [nucl-th]} \BibitemShut {NoStop}%
\bibitem [{\citenamefont {{Morawski}}\ and\ \citenamefont {{Bejger}}(2020)}]{2020A&A...642A..78M}%
  \BibitemOpen
  \bibfield  {author} {\bibinfo {author} {\bibfnamefont {F.}~\bibnamefont {{Morawski}}}\ and\ \bibinfo {author} {\bibfnamefont {M.}~\bibnamefont {{Bejger}}},\ }\href {https://doi.org/10.1051/0004-6361/202038130} {\bibfield  {journal} {\bibinfo  {journal} {\aap}\ }\textbf {\bibinfo {volume} {642}},\ \bibinfo {eid} {A78} (\bibinfo {year} {2020})},\ \Eprint {https://arxiv.org/abs/2006.07194} {arXiv:2006.07194 [astro-ph.HE]} \BibitemShut {NoStop}%
\bibitem [{\citenamefont {{Ventagli}}\ and\ \citenamefont {{Saltas}}(2025)}]{2025JCAP...01..073V}%
  \BibitemOpen
  \bibfield  {author} {\bibinfo {author} {\bibfnamefont {G.}~\bibnamefont {{Ventagli}}}\ and\ \bibinfo {author} {\bibfnamefont {I.~D.}\ \bibnamefont {{Saltas}}},\ }\href {https://doi.org/10.1088/1475-7516/2025/01/073} {\bibfield  {journal} {\bibinfo  {journal} {\jcap}\ }\textbf {\bibinfo {volume} {2025}},\ \bibinfo {eid} {073} (\bibinfo {year} {2025})},\ \Eprint {https://arxiv.org/abs/2405.17908} {arXiv:2405.17908 [astro-ph.HE]} \BibitemShut {NoStop}%
\bibitem [{\citenamefont {{Alford}}\ \emph {et~al.}(1999)\citenamefont {{Alford}}, \citenamefont {{Rajagopal}},\ and\ \citenamefont {{Wilczek}}}]{1999NuPhB.537..443A}%
  \BibitemOpen
  \bibfield  {author} {\bibinfo {author} {\bibfnamefont {M.}~\bibnamefont {{Alford}}}, \bibinfo {author} {\bibfnamefont {K.}~\bibnamefont {{Rajagopal}}},\ and\ \bibinfo {author} {\bibfnamefont {F.}~\bibnamefont {{Wilczek}}},\ }\href {https://doi.org/10.1016/S0550-3213(98)00668-3} {\bibfield  {journal} {\bibinfo  {journal} {Nuclear Physics B}\ }\textbf {\bibinfo {volume} {537}},\ \bibinfo {pages} {443} (\bibinfo {year} {1999})},\ \Eprint {https://arxiv.org/abs/hep-ph/9804403} {arXiv:hep-ph/9804403 [hep-ph]} \BibitemShut {NoStop}%
\bibitem [{\citenamefont {{Pereira}}\ \emph {et~al.}(2023)\citenamefont {{Pereira}}, \citenamefont {{Bejger}}, \citenamefont {{Haensel}},\ and\ \citenamefont {{Zdunik}}}]{2023ApJ...950..185P}%
  \BibitemOpen
  \bibfield  {author} {\bibinfo {author} {\bibfnamefont {J.~P.}\ \bibnamefont {{Pereira}}}, \bibinfo {author} {\bibfnamefont {M.}~\bibnamefont {{Bejger}}}, \bibinfo {author} {\bibfnamefont {P.}~\bibnamefont {{Haensel}}},\ and\ \bibinfo {author} {\bibfnamefont {J.~L.}\ \bibnamefont {{Zdunik}}},\ }\href {https://doi.org/10.3847/1538-4357/acd759} {\bibfield  {journal} {\bibinfo  {journal} {\apj}\ }\textbf {\bibinfo {volume} {950}},\ \bibinfo {eid} {185} (\bibinfo {year} {2023})},\ \Eprint {https://arxiv.org/abs/2210.14048} {arXiv:2210.14048 [astro-ph.HE]} \BibitemShut {NoStop}%
\bibitem [{\citenamefont {{Andersson}}\ and\ \citenamefont {{Kokkotas}}(1998)}]{1998MNRAS.299.1059A}%
  \BibitemOpen
  \bibfield  {author} {\bibinfo {author} {\bibfnamefont {N.}~\bibnamefont {{Andersson}}}\ and\ \bibinfo {author} {\bibfnamefont {K.~D.}\ \bibnamefont {{Kokkotas}}},\ }\href {https://doi.org/10.1046/j.1365-8711.1998.01840.x} {\bibfield  {journal} {\bibinfo  {journal} {\mnras}\ }\textbf {\bibinfo {volume} {299}},\ \bibinfo {pages} {1059} (\bibinfo {year} {1998})},\ \Eprint {https://arxiv.org/abs/gr-qc/9711088} {arXiv:gr-qc/9711088 [gr-qc]} \BibitemShut {NoStop}%
\bibitem [{\citenamefont {{Jim{\'e}nez}}\ and\ \citenamefont {{Fraga}}(2019)}]{2019PhRvD.100k4041J}%
  \BibitemOpen
  \bibfield  {author} {\bibinfo {author} {\bibfnamefont {J.~C.}\ \bibnamefont {{Jim{\'e}nez}}}\ and\ \bibinfo {author} {\bibfnamefont {E.~S.}\ \bibnamefont {{Fraga}}},\ }\href {https://doi.org/10.1103/PhysRevD.100.114041} {\bibfield  {journal} {\bibinfo  {journal} {\prd}\ }\textbf {\bibinfo {volume} {100}},\ \bibinfo {eid} {114041} (\bibinfo {year} {2019})},\ \Eprint {https://arxiv.org/abs/1906.11189} {arXiv:1906.11189 [hep-ph]} \BibitemShut {NoStop}%
\bibitem [{\citenamefont {{Rather}}\ \emph {et~al.}(2023)\citenamefont {{Rather}}, \citenamefont {{Panotopoulos}},\ and\ \citenamefont {{Lopes}}}]{2023EPJC...83.1065R}%
  \BibitemOpen
  \bibfield  {author} {\bibinfo {author} {\bibfnamefont {I.~A.}\ \bibnamefont {{Rather}}}, \bibinfo {author} {\bibfnamefont {G.}~\bibnamefont {{Panotopoulos}}},\ and\ \bibinfo {author} {\bibfnamefont {I.}~\bibnamefont {{Lopes}}},\ }\href {https://doi.org/10.1140/epjc/s10052-023-12223-1} {\bibfield  {journal} {\bibinfo  {journal} {European Physical Journal C}\ }\textbf {\bibinfo {volume} {83}},\ \bibinfo {eid} {1065} (\bibinfo {year} {2023})},\ \Eprint {https://arxiv.org/abs/2307.03703} {arXiv:2307.03703 [astro-ph.HE]} \BibitemShut {NoStop}%
\bibitem [{\citenamefont {{Alford}}\ \emph {et~al.}(2001{\natexlab{b}})\citenamefont {{Alford}}, \citenamefont {{Bowers}},\ and\ \citenamefont {{Rajagopal}}}]{2001PhRvD..63g4016A}%
  \BibitemOpen
  \bibfield  {author} {\bibinfo {author} {\bibfnamefont {M.}~\bibnamefont {{Alford}}}, \bibinfo {author} {\bibfnamefont {J.~A.}\ \bibnamefont {{Bowers}}},\ and\ \bibinfo {author} {\bibfnamefont {K.}~\bibnamefont {{Rajagopal}}},\ }\href {https://doi.org/10.1103/PhysRevD.63.074016} {\bibfield  {journal} {\bibinfo  {journal} {\prd}\ }\textbf {\bibinfo {volume} {63}},\ \bibinfo {eid} {074016} (\bibinfo {year} {2001}{\natexlab{b}})},\ \Eprint {https://arxiv.org/abs/hep-ph/0008208} {arXiv:hep-ph/0008208 [hep-ph]} \BibitemShut {NoStop}%
\bibitem [{\citenamefont {{Mannarelli}}\ \emph {et~al.}(2007)\citenamefont {{Mannarelli}}, \citenamefont {{Rajagopal}},\ and\ \citenamefont {{Sharma}}}]{2007PhRvD..76g4026M}%
  \BibitemOpen
  \bibfield  {author} {\bibinfo {author} {\bibfnamefont {M.}~\bibnamefont {{Mannarelli}}}, \bibinfo {author} {\bibfnamefont {K.}~\bibnamefont {{Rajagopal}}},\ and\ \bibinfo {author} {\bibfnamefont {R.}~\bibnamefont {{Sharma}}},\ }\href {https://doi.org/10.1103/PhysRevD.76.074026} {\bibfield  {journal} {\bibinfo  {journal} {\prd}\ }\textbf {\bibinfo {volume} {76}},\ \bibinfo {eid} {074026} (\bibinfo {year} {2007})},\ \Eprint {https://arxiv.org/abs/hep-ph/0702021} {arXiv:hep-ph/0702021 [hep-ph]} \BibitemShut {NoStop}%
\bibitem [{\citenamefont {{Lau}}\ \emph {et~al.}(2017)\citenamefont {{Lau}}, \citenamefont {{Leung}},\ and\ \citenamefont {{Lin}}}]{2017PhRvD..95j1302L}%
  \BibitemOpen
  \bibfield  {author} {\bibinfo {author} {\bibfnamefont {S.~Y.}\ \bibnamefont {{Lau}}}, \bibinfo {author} {\bibfnamefont {P.~T.}\ \bibnamefont {{Leung}}},\ and\ \bibinfo {author} {\bibfnamefont {L.~M.}\ \bibnamefont {{Lin}}},\ }\href {https://doi.org/10.1103/PhysRevD.95.101302} {\bibfield  {journal} {\bibinfo  {journal} {\prd}\ }\textbf {\bibinfo {volume} {95}},\ \bibinfo {eid} {101302} (\bibinfo {year} {2017})}\BibitemShut {NoStop}%
\bibitem [{\citenamefont {{Lau}}\ \emph {et~al.}(2019)\citenamefont {{Lau}}, \citenamefont {{Leung}},\ and\ \citenamefont {{Lin}}}]{2019PhRvD..99b3018L}%
  \BibitemOpen
  \bibfield  {author} {\bibinfo {author} {\bibfnamefont {S.~Y.}\ \bibnamefont {{Lau}}}, \bibinfo {author} {\bibfnamefont {P.~T.}\ \bibnamefont {{Leung}}},\ and\ \bibinfo {author} {\bibfnamefont {L.~M.}\ \bibnamefont {{Lin}}},\ }\href {https://doi.org/10.1103/PhysRevD.99.023018} {\bibfield  {journal} {\bibinfo  {journal} {\prd}\ }\textbf {\bibinfo {volume} {99}},\ \bibinfo {eid} {023018} (\bibinfo {year} {2019})},\ \Eprint {https://arxiv.org/abs/1808.08107} {arXiv:1808.08107 [astro-ph.HE]} \BibitemShut {NoStop}%
\bibitem [{\citenamefont {{Haskell}}\ \emph {et~al.}(2007)\citenamefont {{Haskell}}, \citenamefont {{Andersson}}, \citenamefont {{Jones}},\ and\ \citenamefont {{Samuelsson}}}]{2007PhRvL..99w1101H}%
  \BibitemOpen
  \bibfield  {author} {\bibinfo {author} {\bibfnamefont {B.}~\bibnamefont {{Haskell}}}, \bibinfo {author} {\bibfnamefont {N.}~\bibnamefont {{Andersson}}}, \bibinfo {author} {\bibfnamefont {D.~I.}\ \bibnamefont {{Jones}}},\ and\ \bibinfo {author} {\bibfnamefont {L.}~\bibnamefont {{Samuelsson}}},\ }\href {https://doi.org/10.1103/PhysRevLett.99.231101} {\bibfield  {journal} {\bibinfo  {journal} {\prl}\ }\textbf {\bibinfo {volume} {99}},\ \bibinfo {eid} {231101} (\bibinfo {year} {2007})},\ \Eprint {https://arxiv.org/abs/0708.2984} {arXiv:0708.2984 [gr-qc]} \BibitemShut {NoStop}%
\bibitem [{\citenamefont {{Pereira}}\ \emph {et~al.}(2021)\citenamefont {{Pereira}}, \citenamefont {{Bejger}}, \citenamefont {{Tonetto}}, \citenamefont {{Lugones}}, \citenamefont {{Haensel}}, \citenamefont {{Zdunik}},\ and\ \citenamefont {{Sieniawska}}}]{2021ApJ...910..145P}%
  \BibitemOpen
  \bibfield  {author} {\bibinfo {author} {\bibfnamefont {J.~P.}\ \bibnamefont {{Pereira}}}, \bibinfo {author} {\bibfnamefont {M.}~\bibnamefont {{Bejger}}}, \bibinfo {author} {\bibfnamefont {L.}~\bibnamefont {{Tonetto}}}, \bibinfo {author} {\bibfnamefont {G.}~\bibnamefont {{Lugones}}}, \bibinfo {author} {\bibfnamefont {P.}~\bibnamefont {{Haensel}}}, \bibinfo {author} {\bibfnamefont {J.~L.}\ \bibnamefont {{Zdunik}}},\ and\ \bibinfo {author} {\bibfnamefont {M.}~\bibnamefont {{Sieniawska}}},\ }\href {https://doi.org/10.3847/1538-4357/abe633} {\bibfield  {journal} {\bibinfo  {journal} {\apj}\ }\textbf {\bibinfo {volume} {910}},\ \bibinfo {eid} {145} (\bibinfo {year} {2021})},\ \Eprint {https://arxiv.org/abs/2011.06361} {arXiv:2011.06361 [astro-ph.HE]} \BibitemShut {NoStop}%
\bibitem [{\citenamefont {{Chetyrkin}}\ and\ \citenamefont {{Khodjamirian}}(2006)}]{2006EPJC...46..721C}%
  \BibitemOpen
  \bibfield  {author} {\bibinfo {author} {\bibfnamefont {K.~G.}\ \bibnamefont {{Chetyrkin}}}\ and\ \bibinfo {author} {\bibfnamefont {A.}~\bibnamefont {{Khodjamirian}}},\ }\href {https://doi.org/10.1140/epjc/s2006-02508-8} {\bibfield  {journal} {\bibinfo  {journal} {European Physical Journal C}\ }\textbf {\bibinfo {volume} {46}},\ \bibinfo {pages} {721} (\bibinfo {year} {2006})},\ \Eprint {https://arxiv.org/abs/hep-ph/0512295} {arXiv:hep-ph/0512295 [hep-ph]} \BibitemShut {NoStop}%
\bibitem [{\citenamefont {{Particle Data Group}}\ \emph {et~al.}(2020)\citenamefont {{Particle Data Group}}, \citenamefont {{Zyla}}, \citenamefont {{Barnett}}, \citenamefont {{Beringer}}, \citenamefont {{Dahl}}, \citenamefont {{Dwyer}}, \citenamefont {{Groom}}, \citenamefont {{Lin}} \emph {et~al.}}]{2020PTEP.2020h3C01P}%
  \BibitemOpen
  \bibfield  {author} {\bibinfo {author} {\bibnamefont {{Particle Data Group}}}, \bibinfo {author} {\bibfnamefont {P.~A.}\ \bibnamefont {{Zyla}}}, \bibinfo {author} {\bibfnamefont {R.~M.}\ \bibnamefont {{Barnett}}}, \bibinfo {author} {\bibfnamefont {J.}~\bibnamefont {{Beringer}}}, \bibinfo {author} {\bibfnamefont {O.}~\bibnamefont {{Dahl}}}, \bibinfo {author} {\bibfnamefont {D.~A.}\ \bibnamefont {{Dwyer}}}, \bibinfo {author} {\bibfnamefont {D.~E.}\ \bibnamefont {{Groom}}}, \bibinfo {author} {\bibfnamefont {C.~J.}\ \bibnamefont {{Lin}}}, \emph {et~al.},\ }\href {https://doi.org/10.1093/ptep/ptaa104} {\bibfield  {journal} {\bibinfo  {journal} {Progress of Theoretical and Experimental Physics}\ }\textbf {\bibinfo {volume} {2020}},\ \bibinfo {eid} {083C01} (\bibinfo {year} {2020})}\BibitemShut {NoStop}%
\bibitem [{\citenamefont {{Particle Data Group}}\ \emph {et~al.}(2024)\citenamefont {{Particle Data Group}}, \citenamefont {{Navas}}, \citenamefont {{Amsler}}, \citenamefont {{Gutsche}}, \citenamefont {{Hanhart}}, \citenamefont {{Hern{\'a}ndez-Rey}}, \citenamefont {{Louren{\c{c}}o}}, \citenamefont {{Masoni}} \emph {et~al.}}]{2024PhRvD.110c0001N}%
  \BibitemOpen
  \bibfield  {author} {\bibinfo {author} {\bibnamefont {{Particle Data Group}}}, \bibinfo {author} {\bibfnamefont {S.}~\bibnamefont {{Navas}}}, \bibinfo {author} {\bibfnamefont {C.}~\bibnamefont {{Amsler}}}, \bibinfo {author} {\bibfnamefont {T.}~\bibnamefont {{Gutsche}}}, \bibinfo {author} {\bibfnamefont {C.}~\bibnamefont {{Hanhart}}}, \bibinfo {author} {\bibfnamefont {J.~J.}\ \bibnamefont {{Hern{\'a}ndez-Rey}}}, \bibinfo {author} {\bibfnamefont {C.}~\bibnamefont {{Louren{\c{c}}o}}}, \bibinfo {author} {\bibfnamefont {A.}~\bibnamefont {{Masoni}}}, \emph {et~al.},\ }\href {https://doi.org/10.1103/PhysRevD.110.030001} {\bibfield  {journal} {\bibinfo  {journal} {\prd}\ }\textbf {\bibinfo {volume} {110}},\ \bibinfo {eid} {030001} (\bibinfo {year} {2024})}\BibitemShut {NoStop}%
\bibitem [{\citenamefont {{Fraga}}\ and\ \citenamefont {{Romatschke}}(2005)}]{2005PhRvD..71j5014F}%
  \BibitemOpen
  \bibfield  {author} {\bibinfo {author} {\bibfnamefont {E.~S.}\ \bibnamefont {{Fraga}}}\ and\ \bibinfo {author} {\bibfnamefont {P.}~\bibnamefont {{Romatschke}}},\ }\href {https://doi.org/10.1103/PhysRevD.71.105014} {\bibfield  {journal} {\bibinfo  {journal} {\prd}\ }\textbf {\bibinfo {volume} {71}},\ \bibinfo {eid} {105014} (\bibinfo {year} {2005})},\ \Eprint {https://arxiv.org/abs/hep-ph/0412298} {arXiv:hep-ph/0412298 [astro-ph]} \BibitemShut {NoStop}%
\bibitem [{\citenamefont {{Restrepo}}\ \emph {et~al.}(2025)\citenamefont {{Restrepo}}, \citenamefont {{Kneur}}, \citenamefont {{Provid{\^e}ncia}},\ and\ \citenamefont {{Benghi Pinto}}}]{2025arXiv250114935R}%
  \BibitemOpen
  \bibfield  {author} {\bibinfo {author} {\bibfnamefont {T.~E.}\ \bibnamefont {{Restrepo}}}, \bibinfo {author} {\bibfnamefont {J.-L.}\ \bibnamefont {{Kneur}}}, \bibinfo {author} {\bibfnamefont {C.}~\bibnamefont {{Provid{\^e}ncia}}},\ and\ \bibinfo {author} {\bibfnamefont {M.}~\bibnamefont {{Benghi Pinto}}},\ }\href {https://doi.org/10.48550/arXiv.2501.14935} {\bibfield  {journal} {\bibinfo  {journal} {arXiv e-prints}\ ,\ \bibinfo {eid} {arXiv:2501.14935}} (\bibinfo {year} {2025})},\ \Eprint {https://arxiv.org/abs/2501.14935} {arXiv:2501.14935 [hep-ph]} \BibitemShut {NoStop}%
\bibitem [{\citenamefont {{Ju}}\ \emph {et~al.}(2025)\citenamefont {{Ju}}, \citenamefont {{Chu}}, \citenamefont {{Wu}},\ and\ \citenamefont {{Liu}}}]{2025EPJC...85...40J}%
  \BibitemOpen
  \bibfield  {author} {\bibinfo {author} {\bibfnamefont {M.}~\bibnamefont {{Ju}}}, \bibinfo {author} {\bibfnamefont {P.}~\bibnamefont {{Chu}}}, \bibinfo {author} {\bibfnamefont {X.}~\bibnamefont {{Wu}}},\ and\ \bibinfo {author} {\bibfnamefont {H.}~\bibnamefont {{Liu}}},\ }\href {https://doi.org/10.1140/epjc/s10052-025-13788-9} {\bibfield  {journal} {\bibinfo  {journal} {European Physical Journal C}\ }\textbf {\bibinfo {volume} {85}},\ \bibinfo {eid} {40} (\bibinfo {year} {2025})},\ \Eprint {https://arxiv.org/abs/2404.14775} {arXiv:2404.14775 [nucl-th]} \BibitemShut {NoStop}%
\bibitem [{\citenamefont {{Chen}}\ and\ \citenamefont {{Lin}}(2023)}]{2023PhRvD.108f4007C}%
  \BibitemOpen
  \bibfield  {author} {\bibinfo {author} {\bibfnamefont {K.}~\bibnamefont {{Chen}}}\ and\ \bibinfo {author} {\bibfnamefont {L.-M.}\ \bibnamefont {{Lin}}},\ }\href {https://doi.org/10.1103/PhysRevD.108.064007} {\bibfield  {journal} {\bibinfo  {journal} {\prd}\ }\textbf {\bibinfo {volume} {108}},\ \bibinfo {eid} {064007} (\bibinfo {year} {2023})},\ \Eprint {https://arxiv.org/abs/2307.01598} {arXiv:2307.01598 [gr-qc]} \BibitemShut {NoStop}%
\bibitem [{\citenamefont {{Oikonomou}}\ and\ \citenamefont {{Moustakidis}}(2023)}]{2023PhRvD.108f3010O}%
  \BibitemOpen
  \bibfield  {author} {\bibinfo {author} {\bibfnamefont {P.~T.}\ \bibnamefont {{Oikonomou}}}\ and\ \bibinfo {author} {\bibfnamefont {C.~C.}\ \bibnamefont {{Moustakidis}}},\ }\href {https://doi.org/10.1103/PhysRevD.108.063010} {\bibfield  {journal} {\bibinfo  {journal} {\prd}\ }\textbf {\bibinfo {volume} {108}},\ \bibinfo {eid} {063010} (\bibinfo {year} {2023})},\ \Eprint {https://arxiv.org/abs/2304.12209} {arXiv:2304.12209 [astro-ph.HE]} \BibitemShut {NoStop}%
\bibitem [{\citenamefont {{Restrepo}}\ \emph {et~al.}(2023)\citenamefont {{Restrepo}}, \citenamefont {{Provid{\^e}ncia}},\ and\ \citenamefont {{Pinto}}}]{2023PhRvD.107k4015R}%
  \BibitemOpen
  \bibfield  {author} {\bibinfo {author} {\bibfnamefont {T.~E.}\ \bibnamefont {{Restrepo}}}, \bibinfo {author} {\bibfnamefont {C.}~\bibnamefont {{Provid{\^e}ncia}}},\ and\ \bibinfo {author} {\bibfnamefont {M.~B.}\ \bibnamefont {{Pinto}}},\ }\href {https://doi.org/10.1103/PhysRevD.107.114015} {\bibfield  {journal} {\bibinfo  {journal} {\prd}\ }\textbf {\bibinfo {volume} {107}},\ \bibinfo {eid} {114015} (\bibinfo {year} {2023})},\ \Eprint {https://arxiv.org/abs/2212.11184} {arXiv:2212.11184 [hep-ph]} \BibitemShut {NoStop}%
\bibitem [{\citenamefont {{Pang}}\ \emph {et~al.}(2021)\citenamefont {{Pang}}, \citenamefont {{Tews}}, \citenamefont {{Coughlin}}, \citenamefont {{Bulla}}, \citenamefont {{Van Den Broeck}},\ and\ \citenamefont {{Dietrich}}}]{2021ApJ...922...14P}%
  \BibitemOpen
  \bibfield  {author} {\bibinfo {author} {\bibfnamefont {P.~T.~H.}\ \bibnamefont {{Pang}}}, \bibinfo {author} {\bibfnamefont {I.}~\bibnamefont {{Tews}}}, \bibinfo {author} {\bibfnamefont {M.~W.}\ \bibnamefont {{Coughlin}}}, \bibinfo {author} {\bibfnamefont {M.}~\bibnamefont {{Bulla}}}, \bibinfo {author} {\bibfnamefont {C.}~\bibnamefont {{Van Den Broeck}}},\ and\ \bibinfo {author} {\bibfnamefont {T.}~\bibnamefont {{Dietrich}}},\ }\href {https://doi.org/10.3847/1538-4357/ac19ab} {\bibfield  {journal} {\bibinfo  {journal} {\apj}\ }\textbf {\bibinfo {volume} {922}},\ \bibinfo {eid} {14} (\bibinfo {year} {2021})},\ \Eprint {https://arxiv.org/abs/2105.08688} {arXiv:2105.08688 [astro-ph.HE]} \BibitemShut {NoStop}%
\bibitem [{\citenamefont {{M{\"u}ller}}\ \emph {et~al.}(2025)\citenamefont {{M{\"u}ller}}, \citenamefont {{Heger}},\ and\ \citenamefont {{Powell}}}]{2025PhRvL.134g1403M}%
  \BibitemOpen
  \bibfield  {author} {\bibinfo {author} {\bibfnamefont {B.}~\bibnamefont {{M{\"u}ller}}}, \bibinfo {author} {\bibfnamefont {A.}~\bibnamefont {{Heger}}},\ and\ \bibinfo {author} {\bibfnamefont {J.}~\bibnamefont {{Powell}}},\ }\href {https://doi.org/10.1103/PhysRevLett.134.071403} {\bibfield  {journal} {\bibinfo  {journal} {\prl}\ }\textbf {\bibinfo {volume} {134}},\ \bibinfo {eid} {071403} (\bibinfo {year} {2025})},\ \Eprint {https://arxiv.org/abs/2407.08407} {arXiv:2407.08407 [astro-ph.HE]} \BibitemShut {NoStop}%
\bibitem [{\citenamefont {{Vinciguerra}}\ \emph {et~al.}(2023)\citenamefont {{Vinciguerra}}, \citenamefont {{Salmi}}, \citenamefont {{Watts}}, \citenamefont {{Choudhury}}, \citenamefont {{Kini}},\ and\ \citenamefont {{Riley}}}]{2023ApJ...959...55V}%
  \BibitemOpen
  \bibfield  {author} {\bibinfo {author} {\bibfnamefont {S.}~\bibnamefont {{Vinciguerra}}}, \bibinfo {author} {\bibfnamefont {T.}~\bibnamefont {{Salmi}}}, \bibinfo {author} {\bibfnamefont {A.~L.}\ \bibnamefont {{Watts}}}, \bibinfo {author} {\bibfnamefont {D.}~\bibnamefont {{Choudhury}}}, \bibinfo {author} {\bibfnamefont {Y.}~\bibnamefont {{Kini}}},\ and\ \bibinfo {author} {\bibfnamefont {T.~E.}\ \bibnamefont {{Riley}}},\ }\href {https://doi.org/10.3847/1538-4357/acf9a0} {\bibfield  {journal} {\bibinfo  {journal} {\apj}\ }\textbf {\bibinfo {volume} {959}},\ \bibinfo {eid} {55} (\bibinfo {year} {2023})},\ \Eprint {https://arxiv.org/abs/2308.08409} {arXiv:2308.08409 [astro-ph.HE]} \BibitemShut {NoStop}%
\bibitem [{\citenamefont {{de Lima}}\ \emph {et~al.}(2020)\citenamefont {{de Lima}}, \citenamefont {{Coelho}}, \citenamefont {{Pereira}}, \citenamefont {{Rodrigues}},\ and\ \citenamefont {{Rueda}}}]{2020ApJ...889..165D}%
  \BibitemOpen
  \bibfield  {author} {\bibinfo {author} {\bibfnamefont {R.~C.~R.}\ \bibnamefont {{de Lima}}}, \bibinfo {author} {\bibfnamefont {J.~G.}\ \bibnamefont {{Coelho}}}, \bibinfo {author} {\bibfnamefont {J.~P.}\ \bibnamefont {{Pereira}}}, \bibinfo {author} {\bibfnamefont {C.~V.}\ \bibnamefont {{Rodrigues}}},\ and\ \bibinfo {author} {\bibfnamefont {J.~A.}\ \bibnamefont {{Rueda}}},\ }\href {https://doi.org/10.3847/1538-4357/ab65f4} {\bibfield  {journal} {\bibinfo  {journal} {\apj}\ }\textbf {\bibinfo {volume} {889}},\ \bibinfo {eid} {165} (\bibinfo {year} {2020})},\ \Eprint {https://arxiv.org/abs/1912.12336} {arXiv:1912.12336 [astro-ph.SR]} \BibitemShut {NoStop}%
\bibitem [{\citenamefont {{de Lima}}\ \emph {et~al.}(2024)\citenamefont {{de Lima}}, \citenamefont {{Pereira}}, \citenamefont {{Coelho}}, \citenamefont {{Nunes}}, \citenamefont {{Stecchini}}, \citenamefont {{Castro}}, \citenamefont {{Gomes}}, \citenamefont {{da Silva}}, \citenamefont {{Rodrigues}}, \citenamefont {{de Araujo}}, \citenamefont {{Bejger}}, \citenamefont {{Haensel}},\ and\ \citenamefont {{Zdunik}}}]{2024JHEAp..42...52D}%
  \BibitemOpen
  \bibfield  {author} {\bibinfo {author} {\bibfnamefont {R.~C.~R.}\ \bibnamefont {{de Lima}}}, \bibinfo {author} {\bibfnamefont {J.~P.}\ \bibnamefont {{Pereira}}}, \bibinfo {author} {\bibfnamefont {J.~G.}\ \bibnamefont {{Coelho}}}, \bibinfo {author} {\bibfnamefont {R.~C.}\ \bibnamefont {{Nunes}}}, \bibinfo {author} {\bibfnamefont {P.~E.}\ \bibnamefont {{Stecchini}}}, \bibinfo {author} {\bibfnamefont {M.}~\bibnamefont {{Castro}}}, \bibinfo {author} {\bibfnamefont {P.}~\bibnamefont {{Gomes}}}, \bibinfo {author} {\bibfnamefont {R.~R.}\ \bibnamefont {{da Silva}}}, \bibinfo {author} {\bibfnamefont {C.~V.}\ \bibnamefont {{Rodrigues}}}, \bibinfo {author} {\bibfnamefont {J.~C.~N.}\ \bibnamefont {{de Araujo}}}, \bibinfo {author} {\bibfnamefont {M.}~\bibnamefont {{Bejger}}}, \bibinfo {author} {\bibfnamefont {P.}~\bibnamefont {{Haensel}}},\ and\ \bibinfo {author} {\bibfnamefont {J.~L.}\ \bibnamefont {{Zdunik}}},\ }\href {https://doi.org/10.1016/j.jheap.2024.04.001} {\bibfield  {journal} {\bibinfo  {journal} {Journal of
  High Energy Astrophysics}\ }\textbf {\bibinfo {volume} {42}},\ \bibinfo {pages} {52} (\bibinfo {year} {2024})},\ \Eprint {https://arxiv.org/abs/2210.06648} {arXiv:2210.06648 [astro-ph.HE]} \BibitemShut {NoStop}%
\bibitem [{\citenamefont {{Holdom}}\ \emph {et~al.}(2018)\citenamefont {{Holdom}}, \citenamefont {{Ren}},\ and\ \citenamefont {{Zhang}}}]{2018PhRvL.120v2001H}%
  \BibitemOpen
  \bibfield  {author} {\bibinfo {author} {\bibfnamefont {B.}~\bibnamefont {{Holdom}}}, \bibinfo {author} {\bibfnamefont {J.}~\bibnamefont {{Ren}}},\ and\ \bibinfo {author} {\bibfnamefont {C.}~\bibnamefont {{Zhang}}},\ }\href {https://doi.org/10.1103/PhysRevLett.120.222001} {\bibfield  {journal} {\bibinfo  {journal} {\prl}\ }\textbf {\bibinfo {volume} {120}},\ \bibinfo {eid} {222001} (\bibinfo {year} {2018})},\ \Eprint {https://arxiv.org/abs/1707.06610} {arXiv:1707.06610 [hep-ph]} \BibitemShut {NoStop}%
\bibitem [{\citenamefont {{Rocha}}\ \emph {et~al.}(2023)\citenamefont {{Rocha}}, \citenamefont {{Horvath}}, \citenamefont {{de S{\'a}}}, \citenamefont {{Chinen}}, \citenamefont {{Bar{\~a}o}},\ and\ \citenamefont {{de Avellar}}}]{2023Univ...10....3R}%
  \BibitemOpen
  \bibfield  {author} {\bibinfo {author} {\bibfnamefont {L.~S.}\ \bibnamefont {{Rocha}}}, \bibinfo {author} {\bibfnamefont {J.~E.}\ \bibnamefont {{Horvath}}}, \bibinfo {author} {\bibfnamefont {L.~M.}\ \bibnamefont {{de S{\'a}}}}, \bibinfo {author} {\bibfnamefont {G.~Y.}\ \bibnamefont {{Chinen}}}, \bibinfo {author} {\bibfnamefont {L.~G.}\ \bibnamefont {{Bar{\~a}o}}},\ and\ \bibinfo {author} {\bibfnamefont {M.~G.~B.}\ \bibnamefont {{de Avellar}}},\ }\href {https://doi.org/10.3390/universe10010003} {\bibfield  {journal} {\bibinfo  {journal} {Universe}\ }\textbf {\bibinfo {volume} {10}},\ \bibinfo {eid} {3} (\bibinfo {year} {2023})},\ \Eprint {https://arxiv.org/abs/2312.13244} {arXiv:2312.13244 [astro-ph.HE]} \BibitemShut {NoStop}%
\end{thebibliography}%

\end{document}